\documentclass[aps,twocolumn,preprintnumbers]{revtex4}

\usepackage{graphicx}  
\usepackage{subfigure}
\usepackage{multirow}

\usepackage{fancyhdr}
\usepackage{longtable}
\usepackage{xcolor}
\usepackage{parskip}
\usepackage[T1]{fontenc}
\usepackage{dcolumn}   
\usepackage{amsmath,amsthm,amssymb}

\usepackage{bm}        
\usepackage{amsfonts}  
\usepackage{amsmath}   
\usepackage{amssymb}   
\usepackage{dsfont}

\usepackage[T1]{fontenc}
\usepackage[utf8]{inputenc}
\usepackage{lmodern}
\usepackage[%
    colorlinks=true,
    pdfborder={0 0 0},
    linkcolor=red
]{hyperref}

\usepackage{natbib}

\newcommand{\bra}[1]{\left\langle #1 \right|}

\newcommand{\pwisein}{\left\{ \begin{array}{ll}}
\newcommand{\pwiseout}{\end{array}\right.}
\newcommand{\ket}[1]{\left| #1 \right\rangle}

\begin{document}

\title{Entanglement and Teleportation in a 1-D Network with Repeaters}

\author{Ganesh Mylavarapu $^{*}$ , Indranil Chakrabarty $^{*}$, Kaushiki Mukherjee $^{**}$ , Minyi Huang$^{+}$, Junde Wu $^{++}$}

\affiliation { $^{*}$ Center for Quantum Science and Technology, International Institute of Information Technology, Hyderabad,}
\affiliation {$^{**}$ Department of Mathematics, Government Girls’ General Degree College, Ekbalpore, Kolkata-700023, India}
\affiliation {$^{+}$ Department of Mathematical Sciences, Zhejiang Sci-Tech University, Hangzhou 310018, PR China}
\affiliation {$^{++}$ School of Mathematical Sciences, Zhejiang University, Hangzhou 310027, PR-China}

\begin{abstract}  
The most simplest form of quantum network is an one dimensional quantum network with a single player in each node. In remote entanglement distribution  each of the players carry out measurement  at the intermediate nodes to produce an entangled state between initial and final node which are remotely separated. It is imperative to say that the flow of information as well as the percolation of entanglement in a network between the source and target node is an important area of study.  This will help us to understand the limits of the resource states as well as the measurements that are carried out in the process of remote entanglement distribution. In this article we investigate  how the concurrence of the final entangled state obtained is connected with the concurrences of the initial entangled states present in a  1-D chain. We extend the works done for the pure entangled states for mixed entangled states like Werner states, Bell diagonal states and for general mixed states. We did not limit ourselves to a situation where the measurements are happening perfectly. We also investigate how these relations change when we consider imperfect swapping. We obtain the limits on the number of swappings as well as  the success probability measurements to ensure the final state to be entangled state after swapping. In addition to these we also investigate on how much  quantum information can be sent from the initial node  to the final node (by computing the teleportation fidelity) when the measurement is perfect and imperfect with the same set of examples. Here also we obtain the limits on the number of swapping and the success probability of measurement to ensure that the final state obtained is capable of transferring the information .  These results have tremendous future applications in sending quantum information between two quantum processors in remote entangled distribution.

\end{abstract}

\maketitle

\section{Introduction}
Quantum entanglement \cite{1}, which had been a part of philosophical interests for many decades, is now widely regarded as an  important resource for various quantum computing and information processing tasks. One of the significant achievements in the domain of quantum information theory is the creation of a quantum network between many quantum processors \cite{2,3,4,5,6,7,8,9,10,11}. Quantum entanglement is the key component of the construction of quantum networks \cite{12,13}.  In particular, in the ground station to ground station networks, quantum entanglement plays the role of a channel. Once the network is established, we can use this network for various information processing tasks. This task includes protocols like teleportation \cite{14,15}, remote state preparation \cite{16}, quantum cryptography \cite{17,18,19,20,21,22,23}, super dense coding \cite{24}, etc. Many researches were done in the context of theoretical and practical implementations of quantum networks. These include network topology involving multiple distant sources. In such a scenario,  the sources distribute physical systems to a subset of distant observers. One of the predominant ways of creating such a network is by broadcasting entangled resources\cite{25}.
If we consider the sources to be  independent of each other($n$-local assumption), we may able to generate  non $n$-local correlations  under appropriate  measurement \cite{26}. The simplest of them is the bi-local network. The non-classicality of such networks was also studied in the presence of noise and also imperfect measurements in $n$-local network scenarios\cite{27}. Further studies were carried out to maximize the overall length or scalability of Quantum Repeater Networks\cite{35}. Also spreading non-locality in Quantum Networks is extended to multipartite entanglement scenarios\cite{36, 37, 38}.\\ 

\noindent Establishing a quantum network between nodes with quantum processors is an active area of research. 1-D chain and 2-D chain networks are created with entanglement as a key resource between two nodes. To set up a quantum network at the very first step, we need to establish entanglement between two nodes, say $A$ and $B$. In general, what happens is that a two-qubit entangled state is created, say in the node $A$. Now one of the two qubits which are created at node $A$ is physically transmitted to node $B$. If node $B$ is remotely located in that case, we use subsequent repeater stations between $A$ and $B$ to facilitate the process of creating the entanglement between the nodes $A$ and $B$. This is achieved by the process of entanglement swapping. If we consider the simplest scenario, one repeater $R$ is between the nodes. Entangled states are shared between the nodes $A$ and $R$ and between the nodes $R$ and $B$. In the ideal scenario, these states are Bell states. Now if a Bell state measurement is carried out at the repeater station $R$, then as a consequence of the Bell measurement, nodes $A$ and $B$ get entangled. It is interesting to see that if we start with Bell states, the state produced between  $A$ and $B$ is also a Bell state. However, we may not have a Bell state to start with but any other entangled state. In that scenario, the final state between the remote nodes is yet another entangled state. At this point, one may argue that instead of doing that entanglement swapping with lesser entangled states, one can convert them to maximally entangled states by entanglement distillation \cite{28}. However, from the information-theoretic point of view, it is not always necessary to have maximally entangled states to demonstrate quantum advantage. In particular, for teleportation and super dense coding, there are states where you can observe quantum advantage in spite of the state not being maximally entangled \cite{29,30,31}. In addition to that, entanglement distillation always comes with a cost.

\noindent In this work, we consider a simple 1-D chain with resources between each party being two qubit mixed states in general. Here we find out the relations that exist in the context of the entanglement distribution between distant nodes by the standard swapping of the entangled resource states. Not only entanglement distribution but also investigate the involvement of the teleportation fidelity of the initial states with the final state obtained after swapping. Such relations were already studied in the context of pure states \cite{11,32,33}. Our work not only extends from pure state to mixed state but also from an ideal network scenario to a nonideal one characterized by imperfect measurements. To further clarify the idea, we start with mixed entangled states as resources in a 1-D chain. Our target is to create remote entanglement between the extreme (initial-final) points of the chain. In that process, our aim is to investigate the interrelationship between the concurrence of a set of initial states with that of the final entangled state created across the network. A similar investigation is also made for teleportation fidelity. This will give us control over the resource states, and we can choose them in such a way that it will give us a better output at the final step. Our results also include the limits on the number of swapping and the success probability of measurements, so the final output state after swapping is an entangled state and is capable of acting as a teleportation channel.

\noindent In section II, we consider the simplest scenario where we have only one repeater and carries out the Bell state measurement in only one station. We carry out both perfect as well as imperfect measurements. We repeat the same measurement types and resources in section III; however, we did not restrict ourselves to a tripartite scenario with a single measurement. Here we look into a multiparty scenario with many repeater stations in between. However, the resource states are all the same. In section IV, we change this criterion and repeat the process with different resource states in a multiparty scenario. In sections IV, V, and VI, we repeat this study for teleportation fidelity. Finally, we conclude in section VII.

\section{Relation of Concurrence for Tripartite Network in Remote Entangled Distribution}
Here we consider the most straightforward situation where Alice and Bob share a mixed entangled state $\rho_{12}$\, and Bob and Charlie share another two qubit mixed state $\rho_{23}.$\ To create entanglement between 1 and 3, Bob measures his qubits at his node. We calculate the relation between the initial and final concurrence of the entangled states before and after swapping. This relation will tell us the entanglement created between initial and final nodes via swapping. In this section, we find the relation between the initial and final states' concurrences. We investigate this for both perfect and imperfect measurement scenarios. First, we take examples like the Werner and Bell Diagonal states; then, we investigate in case of general mixed states. This section can be regarded as an extension of the work done for pure states in the articles \cite{11}.  

\begin{figure}[h]
    \centering
    \includegraphics[scale = 0.8]{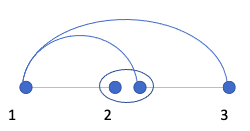}
    \caption{Entanglement swapping at node 2 to obtain an entangled state between initial and final nodes (1, 3)}
    \label{fig: single node swpping}
\end{figure}

\noindent \textbf{Imperfection in Measurements :}
In this work, we consider imperfect scenarios where the measurement devices fail to detect the particles with some probability. Let $\eta \in [0, 1]$ characterize the success probability of perfection in the measurement operator. Thus, The probability of failure is $1 - \eta$. Each of these measurement operators is given by

\begin{eqnarray}
    M_{i, 01}^{noisy} &=& \eta_{i}\ket{\Phi^{+}}\bra{\Phi^{+}} + \frac{1-\eta_{i}}{4}\mathds{I}_{4\times4},\nonumber\\ 
    M_{i, 01}^{noisy} &=& \eta_{i}\ket{\Phi^{-}}\bra{\Phi^{-}} + \frac{1-\eta_{i}}{4}\mathds{I}_{4\times4},\nonumber\\ 
    M_{i, 11}^{noisy} &=& \eta_{i}\ket{\Psi^{+}}\bra{\Psi^{+}} + \frac{1-\eta_{i}}{4}\mathds{I}_{4\times4},\nonumber\\
    M_{i, 11}^{noisy} &=& \eta_{i}\ket{\Psi^{-}}\bra{\Psi^{-}} + \frac{1-\eta_{i}}{4}\mathds{I}_{4\times4}.\label{eq:1}
\end{eqnarray}

\noindent Here $\{ \ket{\Phi^{\pm}}, \ket{\Psi^{\pm}}\}$ are Bell states and $I_{4 \times 4}$ is an identity matrix of order $4$. 

\subsection{Relations for Concurrence for Werner State}

Let $\rho_{12}$ be a Werner state between Alice and Bob and similarly $\rho_{23}$ be a Werner state between Bob and Charlie. These states are given by,
\begin{eqnarray}
\rho_{12} = \frac{1-p_{1}}{4}I + p_{1}\ket{\Psi^{-}}\bra{\Psi^{-}}, {}\nonumber\\
\rho_{23} = \frac{1-p_{2}}{4}I + p_{2}\ket{\Psi^{-}}\bra{\Psi^{-}}. 
\label{eq:2}
\end{eqnarray}
Here $p_{1}$ and $p_{2}$ are visible parameters with $\ket{\Psi^{-}}$ respectively.  The states are entangled when $p_1$ and $p_2$ are both greater than $\frac{1}{3}$. We will consider those entangled parts of the Werner states for our purpose. Since we want to create entanglement between Alice and Charlie, we start with entangled resource states. We can not create entanglement with two non-entangled resource states. Let $C_{12}$ and  $C_{23}$ be the concurrences  of the states $\rho_{12}$ and $\rho_{23}$. Let $C_{13}$ be the concurrence of the final state obtained after swapping. The values  of $C_{12}$ and  $C_{23}$ in terms of the visible parameter $p_1$ and $p_2$ are given by,
\begin{equation}
    C_{12} = \frac{(3p_{1}-1)}{2},
    \label{eq:3}
\end{equation}
\begin{equation}
    C_{23} = \frac{(3p_{2}-1)}{2}.
    \label{eq:4}
\end{equation}

\noindent Next, we consider two scenarios. The first scenario is when the measurement in the repeater node is perfect. In the second scenario, the measurement is imperfect.

\subsubsection{Perfect Measurement for Werner States}

\noindent Here we consider a scenario where the measurement is perfect. Then the final concurrence of the state $\rho_{13}$ in terms of initial concurrences is obtained as \\
\begin{equation}
    C_{13} = 2\frac{(3p_{1}p_{2}-1)}{(3p_{1}-1)(3p_{2}-1)}C_{12}C_{23}.
    \label{eq:5}
\end{equation}\\

\noindent The same expression can be written in terms of the visible parameter as, \\
\begin{equation}
    C_{13} = \frac{(3p_{1}p_{2}-1)}{2}.
    \label{eq:6}
\end{equation}\\

\noindent If we look at the above equation (\ref{eq:6}), we can conclude that if the product of visible parameters $p_{1}$ and $p_{2}$ is greater than $\frac{1}{3}$ only then the value of concurrence is greater than zero. Even if the individual visible parameters are greater than $\frac{1}{3}$, the concurrence is zero after swapping if the product of visible parameters is less than or equal to $\frac{1}{3}$. \\

\begin{figure}[h]
    \centering
    \includegraphics[scale = 0.6]{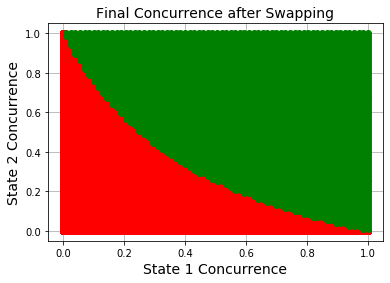}
    \caption{Concurrence for Werner State Single Node Swapping}
    \label{fig:perfect_werner}
\end{figure}

\noindent In Fig.(\ref{fig:perfect_werner}), we provide a numerical estimate of the states whose concurrence is greater than zero in comparison to the states (obtained after swapping) whose concurrence is equal to zero. Here, X-axis represents the concurrence of Werner state $\rho_{12}$, and Y-axis represents the concurrence of Werner state $\rho_{23}$ whose visible parameter values are in the range of zero to one increased by a value of 0.01. The green region represents concurrence which is greater than zero after swapping, whereas the red region represents concurrence which is zero after swapping.

\noindent \textbf{Note:} When the visible parameters $p_{1} = 1$ and $p_{2} = 1$, then initial Werner states become maximally entangled states and the final concurrence after entanglement swapping calculated using above formula is
\begin{eqnarray}
C_{13} = 1 . 
\end{eqnarray}
\noindent It can be observed that the final state after entanglement swapping is also a maximally entangled state with concurrence one, as expected. 
  
\subsubsection{Imperfect Measurement for Werner States}

Here we consider a scenario where the measurement is imperfect as described in equation (\ref{eq:1}) where the measurement operator fails to detect with probability $1 - \eta$.
In such a scenario the final concurrence in terms of probability of success $\eta$ is obtained as,

\begin{equation}
    C_{13} = \frac{2}{(4-3\eta)}\left[\frac{(\eta)(3p_{1}p_{2}-1) - 4(1 - \eta)}{(3p_{1}-1)(3p_{2}-1)}\right]C_{12}C_{23}.
    \label{eq:7}
\end{equation}

\noindent This can be written in terms of $\eta, p_{1}, p_{2}$  as \\
\begin{equation}
    C_{13} = \frac{1}{(4-3\eta)}\left[\frac{(\eta)(3p_{1}p_{2}-1) - 4(1 - \eta)}{2}\right].
    \label{eq:8}
\end{equation}

\noindent Considering equation (\ref{eq:8}) where the visible parameter values to be maximum, Concurrence becomes zero when

\begin{eqnarray}
    \frac{1}{(4-3\eta)}\frac{(\eta)(3p_{1}p_{2}-1) - 4(1 - \eta)}{2} & =  C_{13},{}\nonumber\\
    \frac{1}{(4-3\eta)}\frac{(\eta)(3p_{1}p_{2}-1) - 4(1 - \eta)}{2} & = 0.
\end{eqnarray}

\noindent By solving above equation, $\eta$ resolves to $\frac{2}{3}$. It can be observed that if the value of $\eta$ is less than or equal to $\frac{2}{3}$ irrespective of the visible parameter values, concurrence will always be zero. If the value of $\eta$ is greater than $\frac{2}{3}$, concurrence will be positive. In other words, if the Bell state measurement happens with a success probability less than or equal to $\frac{2}{3}$, then there is no entanglement in the output state. This clearly gives a bound in terms of $\eta$ in the process of establishment of an entanglement network in a 1-D chain.\\

\noindent \textbf{Note:} When $\eta = 1$ the equation reduces to
\begin{equation}
    {C_{13} = \left[\frac{(3p_{1}p_{2}-1)}{2}\right]}.
\end{equation}
\noindent This is equal to the concurrence of the output state in the case of a perfect measurement scenario of the entanglement swapping process. (\ref{eq:6}).

\begin {figure}[h]
    \centering
    \includegraphics[scale = 0.3]{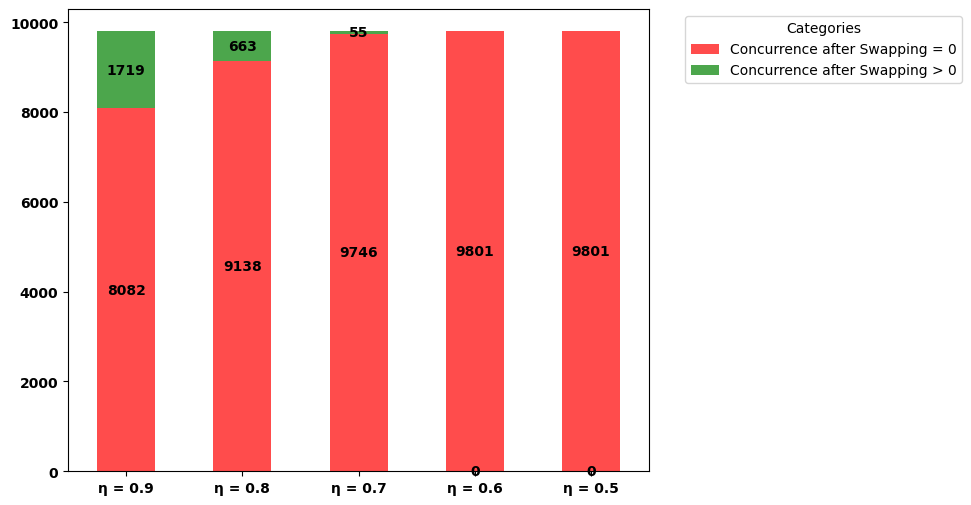}
    \caption{Werner State Single Node Imperfect Swapping}
    \label{fig:Imperfect_Werner}
\end{figure}

\noindent  In Fig.(\ref{fig:Imperfect_Werner}), we provide numerical data to estimate the number of output states of the swapping process whose concurrence is positive in comparison to states with zero concurrences. Here,  X-axis represents the success probability $\eta$  represented in the form of discrete values ranging from 0.0 to 1.0, and Y-axis represents number of distinct Werner states. The figure is made up of bar diagrams corresponding to each success probability. The red regions indicate the total number of states whose final concurrences are zero after swapping, and the green regions indicate the states whose final concurrences are greater than zero after swapping. It can be observed from the figure that as imperfection is introduced concurrence value decreases, and when $\eta = 0.6$ or less, irrespective of the visible parameter values, all states have zero concurrences.

\begin{figure}[h]
    \centering
    \includegraphics[scale = 0.4]{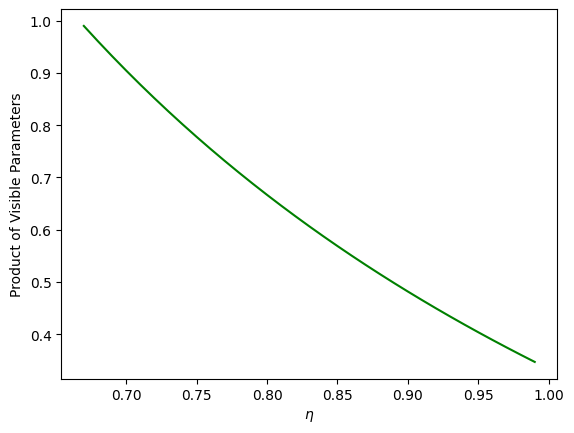}
    \caption{$\eta$ Vs Product of visible state parameters $p_{1}p_{2}$}
\label{fig:werner_imperfect_limits}
\end{figure}

Here we draw another Fig.(\ref{fig:werner_imperfect_limits}) to show the variation of the product of visible parameters with respect to success probability. Here,  X-axis represents $\eta$, and Y-axis represents the product of visible parameters.\\ 


\subsection{Relations on Concurrence for Bell Diagonal States}

Here in this subsection, we consider Bell diagonal states as a second example of mixed resource states. Consider three nodes, a source node (Alice), a repeater (Bob), and a target node (Charlie). Let $\rho_{12}$ be a Bell diagonal state between Alice and Bob and $\rho_{23}$ be a Bell diagonal state between Bob and Charlie. These states are,
\begin{eqnarray}
    \rho_{12} = \frac{1}{4}[I\otimes I + \sum_{i=1}^{3} p_{i}^{1}\sigma_{i} \otimes \sigma_{i}],\nonumber\\
    \rho_{23} = \frac{1}{4}[I\otimes I + \sum_{i=1}^{3} p_{i}^{2}\sigma_{i} \otimes \sigma_{i}],\\
\label{eq:9}
\end{eqnarray}
where $\sigma_{i}$ ($i=1,2,3$) are Pauli matrices, $p_{i}^{j}$ are visible parameters  and  $0 \leqslant |p_{i}^{j}| \leqslant 1$. Just like Werner state, here we consider two scenarios where the measurement in the repeater node is perfect and imperfect.

\subsubsection{Perfect Measurement}

We start with the case when the measurement is perfect. Then the final concurrence is obtained as

\begin{eqnarray}
    &&C_{13} = max
    {}\nonumber\\&&\left[0, \left(\frac{(\Lambda_{1} - \Lambda_{2} - \Lambda_{3} - \Lambda_{4})}{\prod\limits_{j=1}^{2} (\lambda_{1}^{j} - \lambda_{2}^{j} - \lambda_{3}^{j} - \lambda_{4}^{j})}\right)C_{12}C_{23}\right],
    \label{eq:10}
\end{eqnarray}

\noindent where $\Lambda_{1} \ge \Lambda_{2} \ge \Lambda_{3} \ge \Lambda_{4}$ and $\lambda_{1}^{j} \ge \lambda_{2}^{j} \ge \lambda_{3}^{j} \ge \lambda_{4}^{j}$. The values of $\Lambda_{i}$ ($i=1,2,3,4$) are
\begin{eqnarray}
 &&\Lambda_1=   \frac{1 + p_{1}^{1}p_{1}^{2} + p_{2}^{1}p_{2}^{2} + p_{3}^{1}p_{3}^{2}}{4},{}\nonumber\\&&
 \Lambda_2=  \frac{1 - p_{1}^{1}p_{1}^{2} - p_{2}^{1}p_{2}^{2} + p_{3}^{1}p_{3}^{2}}{4},{}\nonumber\\&&
 \Lambda_3=   \frac{1 + p_{1}^{1}p_{1}^{2} - p_{2}^{1}p_{2}^{2} - p_{3}^{1}p_{3}^{2}}{4},{}\nonumber\\&&
 \Lambda_4=   \frac{1 - p_{1}^{1}p_{1}^{2} + p_{2}^{1}p_{2}^{2} - p_{3}^{1}p_{3}^{2}}{4}
\end{eqnarray}

\noindent and the values of $\lambda_{i}$ ($i=1,2,3,4$) are
$\frac{1 + p_{1}^{j} - p_{2}^{j} + p_{3}^{j}}{4}$, 
$\frac{1 - p_{1}^{j} + p_{2}^{j} + p_{3}^{j}}{4}$, 
$\frac{1 + p_{1}^{j} + p_{2}^{j} - p_{3}^{j}}{4}$,
$\frac{1 - p_{1}^{j} - p_{2}^{j} - p_{3}^{j}}{4}$ respectively.
\\
\newline
 \begin{figure}[h]
    \centering
    \includegraphics[scale = 0.6]{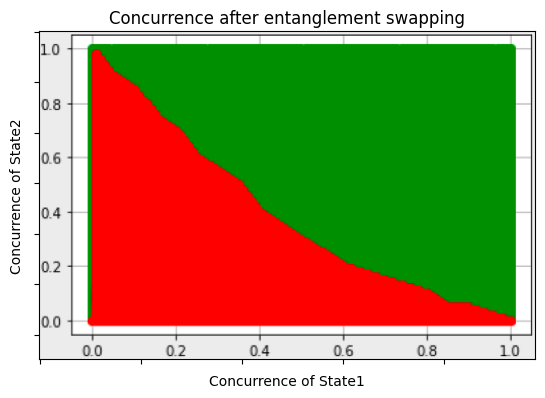}
    \caption{BDS Single Node Perfect Swapping}
    \label{fig:bds_perfect}
\end{figure}
\\

\noindent Fig.(\ref{fig:bds_perfect}) is plotted by considering all the Bell diagonal states where concurrence is greater than zero. Here in the figure, X-axis represents the concurrence of $\rho_{12}$, and Y-axis represents the concurrence of $\rho_{23}$. The values of input parameters $P_i$ ($i=1,2,3$) range from +1 to -1 with a graduation range of 0.1. 
All the initial states whose concurrence is greater than zero are considered as input to the swapping. If the final concurrence after entanglement swapping is zero, it is represented in red, and if the final concurrence is greater than zero, it is represented in green.\\

\noindent It is observed that the final state after entanglement swapping of Bell diagonal states is also a Bell diagonal state. The points of the Bell diagonal states form a Tetrahedron, and it follows that the vertices of the tetrahedron represent pure states. As the section represents states with maximally mixed subsystems,
the four pure states are maximally entangled: They are the 4 Bell states. The pairwise averages of the four corners of the tetrahedron give the six vertices of the octahedron, and all the points inside the octahedron are separable states whose concurrence is zero \cite{34}.

\noindent If the points lie inside the octahedron then $\lvert{t_{1}} \rvert + \lvert{t_{2}} \rvert + \lvert{t_{3}} \rvert \leq 1$ , where
$t_{1} = \frac{p_{00} + p_{01} - p_{10} - p_{11}}{4}$, 
$t_{2} = \frac{ - p_{00} + p_{01} + p_{10} - p_{11}}{4}$, 
$t_{3} = \frac{p_{00} - p_{01} + p_{10} - p_{11}}{4}$ and $p_{00} = \frac{1 + p_{1}^{1}p_{1}^{2} + p_{2}^{1}p_{2}^{2} + p_{3}^{1}p_{3}^{2}}{4}$, 
$p_{01} = \frac{1 + p_{1}^{1}p_{1}^{2} - p_{2}^{1}p_{2}^{2} - p_{3}^{1}p_{3}^{2}}{4}$, 
$p_{10} = \frac{1 - p_{1}^{1}p_{1}^{2} - p_{2}^{1}p_{2}^{2} + p_{3}^{1}p_{3}^{2}}{4}$,
$p_{11} = \frac{1 - p_{1}^{1}p_{1}^{2} + p_{2}^{1}p_{2}^{2} - p_{3}^{1}p_{3}^{2}}{4}$ \\

\subsubsection{Imperfect Measurement}

Here we consider a scenario where the measurement is imperfect, where the measurement is successful with probability $\eta$ and fails with probability $1 - \eta$.  The final concurrence of the state $\rho_{13}$ is given as
\begin{eqnarray}
    &&C_{13} = max {}\nonumber\\&& \left[ 0, 
    \frac{1}{(4-3\eta)}\left(\frac{(\Lambda_{1} - \Lambda_{2} - \Lambda_{3} - \Lambda_{4})}{\prod\limits_{j=1}^{2} (\lambda_{1}^{j} - \lambda_{2}^{j} - \lambda_{3}^{j} - \lambda_{4}^{j})}\right)C_{12}C_{23}\right], {}\nonumber\\&&
\end{eqnarray}
\noindent where $\Lambda_{1} \ge \Lambda_{2} \ge \Lambda_{3} \ge \Lambda_{4}$ and $\lambda_{1}^{j} \ge \lambda_{2}^{j} \ge \lambda_{3}^{j} \ge \lambda_{4}^{j}$. The values of $\Lambda_{i}$ ($i=1,2,3,4$) are
\begin{eqnarray}
&&\Lambda_1={}\nonumber\\&&  \frac{(\eta)(1 + p_{1}^{1}p_{1}^{2} + p_{2}^{1}p_{2}^{2} + p_{3}^{1}p_{3}^{2}) + 4(1-\eta)}{4},  {}\nonumber\\&&
\Lambda_2={}\nonumber\\&&  \frac{(\eta)(1 - p_{1}^{1}p_{1}^{2} - p_{2}^{1}p_{2}^{2} + p_{3}^{1}p_{3}^{2}) + 4(1-\eta)}{4},  {}\nonumber\\&&
\Lambda_3={}\nonumber\\&&  \frac{(\eta)(1 + p_{1}^{1}p_{1}^{2} - p_{2}^{1}p_{2}^{2} - p_{3}^{1}p_{3}^{2} ) + 4(1-\eta)}{4},  {}\nonumber\\&&
\Lambda_4={}\nonumber\\&&   \frac{(\eta)(1 - p_{1}^{1}p_{1}^{2} + p_{2}^{1}p_{2}^{2} - p_{3}^{1}p_{3}^{2} ) + 4(1-\eta)}{4} 
\end{eqnarray}
\noindent and the values of $\lambda_{i}$ ($i=1,2,3,4$) are
$\frac{1 + p_{1}^{j} - p_{2}^{j} + p_{3}^{j}}{4}$, 
$\frac{1 - p_{1}^{j} + p_{2}^{j} + p_{3}^{j}}{4}$, 
$\frac{1 + p_{1}^{j} + p_{2}^{j} - p_{3}^{j}}{4}$,
$\frac{1 - p_{1}^{j} - p_{2}^{j} - p_{3}^{j}}{4}$.\\

\noindent \textbf{Note :} In order to identify the value of $\eta$ where the total concurrence is zero irrespective of the maximum values of visible parameters,

\begin{eqnarray}
&&\Lambda_1=\frac{(\eta)(4) + 4(1-\eta)}{4}, {}\nonumber\\&& 
\Lambda_2=\frac{4(1-\eta)}{4},  {}\nonumber\\&&
\Lambda_3=\frac{4(1-\eta)}{4}, {}\nonumber\\&&
\Lambda_4=\frac{4(1-\eta)}{4}. 
\end{eqnarray}

\noindent Thus the expression $\Lambda_1 - \Lambda_2 - \Lambda_3 - \Lambda_4$  becomes  equal to  $\frac{4\eta - 8(1 - \eta)}{4}$. If the concurrence is zero, then we have the value of $\eta$ as
\begin{eqnarray}
    &&\frac{4\eta - 8(1 - \eta)}{4} = 0, {}\nonumber\\&&
    4\eta - 8 + 8\eta = 0, {}\nonumber\\&&
    \eta = \frac{2}{3}.
\end{eqnarray}

\noindent If the value of $\eta$ is greater than $\frac{2}{3}$, then the final concurrence is greater than zero for some visible parameters. If the value of $\eta$ is less than or equal to $\frac{2}{3}$ irrespective of the values of visible parameters, the final concurrence will always be zero.


\begin{figure}[h]
    \centering
    \includegraphics[scale = 0.33]{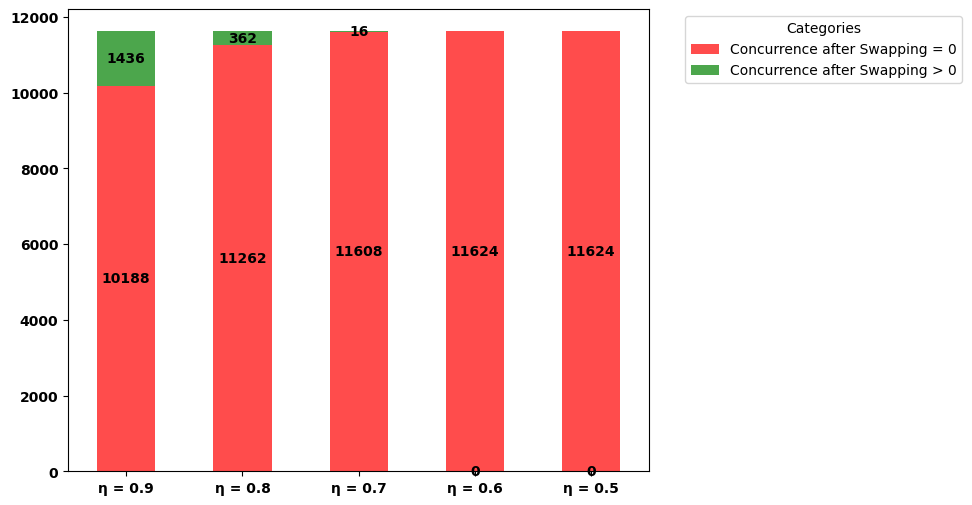}
    \caption{Bell Diagonal State Single Node Imperfect Swapping }
    \label{fig:label_5}
\end{figure}

\noindent In Fig.(\ref{fig:label_5}), we make a similar comparison on Bell diagonal states subject to imperfect measurement scenarios. Here the figure represents the concurrence of states after entanglement swapping for given $\eta$ as X-axis and Y-axis represent the number of states considered for entanglement swapping. Approximately 12$\times 10^{3}$ Bell Diagonal states with concurrence greater than zero are considered. All the red regions in the graph represent the states with concurrence zero after entanglement swapping, and the green regions represent the states with concurrence greater than zero after entanglement swapping. It can be observed from the graph that if the value of $\eta$ (success probability of measurement) is less than 0.7, all the values of concurrence are zero after entanglement swapping, irrespective of the values of visible parameters. This is in accordance with the analytical result for the Bell diagonal states obtained above.

\subsection{Relations on Concurrence for General Mixed State}
In this subsection we consider the most general scenario where each of the resource state between Alice (Source) and Bob (Repeater) and between Bob (Repeater) and Charlie (Target) is a two qubit mixed state given by,
\begin{equation}
 \rho = \frac{1}{4}(I \otimes I + \sum\limits_{i} r_i.\sigma_{i} \otimes I+ \sum\limits_{i} s_{i}.I \otimes \sigma_{i} + \sum\limits_{ij} t_{ij} \sigma_{i} \otimes \sigma_{j})   
 \label{eq:21}
\end{equation}

\noindent where $\sigma_{i} = (\sigma_{1}, \sigma_{2}, \sigma_{3})$ are the Pauli matrices; $r_{i} = (r_{1}, r_{2}, r_{3})$, $s_{i} = (s_{1}, s_{2}, s_{3})$ are the local Bloch vectors and $t_{ij}=Tr(\rho( \sigma_{i} \otimes \sigma_{j} ))$ are the elements of the correlation matrix T = $[t_{ij}]_{3\times3}$.\\

\noindent In this subsection, we did a numerical study to identify the general mixed states which have concurrence values greater than zero after entanglement swapping. This is done in both perfect and imperfect scenarios.

\subsubsection{Perfect Measurement}

Here we consider a scenario where the measurement is perfect. The graph is generated by considering two qubit pairs whose concurrences range from 0 to 1 and then calculating concurrence after single-node entanglement swapping. All the points in the red region are not useful for data transmission from the source node to the target node, whereas all the points in the green region are useful for data transmission, at least in the context of measures of entanglement. \\

\begin{figure}[h]
    \centering
    \includegraphics[scale = 0.6]{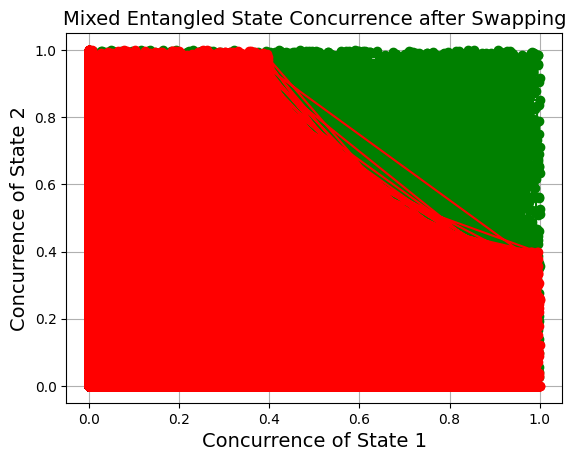}
    \caption{Mixed State Entanglement for Tripartite Network }
    \label{fig:mixed_state_label}
\end{figure}

\noindent In Fig.(\ref{fig:mixed_state_label}) X-axis represents concurrence of $\rho_{12}$ and Y - axis represents concurrence of $\rho_{23}$. The green region represents concurrence greater than zero after entanglement swapping, and the red region represents concurrence equal to zero after entanglement swapping. For the density matrices considered for mixed state swapping, more than twenty percent of the states are useful after entanglement swapping.

\subsubsection{Imperfect Measurement}

\begin{figure}[h]
    \centering
    \includegraphics[scale = 0.3]{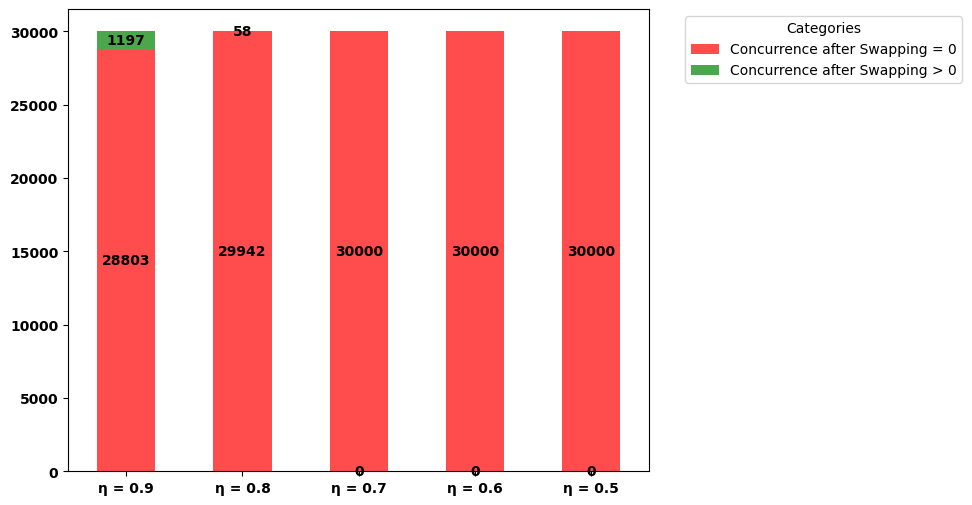}
    \caption{Mixed State Single Node Imperfect Swapping}
    \label{fig:multi_node_imperfect}
\end{figure}

In Fig.(\ref{fig:multi_node_imperfect}), we do the same analysis on the largest class of generalized mixed state for imperfect scenarios. Here, the figure represents concurrence after entanglement swapping for a given $\eta$ labeled on the X-axis. The values of $\eta$ range from 0 to 0.9, and the Y-axis represents a number of distinct general mixed states. Here we consider a scenario where the measurement is imperfect, and this imperfection is characterized by probability $1 - \eta$. There are $30 \times 10^{3}$ distinct mixed states considered, and it can be observed that when the value of success probability $\eta$ is 0.7, all the states have concurrence equal to zero after entanglement swapping.

\section{Relations on Concurrence for Multipartite Network in Remote
Entangled Distribution (RED) with similar states}

This section studies a general situation in a 1-D network where we have more than three nodes. In particular, the entangled states shared by the consecutive nodes are the same. We need to establish entanglement between the initial and final nodes. We consider the process of entanglement swapping as a technique for remote entanglement distribution (RED). Let us assume that we have $(n + 1)$ entangled states with $n + 2$ nodes. In order to obtain an entangled state between initial and final nodes, we carry out $n$ number of entanglement swapping. We consider sequential measurements to create successive entanglements between the nodes $(1, 3)$, $(1, 4)$, and finally between the nodes $(1, n + 2)$ as shown in the figure and obtain the extension of the relationships involving the concurrences of initial and final entangled states.

\begin{figure}[h]
    \centering
    \includegraphics[scale = 0.4]{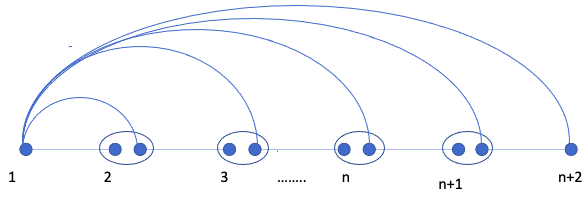}
    \caption{Entanglement swapping is done with sequential measurement at the nodes (2, 3... n+1) to obtain an entangled state between initial and final nodes (1, n+2)}
    \label{fig:my_label}
\end{figure}

\subsection{Relations on Concurrence for Werner States }

In this subsection, we consider a 1-D network of  $n+2$ nodes with a source node, $n$ repeater nodes, and a target node. The interesting part is that all these consecutive nodes share the same Werner state with a visible parameter $p$ between them. Let  us consider $\rho_{i, i+1}$ to be the  Werner state between $i^{th}$ node and $(i+1)^{th}$ node with visible parameters $p$. All other Werner states between the nodes will be the same with the same visible parameter.

\subsubsection{Perfect Measurement}
Here we consider a scenario where the measurement done at the repeater's station is perfect. Then the final concurrence  between the source ($1$) and the target node $(n+2)$ is obtained as,
\begin{equation}
    C_{1, n+2} = 2^{n}\left[\frac{[3(p^{n+1}) - 1](\prod\limits_{i = 1}^{n+1} C_{i, i+1})}{(3p - 1)^{n+1}}\right].
    \label{eq:31}
\end{equation}

The final concurrence in terms of visible parameters can be written as 
\begin{equation}
    C_{1, n+2} = \left[\frac{[3(p^{n+1}) - 1]}{2}\right].
    \label{eq:32}
\end{equation}

When there is only one repeater node $n=1$, the above equation [\ref{eq:32}] reduces to $\frac{3p^{2}-1}{2}$ which is the same as the concurrence for single node swapping with the same visible parameters.


 \begin{figure}[h]
    \centering
    \includegraphics[scale = 0.36]{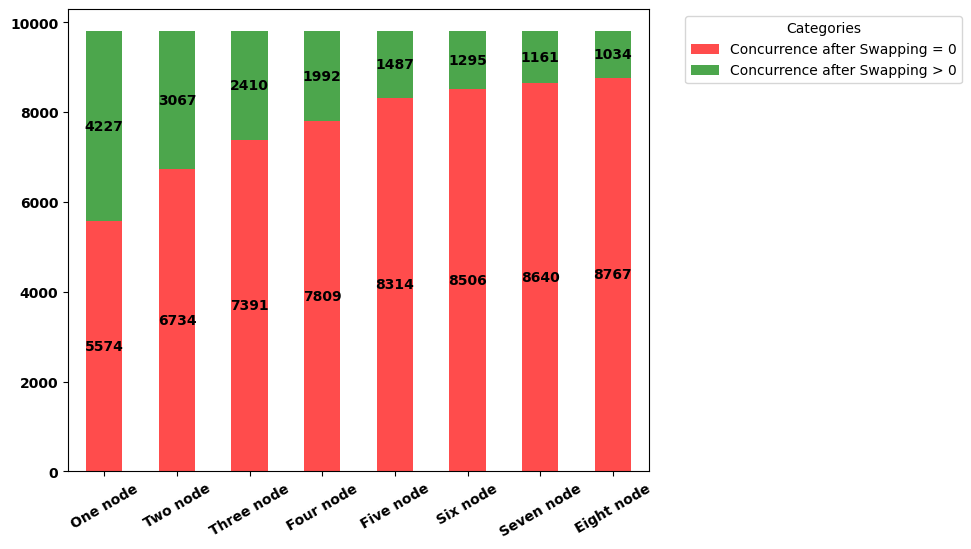}
    \caption{Werner State Multi Node Perfect Swapping}
    \label{fig:werner_multi_mode_perfect}
\end{figure}

\noindent The above Fig.(\ref{fig:werner_multi_mode_perfect}) is plotted for Werner states having the same visible parameter. There are approximately $10 \times 10^{3}$ different values of visible parameters considered to generate the cluster. Here, X-axis represents cases with different numbers of swappings that are considered, and Y-axis represents the total number of states. All the values in the green regions have concurrence greater than zero after swappings, and the numbers in red regions have concurrence equal to zero. 

\subsubsection{Imperfect Measurement}

In this subsection, we study the relationship of the concurrences of the initial resources with the final state when the measurement is not perfect at each repeater station. \\

\noindent \textbf{Case 1:} Here we consider a scenario where the imperfection is the same in all nodes. Let the imperfection parameter is $1 - \eta$ for each node. The final concurrence in terms of initial concurrence is defined as,

\begin{eqnarray}
&& C_{1, n+2} =  \frac{2^{n}}{N}{}\nonumber\\&&
\left[\frac{\eta^{n}[3(p^{n+1}) - 1] - 4(1 - \eta)(\sum\limits_{i=0}^{n-1}(\eta^{n-1-i})(\eta + 4(1 - \eta))^{i})}{(3p - 1)^{n+1}}\right]{}\nonumber\\&&
(\prod\limits_{i = 1}^{n+1} C_{i, i+1}),
\label{eq:22}
\end{eqnarray}

\noindent where $N = (\eta^{n} + 4(1 - \eta)(\sum\limits_{i=0}^{n-1}(\eta^{n-1-i})(\eta + 4(1 - \eta))^{i}))$. \\

\noindent Final concurrence in terms of visible parameter and $\eta$ can be expressed as

\begin{eqnarray}
&&C_{1, n+2} =  {}\nonumber\\&&
\left[\frac{\eta^{n}[3(p^{n+1}) - 1] - 4(1 - \eta)(\sum\limits_{i=0}^{n-1}(\eta^{n-1-i})(\eta + 4(1 - \eta))^{i})}{2N}\right],{}\nonumber\\&&
\end{eqnarray}

\noindent where $N = (\eta^{n} + 4(1 - \eta)(\sum\limits_{i=0}^{n-1}(\eta^{n-1-i})(\eta + 4(1 - \eta))^{i}))$. \\

\noindent \textbf{Note: }It can be observed from the above equation that when the value of $\eta = 1$, the whole equation turns out to be equation (\ref{eq:32}), which is the final concurrence for multi-node perfect swapping.

\begin{figure}[h]
    \centering
    \includegraphics[scale = 1]{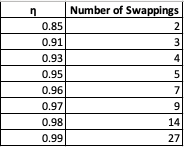}
    \caption{$\eta$ Vs number of swappings}
    \label{fig:eta_no_of_swappings}
\end{figure}

It can be observed from Table (\ref{fig:eta_no_of_swappings}) that if the number of swapping nodes is greater than 27, then it should be perfect swapping to have the non-zero value of the concurrence for the final state. Similarly, if the value of $\eta$ is 0.85, then swapping is restricted to 2 nodes other than the source and target nodes.\\

 \begin{figure}[h]
    \centering
    \includegraphics[scale = 0.24]{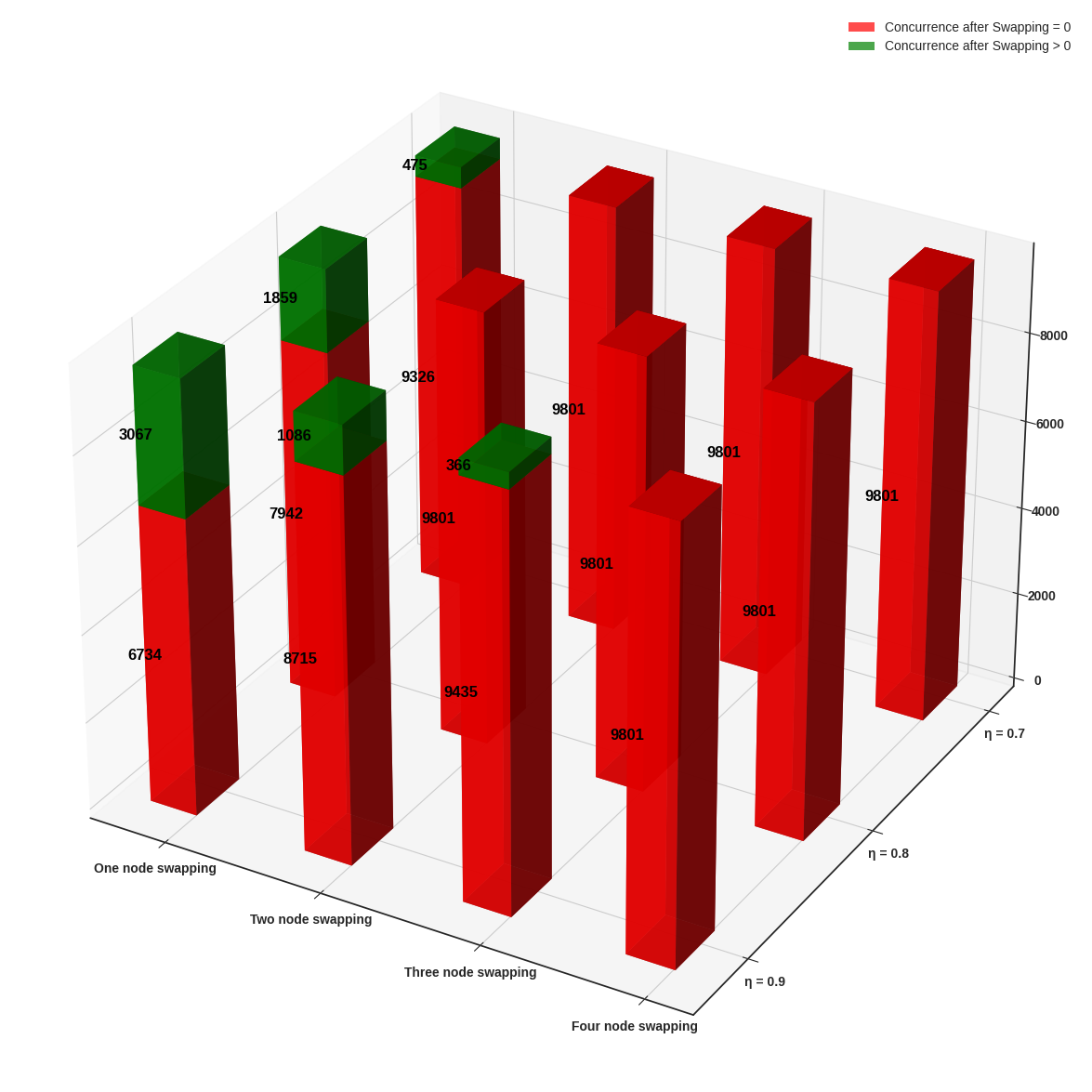}
    \caption{Werner State Multi Node Imperfect Swapping with Same Imperfection}
    \label{fig:werner_multinode_imperfect}
\end{figure}

\noindent In Fig.(\ref{fig:werner_multinode_imperfect}) X-axis represents the number of nodes being swapped, Y-axis represents the imperfection parameter $\eta$ represented as discrete points from 0.0 to 1.0, and Z-axis represents the number of distinct Werner states. It is observed that as the value of $\eta$ decreases below 0.7, concurrence is zero, irrespective of the value of visible parameters. When the value of $\eta$ is 0.9, it can be observed that the value of concurrence is greater than zero till three node swapping, whereas when the value of $\eta$ is 0.8, the value of concurrence is greater than zero till two node swappings and when $\eta$ is 0.7 concurrence is greater than zero only for a single node swapping. The results match with imperfect single node swappings for a given $\eta$ and visible parameter values.\\

\noindent \textbf{Case 2:} Here we consider a scenario where the imperfection is different in different nodes. Let the imperfections are $\eta_{1}, \eta_{2}, ... , \eta_{n}$ in $n$ nodes, then the final concurrence is defined as

\begin{eqnarray}
&&C_{1, n+2} = \frac{2^{n}}{N}{}\nonumber\\&&
\left[\frac{(\prod\limits_{i=1}^{n}\eta_{i})[3(p^{n+1}) - 1] - \sum\limits_{i=1}^{n}(\sum\limits_{cyc}(4^{i})(\prod\limits_{j=1}^{i}(1 - \eta_{i}))(\prod\limits_{k = i+1}^{n}\eta_{k}))}{(3p - 1)^{n+1}}\right]{}\nonumber\\&&
(\prod\limits_{i = 1}^{n+1} C_{i, i+1}),
\end{eqnarray}

\noindent where N = $(\prod\limits_{i=1}^{n}\eta_{i} + \sum\limits_{i=1}^{n}(\sum\limits_{cyc}(4^{i})(\prod\limits_{j=1}^{i}(1 - \eta_{i}))(\prod\limits_{k = i+1}^{n}\eta_{k})))$.\\

\noindent \textbf{Note1: } When all values of $\eta_{i} = \eta$, the whole Equation resolves to Equation (\ref{eq:22}), which is the concurrence for multi-node imperfect swapping with the same imperfection in all nodes.\\

\noindent \textbf{Note2: } When all values of $\eta_{i} = 1$, the whole equation resolves to Equation (\ref{eq:32}), which is the concurrence for multi-node perfect swapping.

 \begin{figure}[h]
    \centering
    \includegraphics[scale = 0.38]{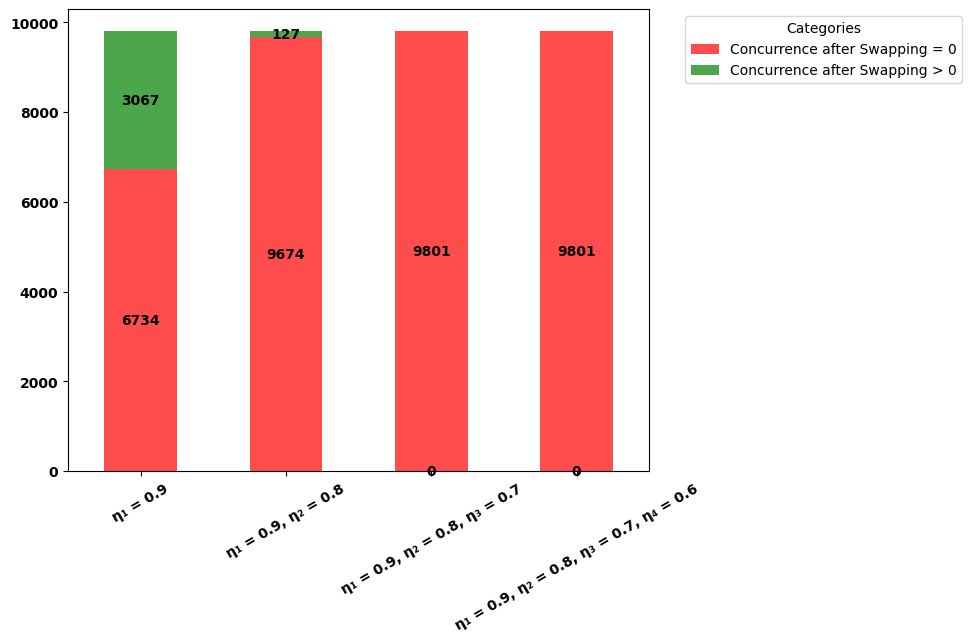}
    \caption{Werner State Multi Node Imperfect Swapping with Different Imperfections }
    \label{fig:label_11}
\end{figure}

\noindent In Fig.(\ref{fig:label_11}), in X-axis, the number of nodes used for swapping is clubbed with different imperfections. A single node represents the value of $\eta$ to be 0.9, and two nodes represent $\eta$ values to be 0.9, 0.8, and so on. Y-axis represents the number of distinct Werner states considered. All the red regions are distinct Werner states whose concurrence is zero after entanglement swappings, and green regions represent distinct Werner states whose concurrence is greater than zero after entanglement swapping.

\subsection{Relations on Concurrence for Bell Diagonal States}

In the next example, we consider the Bell diagonal states in a 1-D network with multiple repeaters scenario. Here also, we have  $n+2$ nodes with a source node, $n$ repeater nodes, and a target node. Let $\rho_{i, i+12}$ is a Bell diagonal state between $i^{th}$ node and $(i+1)^{th}$ node with input parameters $p_{1}$, $p_{2}$, and $p_{3}$ in each node.

\subsubsection{Perfect Measurement}

Here we consider a scenario where the measurement is perfect. Then the final concurrence between the $1^{st}$ and $n+2^{th}$ node is obtained by,

\begin{eqnarray}
        &&C_{1, n+2} = max{}\nonumber\\&&
        \left[0,\left(\frac{(\Lambda_{1} - \Lambda_{2} - \Lambda_{3} - \Lambda_{4})}{\prod\limits_{j=1}^{n+1} (\lambda_{1}^{j} - \lambda_{2}^{j} - \lambda_{3}^{j} - \lambda_{4}^{j})}\right)(\prod\limits_{i = 1}^{n+1}C_{i, i+1})\right],
\end{eqnarray}

\noindent where $\Lambda_{1} \ge \Lambda_{2} \ge \Lambda_{3} \ge \Lambda_{4}$ and $\lambda_{1}^{j} \ge \lambda_{2}^{j} \ge \lambda_{3}^{j} \ge \lambda_{4}^{j}$. The values of $\Lambda_{i}$ are  
\begin{eqnarray}
&&\Lambda_1={}\nonumber\\&&   \frac{1 + p_{1}^{n+1} - (-1)^{n}p_{2}^{n+1} + p_{3}^{n+1}}{4}, {}\nonumber\\&&
\Lambda_2={}\nonumber\\&& \frac{1 - p_{1}^{n+1} + (-1)^{n}p_{2}^{n+1} + p_{3}^{n+1}}{4} ,  {}\nonumber\\&&
\Lambda_3={}\nonumber\\&& \frac{1 + p_{1}^{n+1} + (-1)^{n}p_{2}^{n+1} - p_{3}^{n+1}}{4} ,
{}\nonumber\\&&   
\Lambda_4={}\nonumber\\&&  \frac{1 - p_{1}^{n+1} - (-1)^{n}p_{2}^{n+1} - p_{3}^{n+1}}{4}  
\end{eqnarray}

and the values of $\lambda_{i}$ are$\frac{1 + p_{1} - p_{2} + p_{3}}{4}$, $\frac{1 - p_{1} + p_{2} + p_{3}}{4}$, $\frac{1 + p_{1} + p_{2} - p_{3}}{4}$, $\frac{1 - p_{1} - p_{2} - p_{3}}{4}$.

 \begin{figure}[h]
    \centering
    \includegraphics[scale = 0.33]{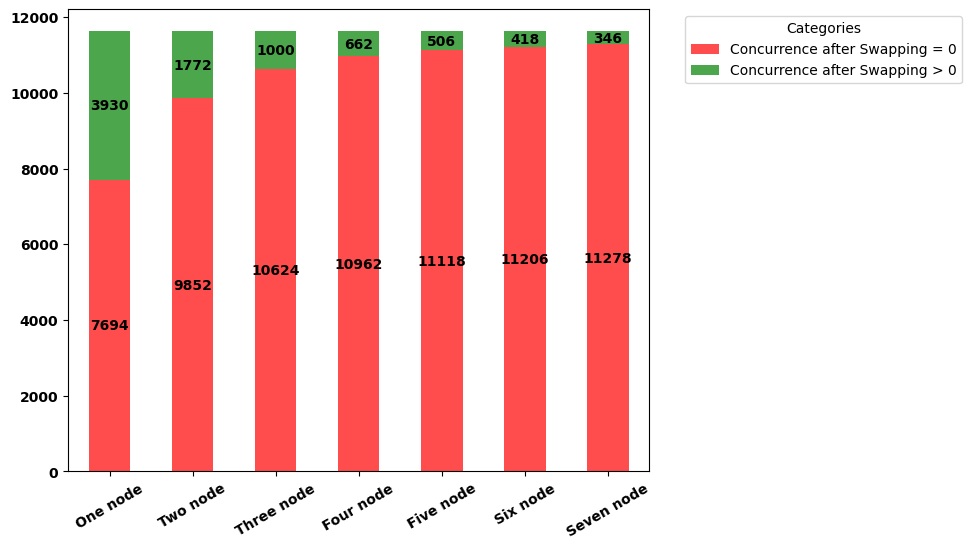}
    \caption{BDS Multi Node Perfect Swapping}
    \label{fig:label_12}
\end{figure}

\noindent In Fig.(\ref{fig:label_12}),  X-axis represents the number of nodes used for swappings, and Y-axis represents the number of distinct Bell diagonal states. There are approximately $12 \times 10^{3}$ states that are considered for entanglement swappings. All the red regions represent the number of output Bell diagonal states whose concurrence is zero after entanglement swapping, and green regions represent the number of output Bell diagonal states whose concurrence is greater than zero after entanglement swapping. \\

\subsubsection{Imperfect Measurement}

\noindent \textbf{Case 1:} As a first case, we consider the scenario where the imperfection is the same in all nodes. Let the imperfection parameter be $1-\eta$ for each node. Then the final concurrence in terms of initial concurrence is given as,

\begin{eqnarray}
        &&C_{1, n+2} = \frac{1}{N}*max {}\nonumber\\&&
        \left[0,\left(\frac{(\Lambda_{1} - \Lambda_{2} - \Lambda_{3} - \Lambda_{4})}{\prod\limits_{j=1}^{n+1} (\lambda_{1}^{j} - \lambda_{2}^{j} - \lambda_{3}^{j} - \lambda_{4}^{j})}\right)(\prod\limits_{i = 1}^{n+1}C_{i, i+1})\right],
\end{eqnarray}

\noindent where $N$ is the normalization factor, and  it is given by,
\begin{equation}
 N= (\eta^{n} + 4(1 - \eta)(\sum\limits_{i=0}^{n-1}(\eta^{n-1-i})(\eta + 4(1 - \eta))^{i})),  
\end{equation}
 \noindent and $\Lambda_{1} \ge \Lambda_{2} \ge \Lambda_{3} \ge \Lambda_{4}$ and $\lambda_{1}^{j} \ge \lambda_{2}^{j} \ge \lambda_{3}^{j} \ge \lambda_{4}^{j}  $. The values of $\Lambda_{i}$ are 
\begin{eqnarray}
 &&\Lambda_1={}\nonumber\\&&  
 \frac{1}{4}\eta^{n}(1 + p_{1}^{(n+1)} - (-1)^{n}p_{2}^{(n+1)} + p_{3}^{(n+1)}) +{}\nonumber\\&&
 4(1 - \eta)(\sum\limits_{i=0}^{n-1}(\eta^{n-1-i})(\eta + 4(1 - \eta))^{i})),
{}\nonumber\\&&
\Lambda_2={}\nonumber\\&&  
 \frac{1}{4}(\eta^{n}(1 - p_{1}^{(n+1)} + (-1)^{n}p_{2}^{(n+1)} + p_{3}^{(n+1)}) + {}\nonumber\\&&
 4(1 - \eta)(\sum\limits_{i=0}^{n-1}(\eta^{n-1-i})(\eta + 4(1 - \eta))^{i})),
{}\nonumber\\&&
\Lambda_3={}\nonumber\\&&  
 \frac{1}{4}(\eta^{n}(1 + p_{1}^{(n+1)} + (-1)^{n}p_{2}^{(n+1)} - p_{3}^{(n+1)}) + {}\nonumber\\&&
 4(1 - \eta)(\sum\limits_{i=0}^{n-1}(\eta^{n-1-i})(\eta + 4(1 - \eta))^{i})),
{}\nonumber\\&&
\Lambda_4={}\nonumber\\&&   
 \frac{1}{4}(\eta^{n}(1 - p_{1}^{(n+1)} - (-1)^{n}p_{2}^{(n+1)} - p_{3}^{(n+1)}) + {}\nonumber\\&&
 4(1 - \eta)(\sum\limits_{i=0}^{n-1}(\eta^{n-1-i})(\eta + 4(1 - \eta))^{i})) 
\end{eqnarray}
\noindent and the values of $\lambda_{i}$ are $\frac{1 + p_{1} - p_{2} + p_{3}}{4}$, $\frac{1 - p_{1} + p_{2} + p_{3}}{4}$, $\frac{1 + p_{1} + p_{2} - p_{3}}{4}$,$\frac{1 - p_{1} - p_{2} - p_{3}}{4}$.

\begin{figure}[h]
    \centering
    \includegraphics[scale = 0.24]{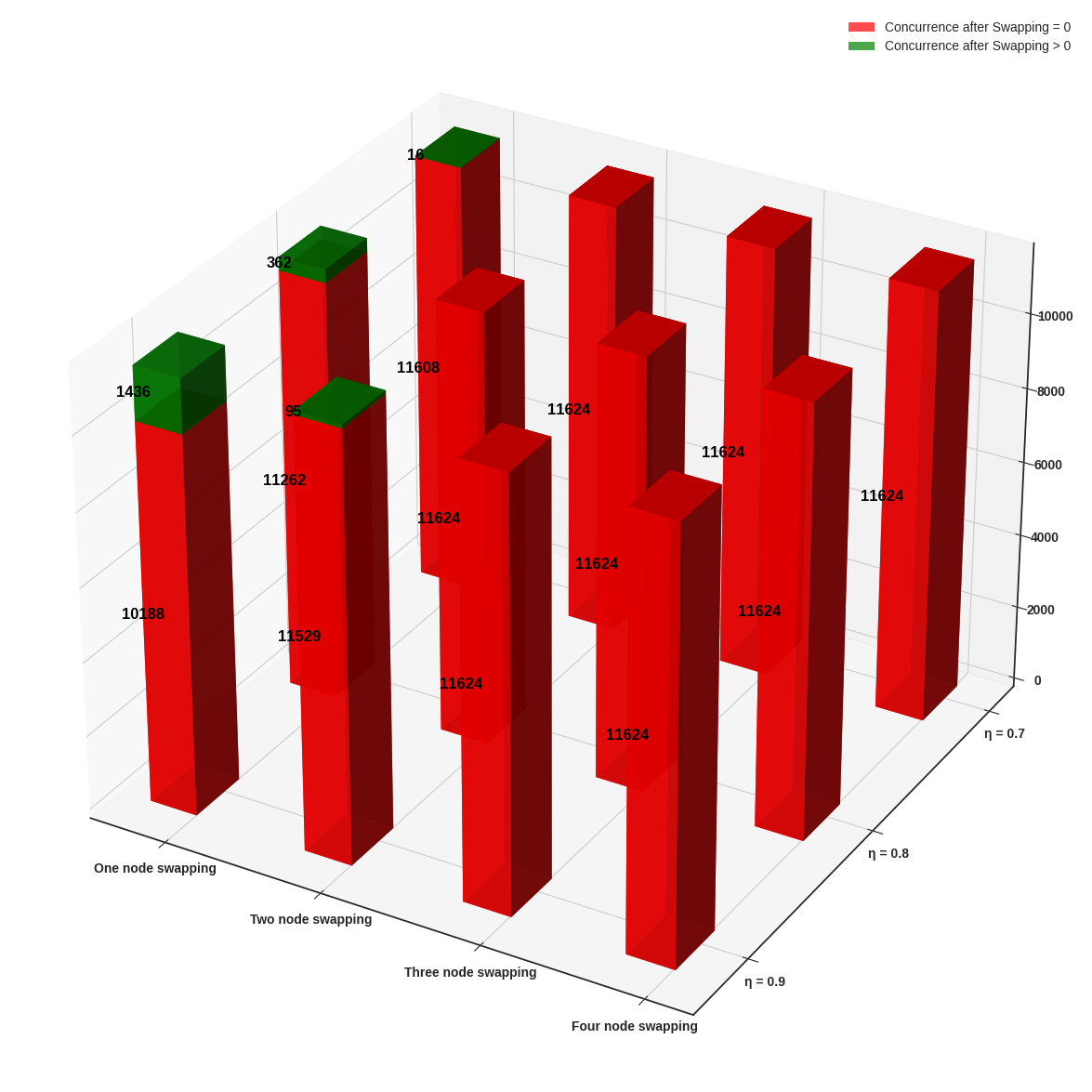}
    \caption{BDS Multi Node Imperfect Swapping with Same Imperfection}
    \label{fig:label_13}
\end{figure}
\noindent In Fig.(\ref{fig:label_13}) X-axis represents the number of nodes being swapped, Y-axis represents the success probability of perfect measurement parameterized by $\eta$. It ranges from 0.0 to 1.0. Here Z-axis represents the number of distinct Bell diagonal states. It is observed that as the value of $\eta$ decreases below 0.7, concurrence is zero, irrespective of the value of visible parameters. When the value of $\eta$ is 0.9, it can be observed that the value of concurrence is greater than zero till two node swapping,  whereas when the value of $\eta$ is 0.8 or 0.7, the value of concurrence is greater than zero for single node swapping only.\\

\noindent \textbf{Case 2:} Here we consider a scenario where the imperfections are different for different nodes. In other words, repeaters carry different imperfect Bell state measurements at their stations. Let the imperfections are $1-\eta_{1}, 1-\eta_{2}, ... , 1-\eta_{n}$ in $n$ nodes, then the final concurrence is defined as
\begin{eqnarray}
    &&C_{1, n+2} = \frac{1}{N}*max {}\nonumber\\&& 
    \left[0,\left(\frac{(\Lambda_{1} - \Lambda_{2} - \Lambda_{3} - \Lambda_{4})}{\prod\limits_{j=1}^{n+1} (\lambda_{1}^{j} - \lambda_{2}^{j} - \lambda_{3}^{j} - \lambda_{4}^{j})}\right)(\prod\limits_{i = 1}^{n+1}C_{i, i+1})\right],
\end{eqnarray}
\noindent where N is the normalization factor and
\begin{equation}
N=(\prod\limits_{i=1}^{n}\eta_{i} + \sum\limits_{i=1}^{n}(\sum\limits_{cyc}(4^{i})(\prod\limits_{j=1}^{i}(1 - \eta_{i}))(\prod\limits_{k = i+1}^{n}\eta_{k}))),    
\end{equation}
and $\Lambda_{1} \ge \Lambda_{2} \ge \Lambda_{3} \ge \Lambda_{4}$ and  $\lambda_{1}^{j} \ge \lambda_{2}^{j} \ge \lambda_{3}^{j} \ge \lambda_{4}^{j}$. The values of $\Lambda_{i}$ are given by
\begin{eqnarray}
&&\Lambda_1= {}\nonumber\\&&
\frac{1}{4}(\prod\limits_{i=1}^{n}\eta_{i}(1 + p_{1}^{(n+1)} - (-1)^{n}p_{2}^{(n+1)} + p_{3}^{(n+1)}) + 
{}\nonumber\\&&
\sum\limits_{i=1}^{n}(\sum\limits_{cyc}(4^{i})(\prod\limits_{j=1}^{i}(1 - \eta_{i}))(\prod\limits_{k = i+1}^{n}\eta_{k}))),
{}\nonumber\\&&
\Lambda_2= {}\nonumber\\&&
\frac{1}{4}(\prod\limits_{i=1}^{n}\eta_{i}(1 - p_{1}^{(n+1)} + (-1)^{n}p_{2}^{(n+1)} + p_{3}^{(n+1)}) + 
{}\nonumber\\&&
\sum\limits_{i=1}^{n}(\sum\limits_{cyc}(4^{i})(\prod\limits_{j=1}^{i}(1 - \eta_{i}))(\prod\limits_{k = i+1}^{n}\eta_{k}))),
{}\nonumber\\&&
\Lambda_3= {}\nonumber\\&&\frac{1}{4}(\prod\limits_{i=1}^{n}\eta_{i}(1 + p_{1}^{(n+1)} + (-1)^{n}p_{2}^{(n+1)} - p_{3}^{(n+1)}) + 
{}\nonumber\\&&
\sum\limits_{i=1}^{n}(\sum\limits_{cyc}(4^{i})(\prod\limits_{j=1}^{i}(1 - \eta_{i}))(\prod\limits_{k = i+1}^{n}\eta_{k}))),
{}\nonumber\\&&
\Lambda_4= {}\nonumber\\&&
\frac{1}{4}(\prod\limits_{i=1}^{n}\eta_{i}(1 - p_{1}^{(n+1)} - (-1)^{n}p_{2}^{(n+1)} - p_{3}^{(n+1)}) + 
{}\nonumber\\&&
\sum\limits_{i=1}^{n}(\sum\limits_{cyc}(4^{i})(\prod\limits_{j=1}^{i}(1 - \eta_{i}))(\prod\limits_{k = i+1}^{n}\eta_{k}))) 
\end{eqnarray}
and the values of $\lambda_{i}$ are $\frac{1 + p_{1} - p_{2} + p_{3}}{4}$, $\frac{1 - p_{1} + p_{2} + p_{3}}{4}$,$\frac{1 + p_{1} + p_{2} - p_{3}}{4}$, $\frac{1 - p_{1} - p_{2} - p_{3}}{4}$.\\

\begin{figure}[h]
    \centering
    \includegraphics[scale = 0.34]{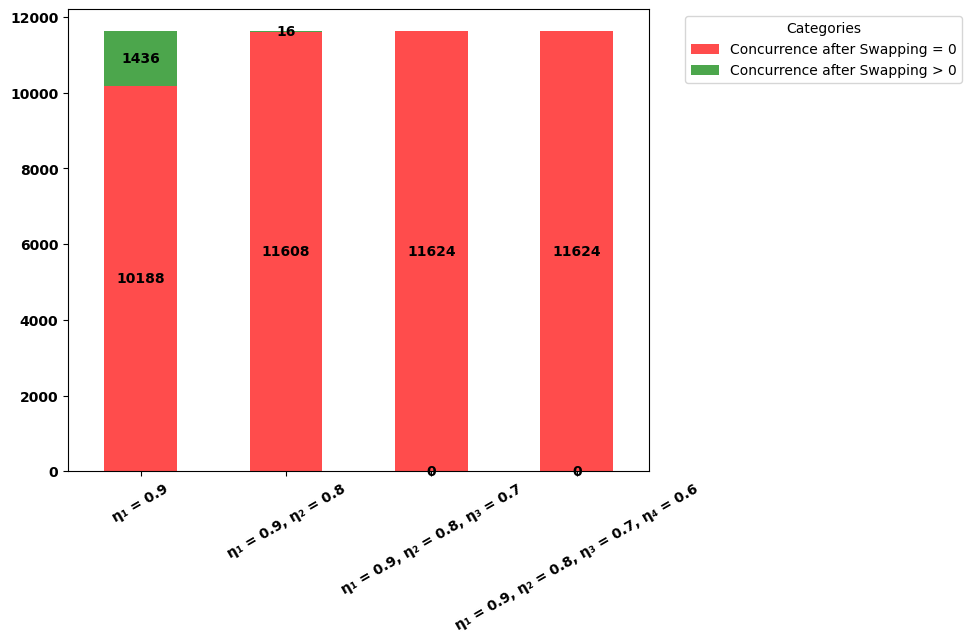}
    \caption{BDS Multi Node with Different Imperfect Swappings}
    \label{fig:label_14}
\end{figure}

\noindent In Fig.(\ref{fig:label_14}) X-axis represents multi-node swappings with different imperfections. A single node represents the value of $\eta$ to be 0.9, and two nodes represent $\eta$ values to be 0.9, 0.8, and so on. Y-axis represents the number of distinct Bell diagonal states considered. All the red regions are distinct Bell diagonal states whose concurrence is zero after entanglement swappings, and green regions represent distinct Bell diagonal states whose concurrence is greater than zero after entanglement swapping.

\subsection{Relations on Concurrence for General Mixed State}

\noindent Finally, we take the general mixed states as a resource to the 1-D network. In the case of multi-node general mixed states, we did a numerical study considering $30 \times 10^{3}$ different mixed entangled states and studied final concurrence after multiple swappings.
 \subsubsection{Perfect Measurement}

\begin{figure}[h]
    \centering
    \includegraphics[scale = 0.35]{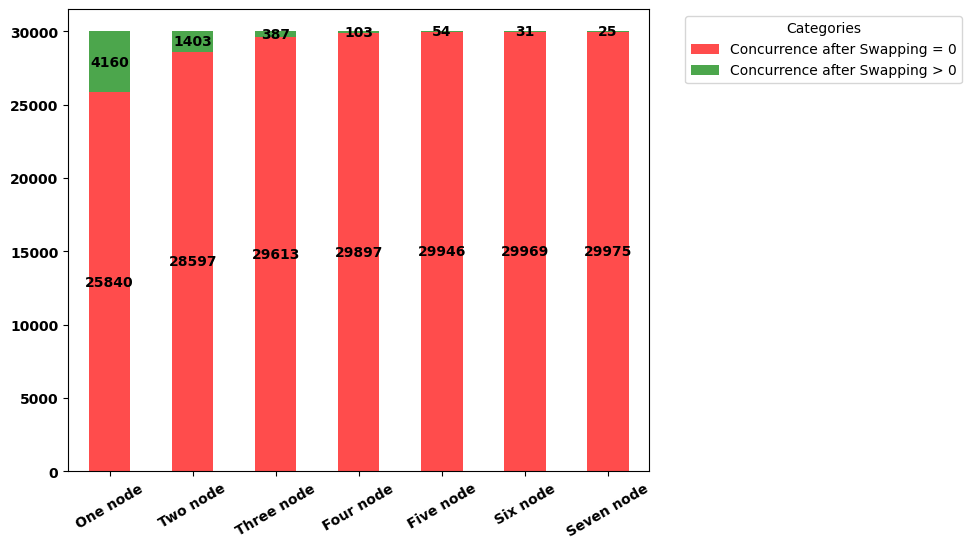}
    \caption{General Mixed States with Perfect Swappings}
    \label{fig:label_15}
\end{figure}

\noindent In Fig.(\ref{fig:label_15}), X-axis represents the number of nodes used in the swapping, and Y-axis represents $30 \times 10^{3}$ mixed states. Red regions represent the number of output states with zero concurrences after swapping, while the green region represents output states whose concurrence is greater than zero after entanglement swapping. In this case, the measurement is perfect, and it is observed from the below graph that we observed concurrence up to seven node swappings, and after that, there is no concurrence in output states.\\
\subsubsection{Imperfect Measurement}

\noindent \textbf{Case1:} In this scenario the imperfection ($1-\eta$) of the measurement is same in all the nodes while swapping process.\\ 

 \begin{figure}[h]
    \centering
    \includegraphics[scale = 0.28]{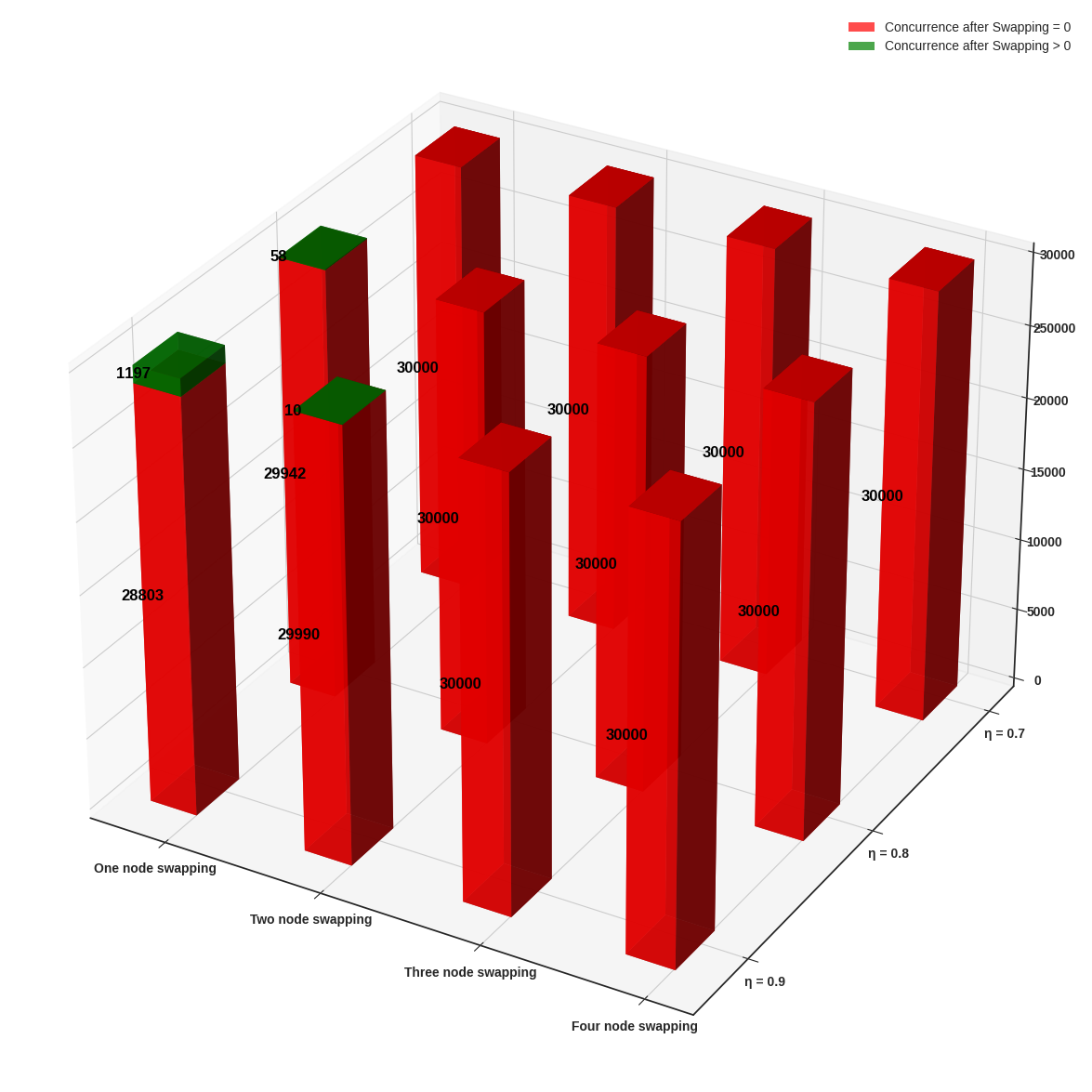}
    \caption{General Mixed States with Imperfect Swapping}
    \label{fig: mixed_state_multi_node_imp_conc}
\end{figure}

\noindent In Fig.(\ref{fig: mixed_state_multi_node_imp_conc}) X-axis represents the number of nodes used for swapping, Y-axis represents $\eta$, and Z-axis represents the number of states considered for swapping. There are $30 \times 10^{3}$ states that are considered for swapping, and the red region represents those output states whose concurrence is equal to zero after swapping, and the green region represents concurrence greater than zero after entanglement swapping. It can be observed from the graph that for a single node swapping, there are 1197 states whose concurrence is greater than zero when $\eta = 0.9$ and 58 states whose concurrence is greater than zero when $\eta = 0.8$. For the rest, the states have zero concurrences after entanglement swapping.\\

\noindent \textbf{Case2:} In this scenario, imperfection is different in different nodes at the time of swapping.

 \begin{figure}[h]
    \centering
    \includegraphics[scale = 0.35]{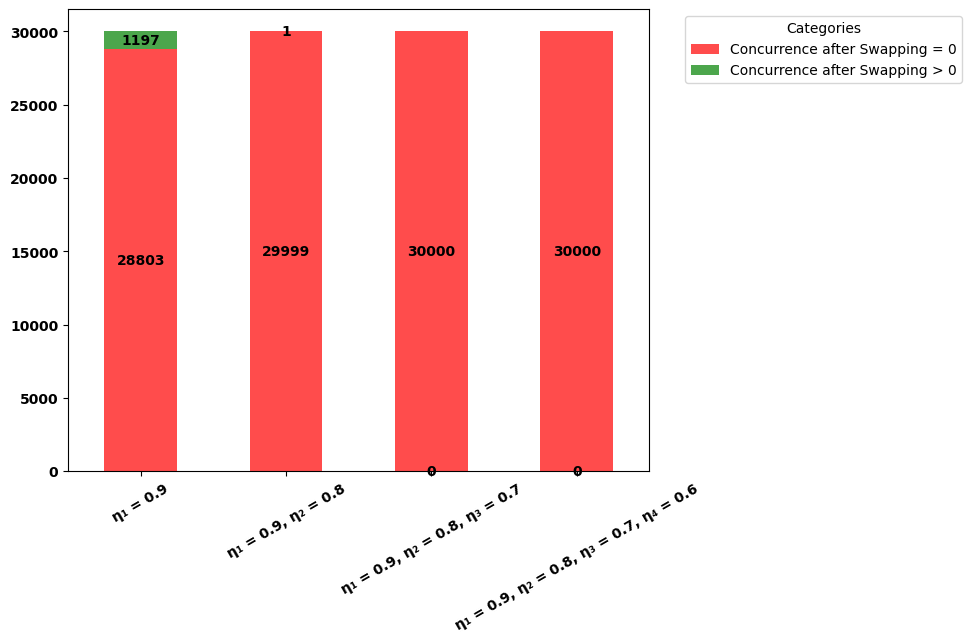}
    \caption{General Mixed States with Imperfect Swapping} \label{fig:mixed_state_multi_node_imp_diff_eta}
\end{figure}

\noindent In Fig.(\ref{fig:mixed_state_multi_node_imp_diff_eta}), X-axis is clubbed with the values of imperfections with the number of swappings. The values of imperfections are taken on an exemplary basis. Here Y-axis represents the number of mixed states considered. The red region represents all the states whose concurrence is zero after entanglement swapping, and the green region represents all the states whose concurrence is greater than zero after entanglement swapping. It can be observed from the graph that after two node swappings ($\eta  = 0.9, 0.8$), all the other nodes have zero concurrences.

\section{Relations on Concurrence for Multipartite Network in Remote
Entangled Distribution (RED) with different states}

\noindent This section studies a general situation with more than three nodes, where consecutive nodes with states with different input parameters at each node are entangled. We need to establish entanglement between the initial and final nodes. We consider the process of entanglement swapping as a technique for remote entanglement distribution (RED). Let us assume that we have $(n + 1)$ entangled states with $n + 2$ nodes. In order to obtain an entangled state between initial and final nodes, we carry out $n$ number of entanglement swapping. We consider sequential measurements to create successive entanglements between the nodes (1, 3), (1, 4), and finally between the nodes (1, n + 2) and obtain the extension of the relationships involving the concurrences of initial and final entangled states.\\

\subsection{Relation of Concurrences for Werner States}
\noindent Consider $n+2$ nodes, with a source node, $n$ repeater nodes and a target node . Let $\rho_{i, i+2}$ be a Werner state between $i^{th}$ node and $(i+1)^{th}$ node with visible parameters $p_{i}$ .

\subsubsection{Perfect Measurement}

\noindent Here we consider a scenario where the measurement is perfect. Then the concurrence of the final state is obtained as,

\begin{equation}
    C_{1, n+2} = 2^{n}\left[\frac{[3(\prod\limits_{i=1}^{n+1} p_{i}) - 1](\prod\limits_{i = 1}^{n+1} C_{i, i+1})}{\prod\limits_{i=1}^{n+1} (3p_{i} - 1)}\right].
    \label{eq:33}
\end{equation}

The concurrence can be expressed in terms of visible parameters as

\begin{equation}
    C_{1, n+2} = \left[\frac{[3(\prod\limits_{i=1}^{n+1} p_{i}) - 1]}{2}\right].
    \label{eq:34}
\end{equation}

\noindent \textbf{Note:} When all the visible parameters are equal then equation (\ref{eq:33}) becomes (\ref{eq:31}) and equation (\ref{eq:34}) becomes (\ref{eq:32}).\\

\begin{figure}[h]
    \centering
    \includegraphics[scale = 0.36]{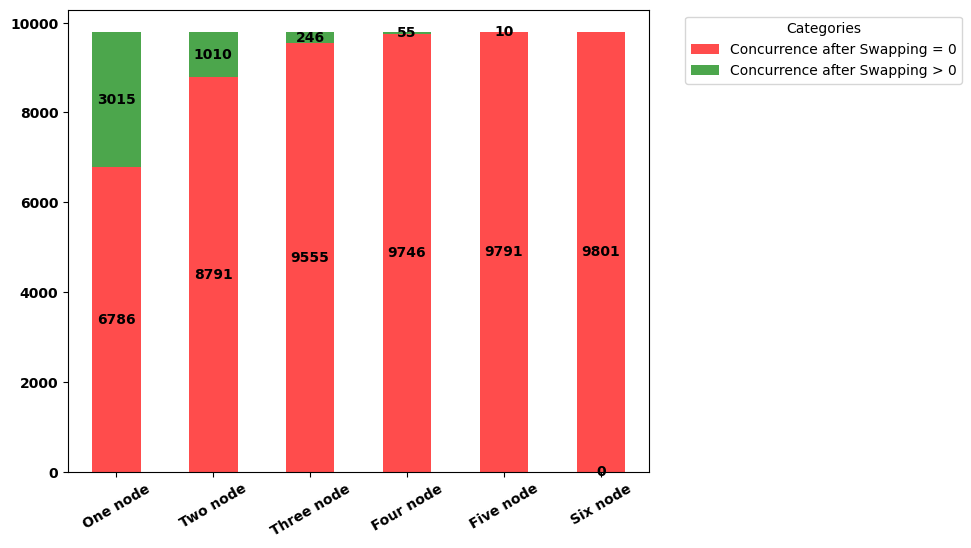}
    \caption{Werner State Multi Node Perfect Swappings}
    \label{fig:label_16}
\end{figure} 

\noindent In Fig.(\ref{fig:label_16}), X-axis represents the number of nodes used for swappings, and Y-axis represents the number of states considered. Approximately $10 \times 10^{3}$ states are considered, and all the red regions represent the number of states with zero concurrences after entanglement swapping, and green regions represent the number of output states with concurrence greater than zero after entanglement swapping. Here we consider a scenario where the measurement is perfect. Then the concurrence of the final state is obtained as,

\subsubsection{Imperfect Measurement}

\noindent \textbf{Case 1:} Here we consider a scenario where the imperfection in the measurement is the same in all nodes. Let $\eta$ be the parameter to represent the success probability associated with the measurement process for each node. The final concurrence in terms of initial concurrence is defined as\\

\begin{eqnarray}
    &&C_{1, n+2} = \frac{2^{n}}{N}{}\nonumber\\&&
    \left[\frac{\eta^{n}[3(\prod\limits_{i=1}^{n+1}p_{i}) - 1] - 4(1 - \eta)(\sum\limits_{i=0}^{n+1}(\eta^{n-1-i})(\eta + 4(1 - \eta))^{i})}{\prod\limits_{i=1}^{n+1}(3p_{i} - 1)}\right]{}\nonumber\\&&
    (\prod\limits_{i = 1}^{n+1} C_{i, i+1}).
\end{eqnarray}

\noindent where the normalization constant $N$ is given by,
\begin{equation}
 N = (\eta^{n} + 4(1 - \eta)(\sum\limits_{i=0}^{n-1}(\eta^{n-1-i})(\eta + 4(1 - \eta))^{i})).   
\end{equation}

\begin{figure}[h]
    \centering
    \includegraphics[scale = 0.26]{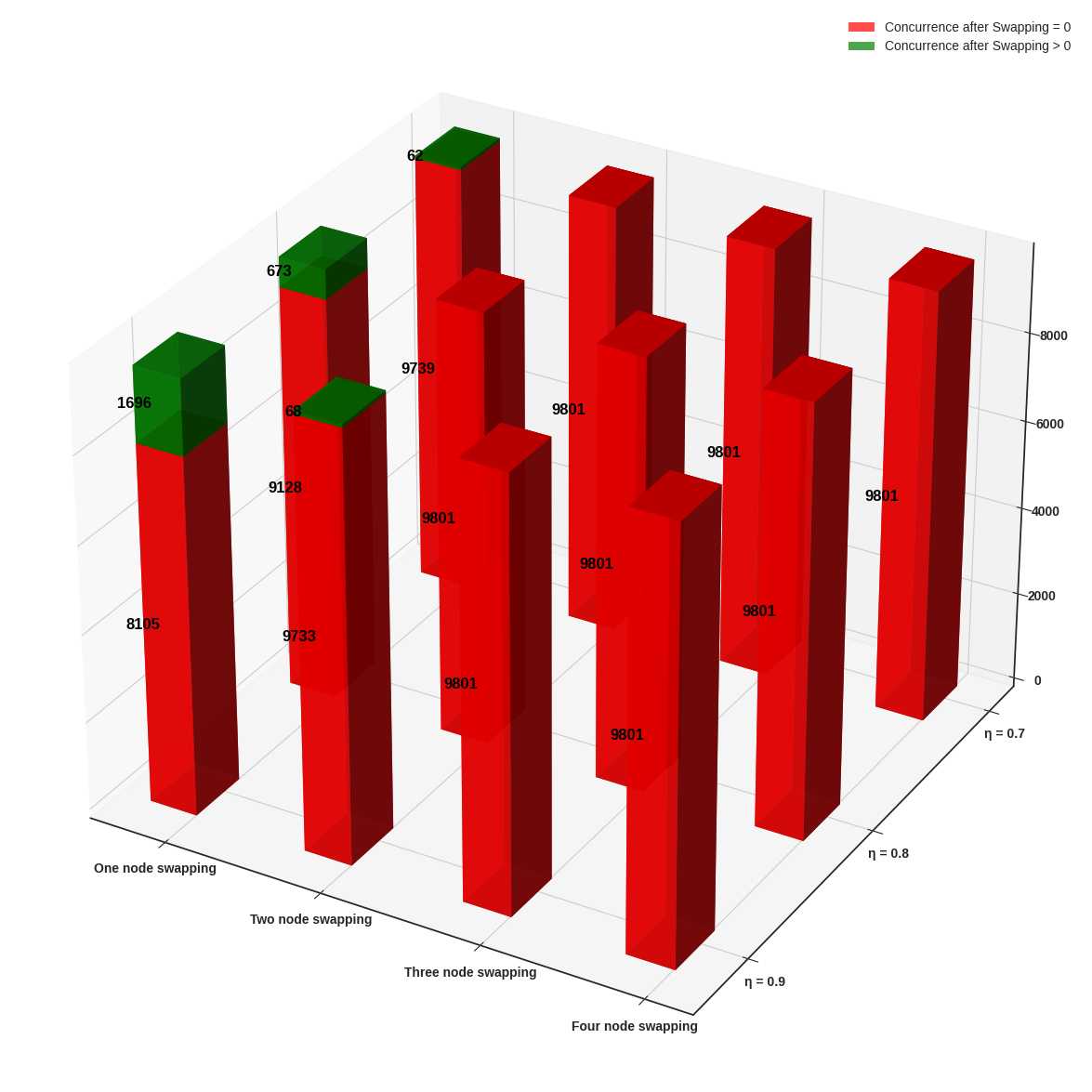}
    \caption{Werner State Multi Node Imperfect Swappings with Same Imperfections}
    \label{fig:label_17}
\end{figure}

\noindent In Fig.(\ref{fig:label_17}) X-axis represents the number of nodes used for swappings, Y-axis represents $\eta$ from 0.0 to 0.9, and Z-axis represents the number of states considered. The red region represents the number of output states whose concurrence is zero after entanglement swapping, and the green region represents the number of output states whose concurrence is greater than zero after entanglement swapping.\\

\noindent \textbf{Case 2:} Here we consider a scenario where the imperfection is different in different nodes. Let the imperfections are $1-\eta_{i}$ in $n$ nodes, then the final concurrence is defined as,

\begin{eqnarray}
    &&C_{1, n+2} = \frac{2^{n}}{N}\left[\frac{\prod\limits_{i=1}^{n}\eta_{i}[3(\prod\limits_{i=1}^{n+1}p_{i}) - 1] - \sum\limits_{i=1}^{n}
    (\sum\limits_{cyc}(4^{i})(\prod\limits_{j=1}^{i}(1 - \eta_{i}))(\prod\limits_{k = i+1}^{n}\eta_{k}))}{\prod\limits_{i=1}^{n+1}(3p_{i} - 1)}\right]\\&&(\prod\limits_{i = 1}^{n+1} C_{i, i+1})
\end{eqnarray}

\noindent Here $N = (\prod\limits_{i=1}^{n}\eta_{i} + \sum\limits_{i=1}^{n}(\sum\limits_{cyc}(4^{i})(\prod\limits_{j=1}^{i}(1 - \eta_{i}))(\prod\limits_{k = i+1}^{n}\eta_{k})))$

 \begin{figure}[h]
    \centering
    \includegraphics[scale = 0.36]{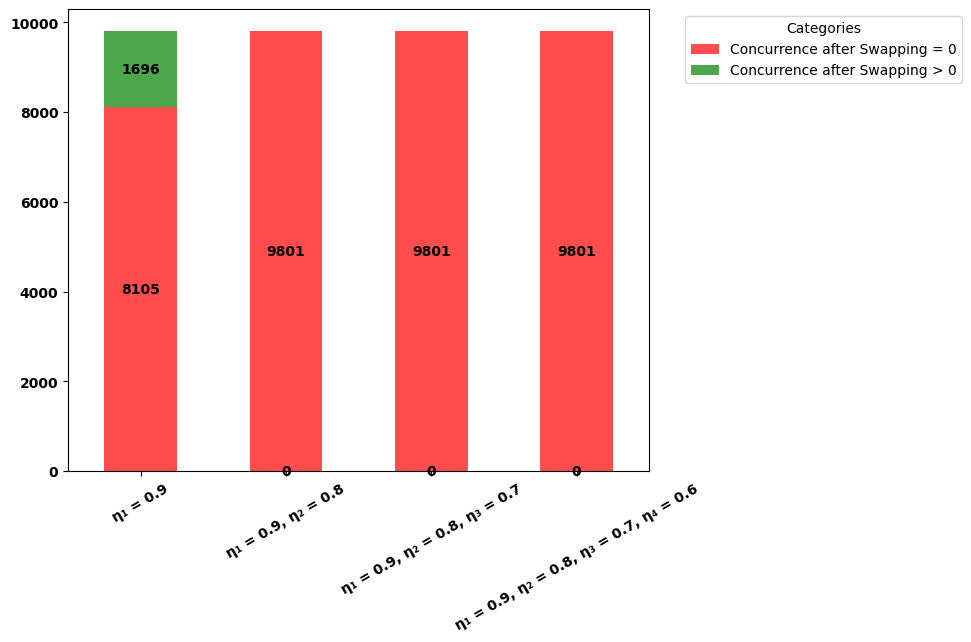}
    \caption{Werner Multi Node with different Visible Parameters different Imperfect Swappings}
    \label{fig:Werner_MultiNode_Diff_Vis_diff_eta_imperfect}
\end{figure}
\noindent In Fig.(\ref{fig:Werner_MultiNode_Diff_Vis_diff_eta_imperfect}) X-axis represents multinode swappings with different imperfections, and Y-axis represents the number of states considered after entanglement swapping. All the red regions represent states with concurrence equal to zero after swapping, and all green regions represent states with concurrence greater than zero after entanglement swapping. It can be observed from the graph that there are states with concurrence greater than zero only for single-node swappings.


\subsection{Relation on Concurrences for Bell Diagonal States}

\noindent Consider $n+2$ nodes with a source node, n repeater nodes, and a target node. Let $\rho_{i, i+1}$ is a Bell diagonal state between $i^{th}$ node and $(i+1)^{th}$ node with visible parameters be $p_{1}^{i}, p_{2}^{i},p_{3}^{i}$ at $i^{th}$ node.

\subsubsection{Perfect Measurement}

Here we consider a scenario where measurement is perfect. Then the final concurrence between the source and the target node is obtained as,

\begin{eqnarray}
        &&C_{1, n+2} = max{}\nonumber\\&&
        \left[0,\left(\frac{(\Lambda_{1} - \Lambda_{2} - \Lambda_{3} - \Lambda_{4})}{\prod\limits_{j=1}^{n+1} (\lambda_{1}^{j} - \lambda_{2}^{j} - \lambda_{3}^{j} - \lambda_{4}^{j})}\right)(\prod\limits_{i = 1}^{n+1}C_{i, i+1})\right],
\end{eqnarray}
\noindent where $\Lambda_{1} \ge \Lambda_{2} \ge \Lambda_{3} \ge \Lambda_{4}$ and $\lambda_{1}^{j} \ge \lambda_{2}^{j} \ge \lambda_{3}^{j} \ge \lambda_{4}^{j}$.\\

The values of $\Lambda_{i}$ are 
\begin{eqnarray}
&&\Lambda_1={}\nonumber\\&&    \frac{(1 + \prod\limits_{i = 1}^{n+1}p_{1}^{i} - (-1)^{n}\prod\limits_{i = 1}^{n+1}p_{2}^{i} + \prod\limits_{i = 1}^{n+1}p_{3}^{i})}{4},
{}\nonumber\\&&
\Lambda_2={}\nonumber\\&&    \frac{(1 - \prod\limits_{i = 1}^{n+1}p_{1}^{i} + (-1)^{n}\prod\limits_{i = 1}^{n+1}p_{2}^{i} + \prod\limits_{i = 1}^{n+1}p_{3}^{i})}{4},
{}\nonumber\\&&
\Lambda_3={}\nonumber\\&&    \frac{(1 + \prod\limits_{i = 1}^{n+1}p_{1}^{i} + (-1)^{n}\prod\limits_{i = 1}^{n+1}p_{2}^{i} - \prod\limits_{i = 1}^{n+1}p_{3}^{i})}{4},
{}\nonumber\\&&
\Lambda_4={}\nonumber\\&&    \frac{(1 - \prod\limits_{i = 1}^{n+1}p_{1}^{i} - (-1)^{n}\prod\limits_{i = 1}^{n+1}p_{2}^{i} - \prod\limits_{i = 1}^{n+1}p_{3}^{i})}{4},
\end{eqnarray}

\noindent and the values of $\lambda_{i}$ are
$\frac{1 + p_{1}^{j} - p_{2}^{j} + p_{3}^{j}}{4}$, 
$\frac{1 - p_{1}^{j} + p_{2}^{j} + p_{3}^{j}}{4}$, 
$\frac{1 + p_{1}^{j} + p_{2}^{j} - p_{3}^{j}}{4}$,
$\frac{1 - p_{1}^{j} - p_{2}^{j} - p_{3}^{j}}{4}$.

 \begin{figure}[h]
    \centering
    \includegraphics[scale = 0.36]{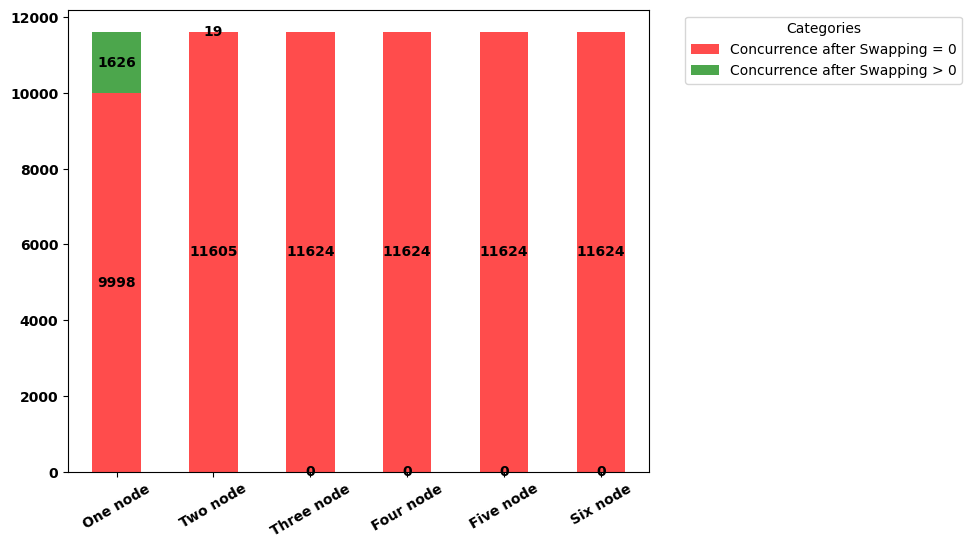}
    \caption{BDS Multi Node with Different Visible Parameters Perfect Swappings}
    \label{fig:label_18}
\end{figure}

\noindent In Fig.(\ref{fig:label_18}) X-axis represents the number of nodes used for entanglement swapping, and Y-axis represents the number of states considered for entanglement swapping. Approximately $12 \times 10^{3}$ states are considered for swapping; red regions represent states whose concurrence is zero after swapping, while green region represents states whose concurrence is greater than zero after swapping.

\subsubsection{Imperfect Measurement}

\noindent \textbf{Case 1:} Here we consider a scenario where the imperfection is the same in all nodes. Let the imperfection parameter be $1-\eta$ for each node. The final concurrence in terms of initial concurrence is defined as\\
\begin{eqnarray}
        &&C_{1, n+2} = \frac{1}{N}*max{}\nonumber\\&&
        \left[0,\left(\frac{(\Lambda_{1} - \Lambda_{2} - \Lambda_{3} - \Lambda_{4})}{\prod\limits_{j=1}^{n+1} (\lambda_{1}^{j} - \lambda_{2}^{j} - \lambda_{3}^{j} - \lambda_{4}^{j})}\right)(\prod\limits_{i = 1}^{n+1}C_{i, i+1})\right],
\end{eqnarray}
\noindent where $N$ is the normalization factor and is given by, 
\begin{equation}
(\eta^{n} + 4(1 - \eta)(\sum\limits_{i=0}^{n-1}(\eta^{n-1-i})(\eta + 4(1 - \eta))^{i})),    
\end{equation}
\noindent and we have $\Lambda_{1} \ge \Lambda_{2} \ge \Lambda_{3} \ge \Lambda_{4}$.  Here $\lambda_{1}^{j} \ge \lambda_{2}^{j} \ge \lambda_{3}^{j} \ge \lambda_{4}^{j}$. The values of $\Lambda_{i}$ are given by,
\begin{eqnarray}
 && \Lambda_1 ={}\nonumber\\&& \frac{1}{4}(\eta)^{n}(1 + \prod\limits_{i = 1}^{n+1}p_{1}^{i} - (-1)^{n}\prod\limits_{i = 1}^{n+1}p_{2}^{i} +{}\nonumber\\&&
 \prod\limits_{i = 1}^{n+1}p_{3}^{i}) + 4(1 - \eta)(\sum\limits_{i=0}^{n-1}(\eta^{n-1-i})(\eta + 4(1 - \eta))^{i}))),{}\nonumber\\&&
 \Lambda_2 ={}\nonumber\\&&   \frac{1}{4}((\eta)^{n}(1 - \prod\limits_{i = 1}^{n+1}p_{1}^{i} + (-1)^{n}\prod\limits_{i = 1}^{n+1}p_{2}^{i} + {}\nonumber\\&& \prod\limits_{i = 1}^{n+1}p_{3}^{i}) + 4(1 - \eta)(\sum\limits_{i=0}^{n-1}(\eta^{n-1-i})(\eta + 4(1 - \eta))^{i})),{}\nonumber\\&&
{}\nonumber\\&&
\Lambda_3 ={}\nonumber\\&&  \frac{1}{4}((\eta)^{n}(1 + \prod\limits_{i = 1}^{n+1}p_{1}^{i} + (-1)^{n}\prod\limits_{i = 1}^{n+1}p_{2}^{i} - {}\nonumber\\&& \prod\limits_{i = 1}^{n+1}p_{3}^{i}) + 4(1 - \eta)(\sum\limits_{i=0}^{n-1}(\eta^{n-1-i})(\eta + 4(1 - \eta))^{i}))), {}\nonumber\\&&
{}\nonumber\\&&
\Lambda_4 ={}\nonumber\\&&   \frac{1}{4}((\eta)^{n}(1 - \prod\limits_{i = 1}^{n+1}p_{1}^{i} - (-1)^{n}\prod\limits_{i = 1}^{n+1}p_{2}^{i} - {}\nonumber\\&& \prod\limits_{i = 1}^{n+1}p_{3}^{i}) + 4(1 - \eta)(\sum\limits_{i=0}^{n-1}(\eta^{n-1-i})(\eta + 4(1 - \eta))^{i}))) {}\nonumber\\&&
 \end{eqnarray}

\noindent and the values of $\lambda_{i}$ are
$\frac{1 + p_{1}^{j} - p_{2}^{j} + p_{3}^{j}}{4}$, 
$\frac{1 - p_{1}^{j} + p_{2}^{j} + p_{3}^{j}}{4}$, 
$\frac{1 + p_{1}^{j} + p_{2}^{j} - p_{3}^{j}}{4}$,
$\frac{1 - p_{1}^{j} - p_{2}^{j} - p_{3}^{j}}{4}$.

 \begin{figure}[h]
    \centering
    \includegraphics[scale = 0.26]{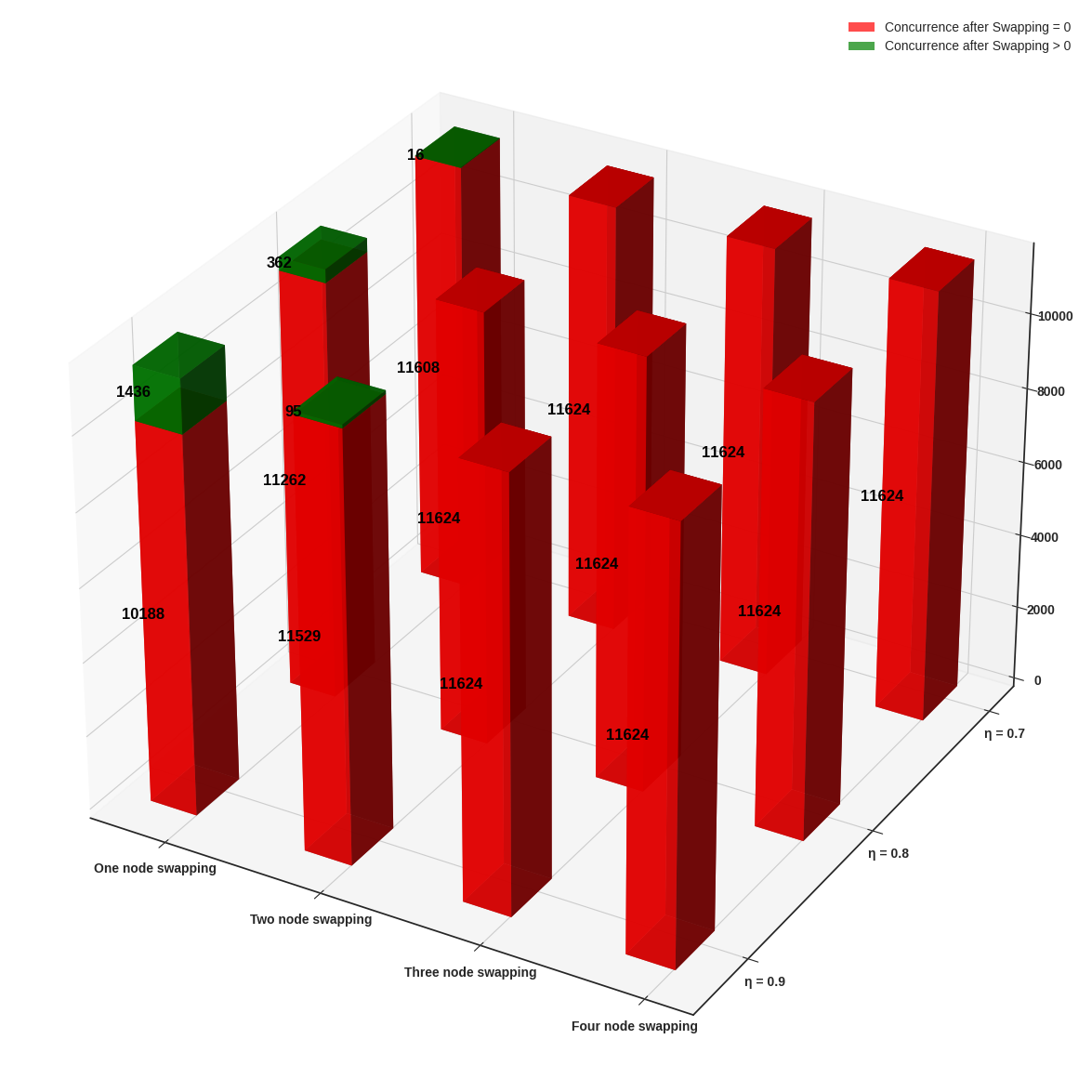}
    \caption{BDS Multi Node Imperfect Swappings with Same Imperfections}
    \label{fig:label_19}
\end{figure}

\noindent In Fig.(\ref{fig:label_19}) X-axis represents the number of nodes used for swappings, Y-axis represents $\eta$ ranging from 0.0 to 0.9, and Z-axis represents the number of states. In the figure, approximately $12 \times 10^{3}$ states are considered. The red region represents all the states whose concurrence is equal to zero after entanglement swapping, and the green region represents all the states whose concurrence is greater than zero after entanglement swapping.\\

\newpage

\noindent \textbf{Case 2:} Here we consider a scenario where the imperfection is different in different nodes. Let the probabilities of success for measurements be $\eta_{1}, \eta_{2}, ... , \eta_{n}$ in $n$ nodes, then the final concurrence is defined as\\

\begin{eqnarray}
    &&C_{1, n+2} = \frac{1}{N}*max{}\nonumber\\&&
    \left[0,\left(\frac{(\Lambda_{1} - \Lambda_{2} - \Lambda_{3} - \Lambda_{4})}{\prod\limits_{j=1}^{n+1} (\lambda_{1}^{j} - \lambda_{2}^{j} - \lambda_{3}^{j} - \lambda_{4}^{j})}\right)(\prod\limits_{i = 1}^{n+1}C_{i, i+1})\right],
\end{eqnarray}
\noindent where N is the normalization factor, which is given by, 
\begin{equation}
N=(\prod\limits_{i=1}^{n}\eta_{i} + \sum\limits_{i=1}^{n}(\sum\limits_{cyc}(4^{i})(\prod\limits_{j=1}^{i}(1 - \eta_{i}))(\prod\limits_{k = i+1}^{n}\eta_{k}))).    
\end{equation}
Here, $\Lambda_{1} \ge \Lambda_{2} \ge \Lambda_{3} \ge \Lambda_{4}$ and $\lambda_{1}^{j} \ge \lambda_{2}^{j} \ge \lambda_{3}^{j} \ge \lambda_{4}^{j}$.\\

The values of $\Lambda_{i}$ are given by,
\begin{eqnarray}
&&\Lambda_1={}\nonumber\\&&    \frac{1}{4}(\prod\limits_{i=1}^{n}\eta_{i}(1 + \prod\limits_{i = 1}^{n+1}p_{1}^{i} - (-1)^{n}\prod\limits_{i = 1}^{n+1}p_{2}^{i}{}\nonumber\\&&
+ \prod\limits_{i = 1}^{n+1}p_{3}^{i}) + \sum\limits_{i=1}^{n}(\sum\limits_{cyc}(4^{i})(\prod\limits_{j=1}^{i}(1 - \eta_{i}))(\prod\limits_{k = i+1}^{n}\eta_{k}))),{}\nonumber\\&&
\Lambda_2={}\nonumber\\&&   \frac{1}{4}(\prod\limits_{i=1}^{n}\eta_{i}(1 - \prod\limits_{i = 1}^{n+1}p_{1}^{i} + (-1)^{n}\prod\limits_{i = 1}^{n+1}p_{2}^{i}{}\nonumber\\&&
+ \prod\limits_{i = 1}^{n+1}p_{3}^{i}) + \sum\limits_{i=1}^{n}(\sum\limits_{cyc}(4^{i})(\prod\limits_{j=1}^{i}(1 - \eta_{i}))(\prod\limits_{k = i+1}^{n}\eta_{k}))), {}\nonumber\\&&
\Lambda_3={}\nonumber\\&&    \frac{1}{4}(\prod\limits_{i=1}^{n}\eta_{i}(1 + \prod\limits_{i = 1}^{n+1}p_{1}^{i} + (-1)^{n}\prod\limits_{i = 1}^{n+1}p_{2}^{i} {}\nonumber\\&&
- \prod\limits_{i = 1}^{n+1}p_{3}^{i}) + \sum\limits_{i=1}^{n}(\sum\limits_{cyc}(4^{i})(\prod\limits_{j=1}^{i}(1 - \eta_{i}))(\prod\limits_{k = i+1}^{n}\eta_{k}))),{}\nonumber\\&&
\Lambda_4={}\nonumber\\&&   \frac{1}{4}(\prod\limits_{i=1}^{n}\eta_{i}(1 - \prod\limits_{i = 1}^{n+1}p_{1}^{i} - (-1)^{n}\prod\limits_{i = 1}^{n+1}p_{2}^{i} {}\nonumber\\&&
- \prod\limits_{i = 1}^{n+1}p_{3}^{i}) + \sum\limits_{i=1}^{n}(\sum\limits_{cyc}(4^{i})(\prod\limits_{j=1}^{i}(1 - \eta_{i}))(\prod\limits_{k = i+1}^{n}\eta_{k}))),{}\nonumber\\&&
\end{eqnarray}

\noindent and the values of $\lambda_{i}$ are
$\frac{1 + p_{1}^{j} - p_{2}^{j} + p_{3}^{j}}{4}$, 
$\frac{1 - p_{1}^{j} + p_{2}^{j} + p_{3}^{j}}{4}$, 
$\frac{1 + p_{1}^{j} + p_{2}^{j} - p_{3}^{j}}{4}$,
$\frac{1 - p_{1}^{j} - p_{2}^{j} - p_{3}^{j}}{4}$.\\

\begin{figure}[h]
    \centering
    \includegraphics[scale = 0.37]{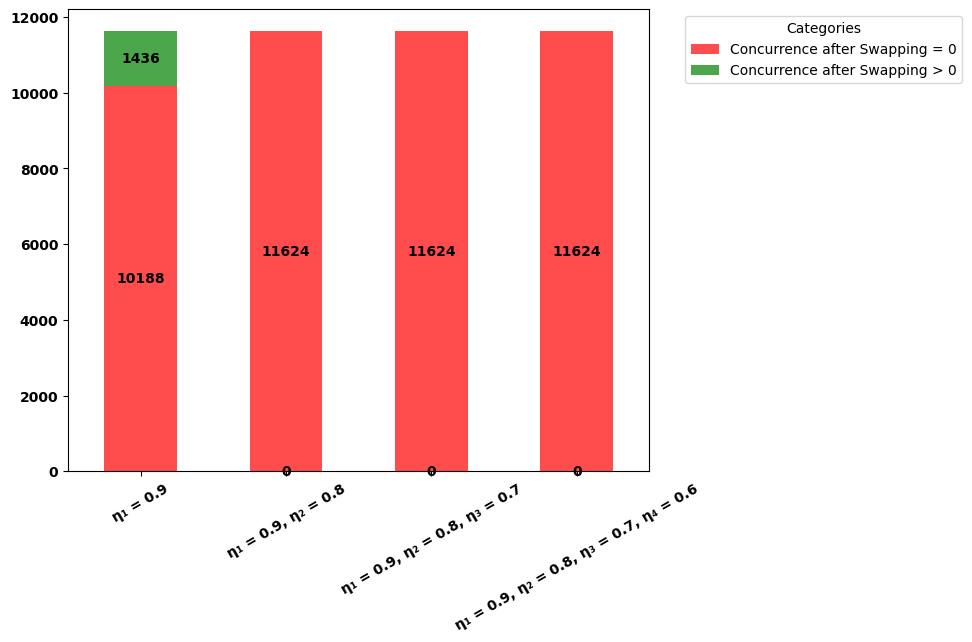}
    \caption{BDS multi node different visible parameters and different imperfection swappings}    \label{fig:BDS_state_multi_diff_param_diff_imperfect_conc}
\end{figure}

\noindent In Fig.(\ref{fig:BDS_state_multi_diff_param_diff_imperfect_conc})X-axis represents multinode swappings with different imperfections, and Y-axis represents the number of output states considered after entanglement swapping. All the red regions represent states with concurrence equal to zero after swapping, and all green regions represent states with concurrence greater than zero after entanglement swapping. It can be observed from the graph that there are states with concurrence greater than zero only for single-node swappings.


\subsection{Relations on Concurrence for General Mixed States}

Consider $n+2$ nodes with a source node, $n$ repeater nodes, and a target node. Let $\rho_{i, i+1}$ be a general mixed state between $i^{th}$ node and $(i+1)^{th}$ node, and all the nodes have different general mixed states. In this subsection, we only give numerical evidence of how the final concurrence behaves with initial concurrences.

\subsubsection{Perfect Measurement}
In this scenario, we consider the measurement to be perfect in all the nodes. The final concurrence after multi-node swappings can be observed from the following graph.\\

\begin{figure}[h]
    \centering
    \includegraphics[scale = 0.37]{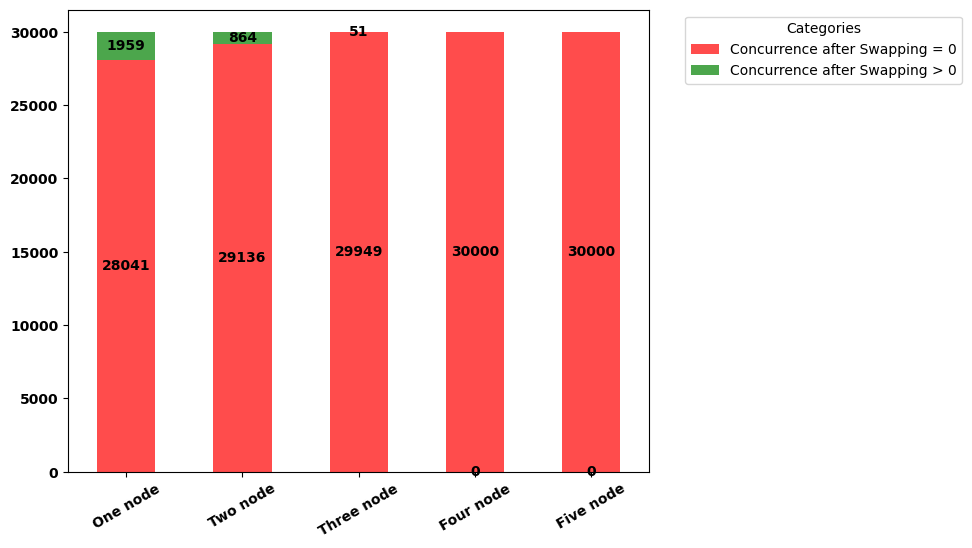}
    \caption{General mixed state multi node perfect swapping}    \label{fig:mixed_state_multi_diff_param_perfect_conc}
\end{figure}

\noindent In Fig.(\ref{fig:mixed_state_multi_diff_param_perfect_conc}) X-axis represents multi-node swapping with the number of nodes used as points on the X-axis, and Y-axis represents the number of general mixed states considered for entanglement swapping. There are $30 \times 10^{3}$ distinct mixed states considered, and the red region represents the states whose concurrence is zero after entanglement swapping. The green region represents the number of output states whose concurrence is greater than zero after entanglement swapping. It can be observed that the data which is considered concurrence can be observed only for three node swappings for perfect measurement.

\subsubsection{Imperfect Measurement}

\noindent\textbf{Case I:} Here we consider a scenario where the imperfection is the same in all nodes. Let the imperfection parameter is $1-\eta$ for each node. 

\begin{figure}[h]
    \centering
    \includegraphics[scale = 0.24]{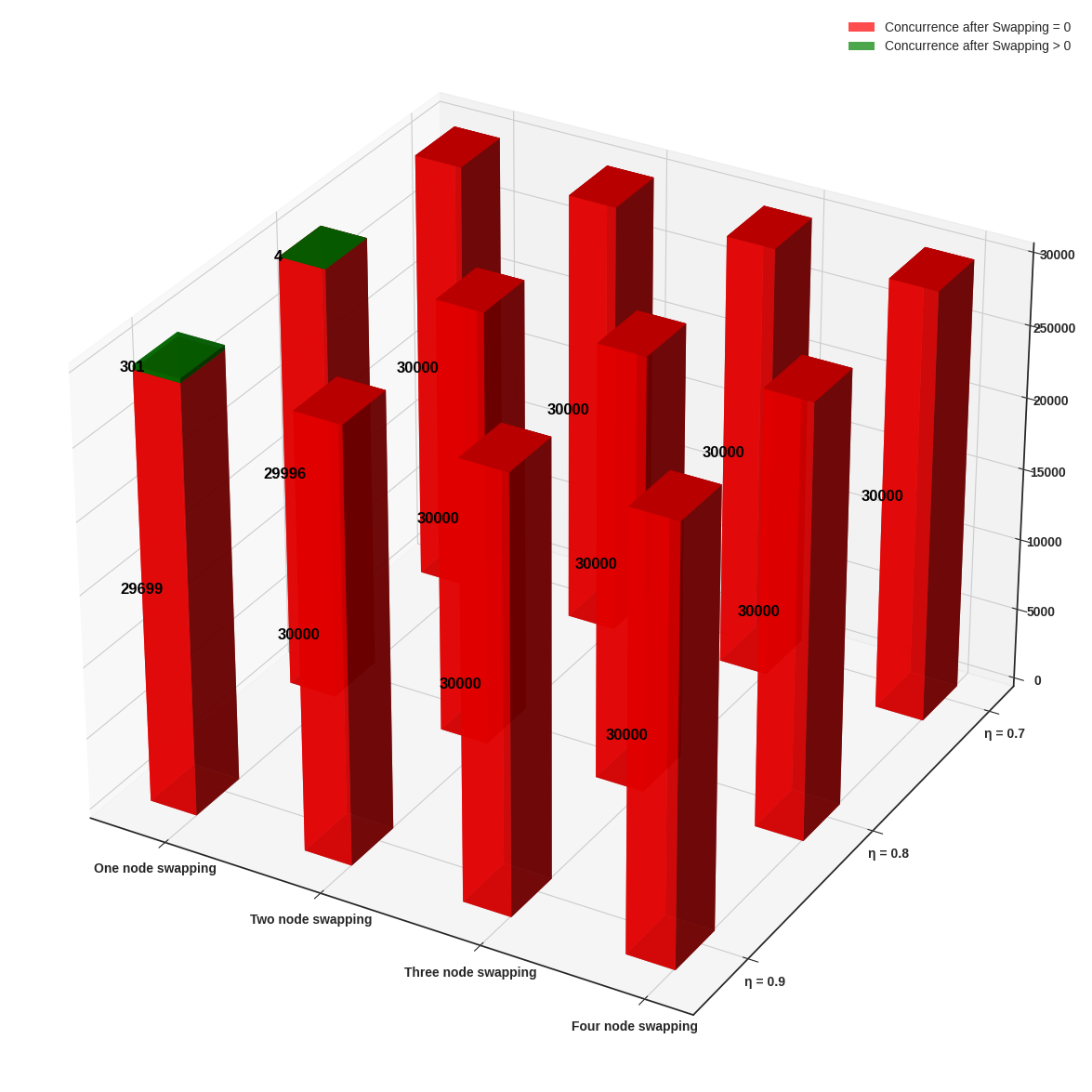}
    \caption{General mixed state multi node imperfect swapping with same imperfection}    \label{fig:mixed_state_multi_same_imp_conc}
\end{figure}

\noindent In Fig.(\ref{fig:mixed_state_multi_same_imp_conc}) X-axis represents multi node swapping with number of nodes as points on X-axis, Y-axis represents $\eta$ and Z-axis represents number of general mixed states. There are $30 \times 10^{3}$ states are considered and it can be observed from the graph that when $\eta = 0.9$ there are 301 states whose concurrence is greater than zero for single node swapping and when $\eta = 0.8$ there are 4 states whose concurrence is greater than zero for single node swapping. For the rest of swappings and imperfections all the states have zero concurrence.\\

\noindent\textbf{Case II:} Here we consider a scenario where the imperfection is different in different nodes. Let the imperfection parameter is $\eta_{i}$ in $i^{th}$ node.\\

\begin{figure}[h]
    \centering
    \includegraphics[scale = 0.37]{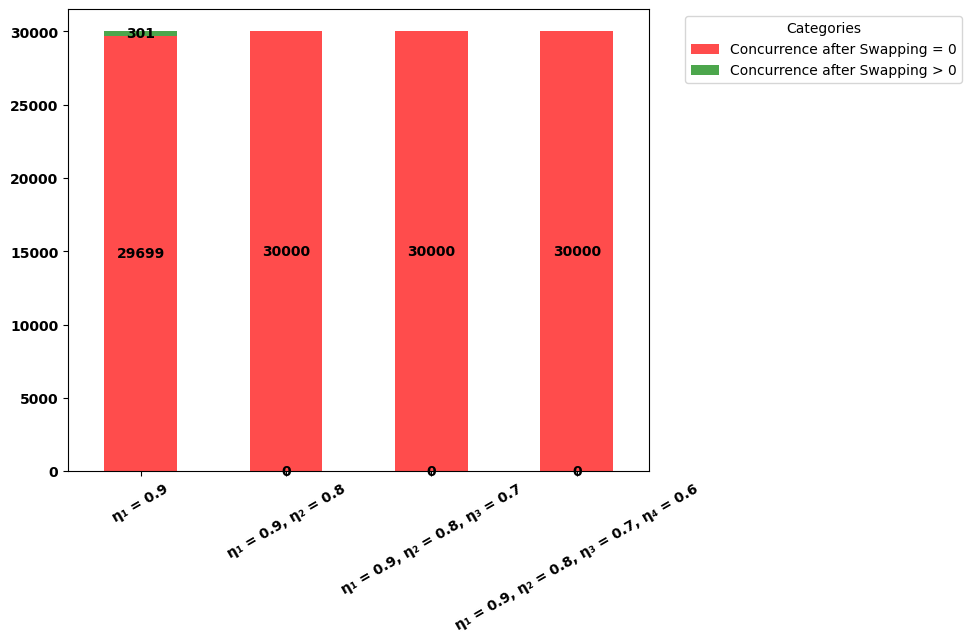}
    \caption{General mixed state multi node imperfect swapping with different imperfection}    \label{fig:mixed_state_multi_diff_eta_conc}
\end{figure}

\noindent In Fig.(\ref{fig:mixed_state_multi_diff_eta_conc}), X-axis is clubbed with the number of swappings and $\eta$, Y-axis represents the number of general mixed states considered. It can be observed from the figure that only when $\eta = 0.9$ are there 301 states whose concurrence is greater than zero, represented by a small green patch.  

\section{Relations On Teleportation Fidelity In Tripartite Remote Entangled Distribution (RED)}

\noindent Quantum teleportation is about sending information in an unknown quantum state from one place to another. This is done with the help of an entangled state. In other words, the entangled state acts as a resource for quantum teleportation. An entangled state is said to be useful for teleportation when the value of the teleportation fidelity ( given by the quantity $F_{max}$) is more than the classically achievable limit, which is $\frac{2}{3}$ \cite{15}. We can write the teleportation fidelity of an arbitrary two-qubit mixed state,
\begin{eqnarray}
 \rho= 1/4[I_{4 \times 4}+\sum_i \sigma_i \otimes I +\sum_j I \otimes \sigma_j +\sum_{i,j} t_{ij} \sigma_i \otimes \sigma_j] {}\nonumber\\&&  
\end{eqnarray}
(where $I$ is the identity matrix, $\sigma_i$ are Pauli matrices and $T=Tr(\rho \sigma_i \otimes \sigma_j)$ is the correlation matrix), as
\begin{equation}
 F_{max}(\rho) = \frac{1}{2}\left[1 + \frac{1}{3}(N(\rho)\right].   
\end{equation}
Here $N(\rho)$ = Trace($\sqrt{T\textsuperscript{\textdagger}T}$).  In such cases, only we can say that there is a definitive quantum advantage that we are gaining in the process of teleportation with the help of the entangled state. Hence these entangled states for which $F_{max}>2/3$ will be called a resource. The entangled Werner state in $2 \bigotimes 2$ dimensions is one example of a useful resource for teleportation for a certain range of classical probability of mixing. Other examples of mixed entangled states as a resource for teleportation are also there. \\

\noindent A quantum network is a collection of nodes interconnected by entangled states that allow information sharing. Here in this section, we consider the simplest scenario where we have two parties and a repeater between them. The resource states between the nodes and the repeater are mixed entangled states. This section obtains the relations for teleportation fidelity of a final state obtained as a result of intermediate swapping with that of the initial resource states. These relations are important in the context of determining the amount of information that flows in a quantum network. In other words, we connect the teleportation fidelities of entangled states before and after entanglement swapping. It significantly tells us about a
quantum network’s capacity to send quantum information between two desired nodes.

\subsection{Relations on Teleportation Fidelity for Werner States}

In this subsection, we consider the Werner state as an example. We also deal with a three-party scenario where the second party being the repeater, carries out Bell state measurement.

\subsubsection{Perfect Measurement}

\noindent Consider three nodes, a source node (Alice), a repeater node (Bob), and a target node (Charlie). Let $\rho_{12}$ be a Werner state between Alice and Bob, and $\rho_{23}$ be a Werner state between Bob and Charlie. After performing entanglement swapping, the final fidelity between Alice and Charlie in terms of initial fidelities can be defined as \\

\begin{equation}
    F_{max}(\ket{\psi_{13}}) = 2\left[\frac{\left(1 + {p_{1}p_{2}}\right)}{\left(1 + p_{1} \right)\left(1 + p_{2} \right)}\right]F(\ket{\psi_{12}})F(\ket{\psi_{23}}).
\end{equation}

\noindent Teleportation fidelity in terms of $p_{1}, p_{2}$ the visible parameters can be defined as\\

\begin{equation}
    F_{max}(\ket{\psi_{13}}) = \left[\frac{\left(1 + {p_{1}p_{2}}\right)}{2}\right].
\end{equation}

\noindent For final Teleportation fidelity to be greater than $\frac{2}{3}$, we should have
$\left[\frac{\left(1 + {p_{1}p_{2}}\right)}{2}\right] \geq \frac{2}{3}$. By solving it, we get the condition that the product of $p_{1}, p_{2}$ should be greater than $\frac{1}{3}$ and also the individual values of visible parameters should also be greater than $\frac{1}{3}$.\\

\begin{figure}[h]
    \centering
    \includegraphics[scale = 0.6]{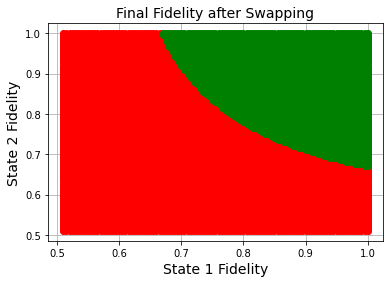}
    \caption{Werner State Single Node Perfect Swapping}
    \label{fig:label_20}
\end{figure}

\noindent In Fig.(\ref{fig:label_20}) X-axis represents fidelity of $\rho_{12}$ and Y-axis represent fidelity of $\rho_{23}$. All the red region in the graph represents the fidelities of output states $\leq \frac{2}{3}$, and the green region represents the fidelities $> \frac{2}{3}$ after entanglement swapping.

\subsubsection{Imperfect Measurements}

\noindent Similar to the previous scenario, here also starts with three nodes: a source node (Alice), a repeater node (Bob), and a target node (Charlie). Let $\rho_{12}$ be a Werner state between Alice and Bob, and $\rho_{23}$ be a Werner state between Bob and Charlie. After performing entanglement swapping with an imperfection $1-\eta$, the final fidelity can be written  in terms of initial fidelities as
\begin{equation}
    F_{max}(\ket{\psi_{13}}) = 2\left[\frac{\left(1 + \frac{\eta{p_{1}p_{2}}}{(\eta + 4(1 - \eta))}\right)}{\left(1 + p_{1} \right)\left(1 + p_{2} \right)}\right]F(\ket{\psi_{12}})F(\ket{\psi_{23}}).
\end{equation}
\noindent We can rewrite the final teleportation fidelity between the source and target node in terms of the initial resource visible parameters as 
\begin{equation}
    F_{max}(\ket{\psi_{13}}) = \frac{1}{2}\left[\left(1 + \frac{\eta{p_{1}p_{2}}}{(\eta + 4(1 - \eta))}\right)\right].
\end{equation}

\begin{figure}[h]
    \centering
    \includegraphics[scale = 0.38]{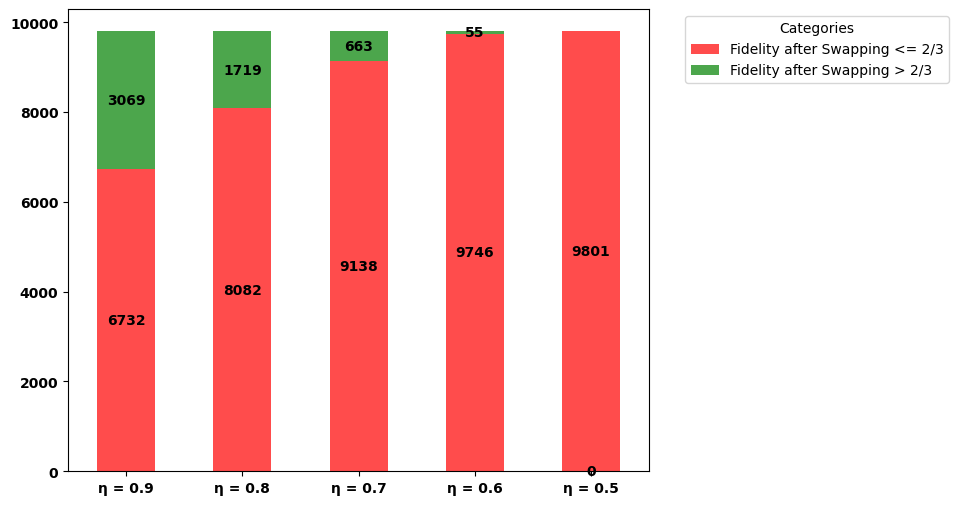}
    \caption{Werner State Single Node Imperfect Swapping}
    \label{fig:label_21}
\end{figure}

\noindent In Fig.(\ref{fig:label_21}), approximately $10 \times 10^{3}$ Werner state density matrices are considered after entanglement swappings. X-axis represents $\eta$ from 0.0 to 1.0, and Y-axis represents the number of Werner states. The counts in the green regions represent the number of output density matrices whose teleportation fidelity is greater than $\frac{2}{3}$, and the counts in the red region represent the number of density matrices whose teleportation fidelity is less than or equal to $\frac{2}{3}$ based on $\eta, p1, p2$. It is observed from the figure that if the $\eta$ value is less than or equal to 0.6, fidelity will be less than zero, irrespective of the values of visible parameters.

\subsection{Relations on Teleportation Fidelity for Bell Diagonal States}

As a  next example next, we consider the Bell diagonal state as a resource state between the nodes to start with. Then we carry out the usual process of entanglement swapping at the repeater node. 

\subsubsection{Perfect Measurement}

\noindent Consider three nodes, a source node (Alice), a repeater
node (Bob), and a target node (Charlie). In this case, $\rho_{12}$ is a Bell diagonal State between Alice and Bob, and $\rho_{23}$ is also a Bell diagonal State between Bob and Charlie. After performing entanglement swapping, the final teleportation fidelity of the state between Alice and Charlie when the measurement is perfect is given as, 
\begin{eqnarray}
    F_{max}(\ket{\psi_{13}}) && = 2\left[\frac{\left(1 + \frac{p_{1}^{1}p_{1}^{2}+ p_{2}^{1}p_{2}^{2} +p_{3}^{1}p_{3}^{2}}{3}\right)}{\left(1 + \frac{p_{1}^{1} +  p_{2}^{1} +p_{3}^{1}}{3}\right)\left(1 + \frac{p_{1}^{2} +  p_{2}^{2} +p_{3}^{2}}{3}\right)}\right]{}\nonumber\\&& F(\ket{\psi_{12}})F(\ket{\psi_{23}}).
\end{eqnarray}

\begin{figure}[h]
    \centering
    \includegraphics[scale = 0.5]{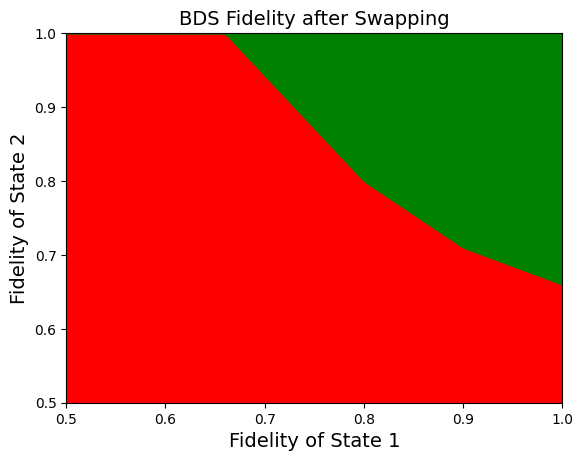}
    \caption{BDS Single Node Fidelity}
    \label{fig:label_bds_single_node_fidelity}
\end{figure}

\noindent In Fig.(\ref{fig:label_bds_single_node_fidelity}), we plot the teleportation fidelity of the output state after a single node perfect swapping. The green region represents the states whose teleportation fidelity value is greater than $\frac{2}{3}$. The complementary region is given by red color. Here X-axis and Y-axis represent the teleportation fidelity of two initial Bell diagonal states before swapping.

\subsubsection{Imperfect Measurements}
Consider three nodes, a source node (Alice), a repeater
node (Bob), and a target node (Charlie). Let $\rho_{12}$ be a Bell diagonal state between Alice and Bob, and $\rho_{23}$ be a Bell diagonal state between Bob and Charlie. After performing entanglement swapping, the final fidelity between Alice and Charlie, when the measurement is imperfect (the probability of success of measurement being $\eta$), is given as
\begin{eqnarray}
    &&F_{max}(\ket{\psi_{13}}) = 2{}\nonumber\\&&\left[\frac{\left(1 + \eta(\frac{p_{1}^{1}p_{1}^{2}+ p_{2}^{1}p_{2}^{2} +p_{3}^{1}p_{3}^{2}}{\eta + 4(1 - \eta)})\right)}{\left(1 + \frac{p_{1}^{1} -  p_{2}^{1} + p_{3}^{1}}{3}\right)\left(1 + \frac{p_{1}^{2} -  p_{2}^{2} +p_{3}^{2}}{3}\right)}\right]{}\nonumber\\&&F(\ket{\psi_{12}})F(\ket{\psi_{23}}).
\end{eqnarray}

 \begin{figure}[h]
    \centering
    \includegraphics[scale = 0.38]{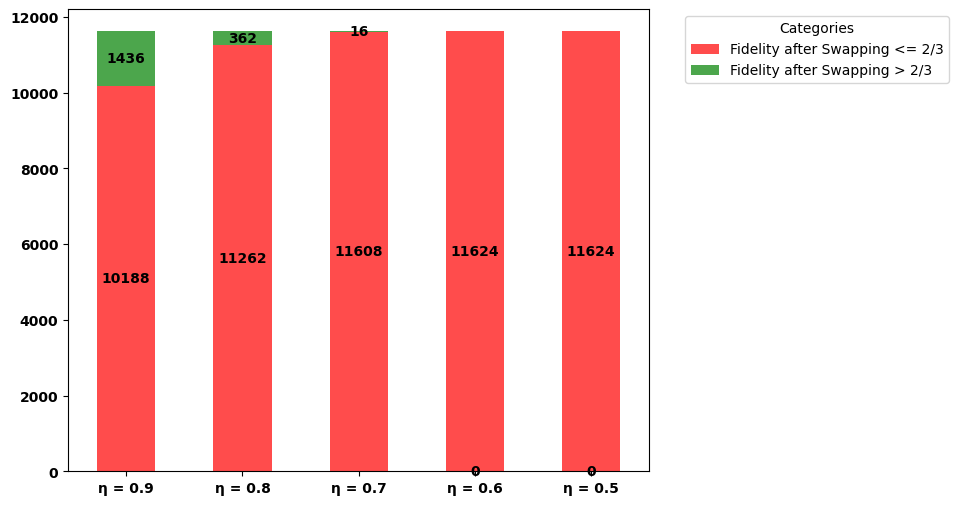}
    \caption{BDS Single Node Imperfect Swappings}
    \label{fig:label_22}
\end{figure}

\noindent In Fig.(\ref{fig:label_22}), the X-axis represents the probability of success of the Bell state measurement, and Y-axis represents the number of distinct Bell diagonal states considered for swapping. There are approximately $12 \times 10^{3}$ considered for entanglement swappings. All the red regions represent the number of Bell diagonal states whose fidelity is less than and equal to $\frac{2}{3}$ after entanglement swapping, and green regions represent the number of Bell diagonal states whose fidelity is greater than $\frac{2}{3}$ after entanglement swapping. We can see from the figure that after $\eta$ lower than 0.7, all the output states are not useful for teleportation. \\

\subsection{Relations on Teleportation Fidelity for General Mixed States}

\noindent In this subsection, we consider the general mixed state as a resource between Alice- Bob and also between Bob-Charlie. Then we carry out both perfect and imperfect measurements to see how the teleportation fidelity after swapping behaves with initial fidelity. 

\subsubsection{Perfect Measurements}

Consider three nodes, a source node (Alice), a repeater
node (Bob), and a target node (Charlie). Let $\rho_{12}$ be a general mixed state between Alice and Bob, and $\rho_{23}$ be a general mixed state between Bob and Charlie. After performing entanglement swapping, the final fidelity between Alice and Charlie when the measurement is perfect.

 \begin{figure}[h]
    \centering
    \includegraphics[scale = 0.52]{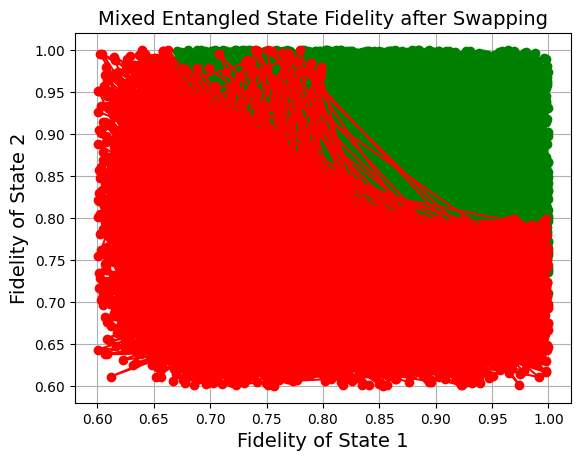}
    \caption{Mixed state Single Node Imperfect Swappings}
    \label{fig:mixed_state_single_node_perfect_tf}
\end{figure}

\noindent In Fig.(\ref{fig:mixed_state_single_node_perfect_tf}), X-axis represents teleportation fidelities of general mixed states $\rho_{12}$, and Y-axis represents teleportation fidelities of general mixed states $\rho_{23}$; input states considered for swapping. There are  $30 \times 10^{3}$ considered for entanglement swappings. All the red regions represent the number of general mixed states whose fidelity is less than or equal to $\frac{2}{3}$ after entanglement swapping, and green regions represent the number of general mixed states whose fidelity is greater than $\frac{2}{3}$ after entanglement swapping.

\subsubsection{Imperfect Measurements}
Consider three nodes, a source node (Alice), a repeater
node (Bob), and a target node (Charlie). Let $\rho_{12}$ be a general mixed state between Alice and Bob, and $\rho_{23}$ be a general mixed state between Bob and Charlie. After performing entanglement swapping, the final fidelity between Alice and Charlie when the measurement is imperfect.

 \begin{figure}[h]
    \centering
    \includegraphics[scale = 0.38]{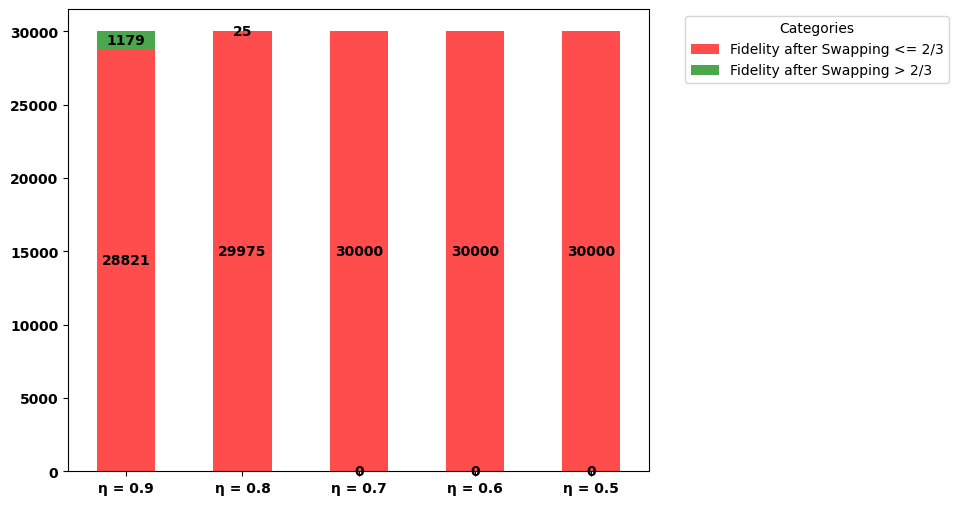}
    \caption{Mixed state Single Node Imperfect Swappings}
    \label{fig:mixed_state_single_node_imperfect_tf}
\end{figure}

\noindent In Fig.(\ref{fig:mixed_state_single_node_imperfect_tf}), X-axis represents the probability of success of the measurement, and Y-axis represents the number of distinct general mixed states considered for swapping. There are  $30 \times 10^{3}$ considered for entanglement swappings. All the red regions represent the number of general mixed states as output whose fidelity is less than and equal to $\frac{2}{3}$ after entanglement swapping, and green regions represent the number of general mixed states whose fidelity is greater than $\frac{2}{3}$ after entanglement swapping.

\section{Relations On Teleportation Fidelity In Multi partite with similar states Remote Entangled Distribution (RED)}

\noindent In this section, we begin with a 1-D network with $(n + 2)$ nodes. There are $n$ repeater stations between the source and target nodes. The main objective is to see whether, after $n$ node swapping, how the teleportation fidelity of the final state between source and target will depend upon the teleportation fidelity of the resource states. Here we consider resource states with the same parameters. We also consider both perfect and imperfect measurements.

\subsection{Relations on Teleportation Fidelity for Werner States}

\noindent As a first example, let us consider the Werner states as resource states between nodes. We have taken all the resource Werner states to be the same as far as the input state parameters of the state are considered.\\ 

\subsubsection{Perfect Measurements}
Consider $n+2$ nodes, a source node, $n$ repeater nodes, and a target node. After performing entanglement swapping at all the intermediate nodes, the final fidelity between the source and target node, when the measurement is perfect, is given as, 
\begin{eqnarray}
    &&F(\ket{\psi_{1, n+2}}) = 2^{n}\left[\frac{1 +  \left(p\right)^{n+1}}{\left({(1+p)^{n+1}}\right)}\right]{}\nonumber\\&&\prod\limits_{i = 1}^{n+1}F(\ket{\psi_{i, i+1}}).
\end{eqnarray}

\begin{figure}[h]
    \centering
    \includegraphics[scale = 0.38]{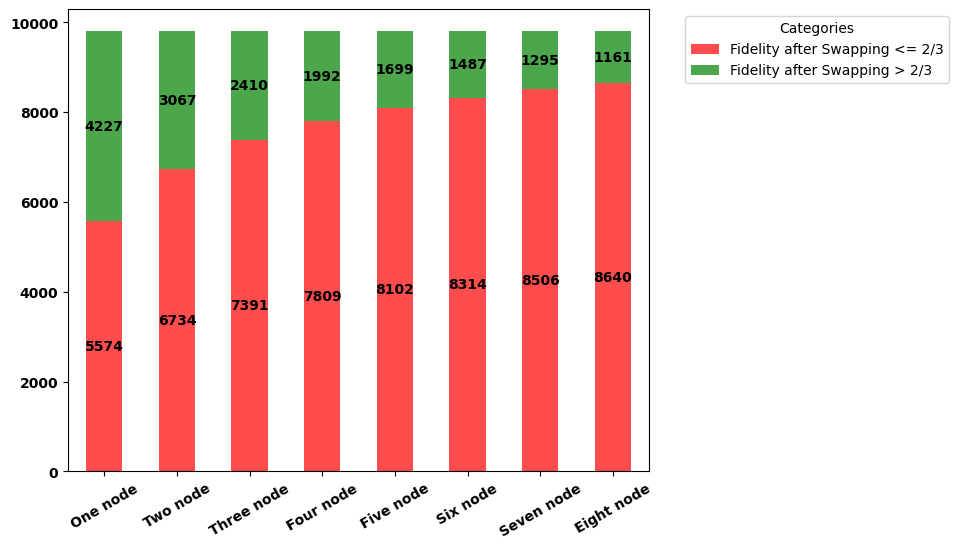}
    \caption{Werner State Multi Node Perfect Swappings with Same Visible Parameters}
    \label{fig:label_23}
\end{figure}
\noindent In Fig.(\ref{fig:label_23}), X-axis represents the number of nodes being swapped, and Y-axis represents the number of distinct Werner States considered for swapping. There are approximately $10 \times 10^{3}$ considered for entanglement swappings. All the red regions represent the number of output Werner States whose fidelity is less than or equal to $\frac{2}{3}$ after entanglement swapping, and green regions represent the number of Werner States whose fidelity is greater than $\frac{2}{3}$ after entanglement swapping. \\

\subsubsection{Imperfect Measurements}

\noindent\textbf{Case I:} Consider $n+2$ nodes: a source node, $n$ repeater nodes, and a target node. After performing entanglement swapping at all the intermediate nodes, the final fidelity between the source and target node when the measurement is imperfect ($1-\eta$ being the imperfection) with the imperfection being the same in all nodes, is given as \\

\begin{eqnarray}
    &&F(\ket{\psi_{1, n+2}}) = 2^{n}{}\nonumber\\&&\left[\frac{1 +  \frac{\eta^{n}\left(p\right)^{n+1}}{N}}{\left({(1+p)^{n+1}}\right)}\right]\prod\limits_{i = 1}^{n+1}F(\ket{\psi_{i, i+1}}).
\end{eqnarray}

\noindent The normalization constant $N$ is  given by, $(\eta^{n} + 4(1 - \eta)(\sum\limits_{i=0}^{n-1}(\eta^{n-1-i})(\eta + 4(1 - \eta))^{i}))$.

 \begin{figure}[h]
    \centering
    \includegraphics[scale = 0.24]{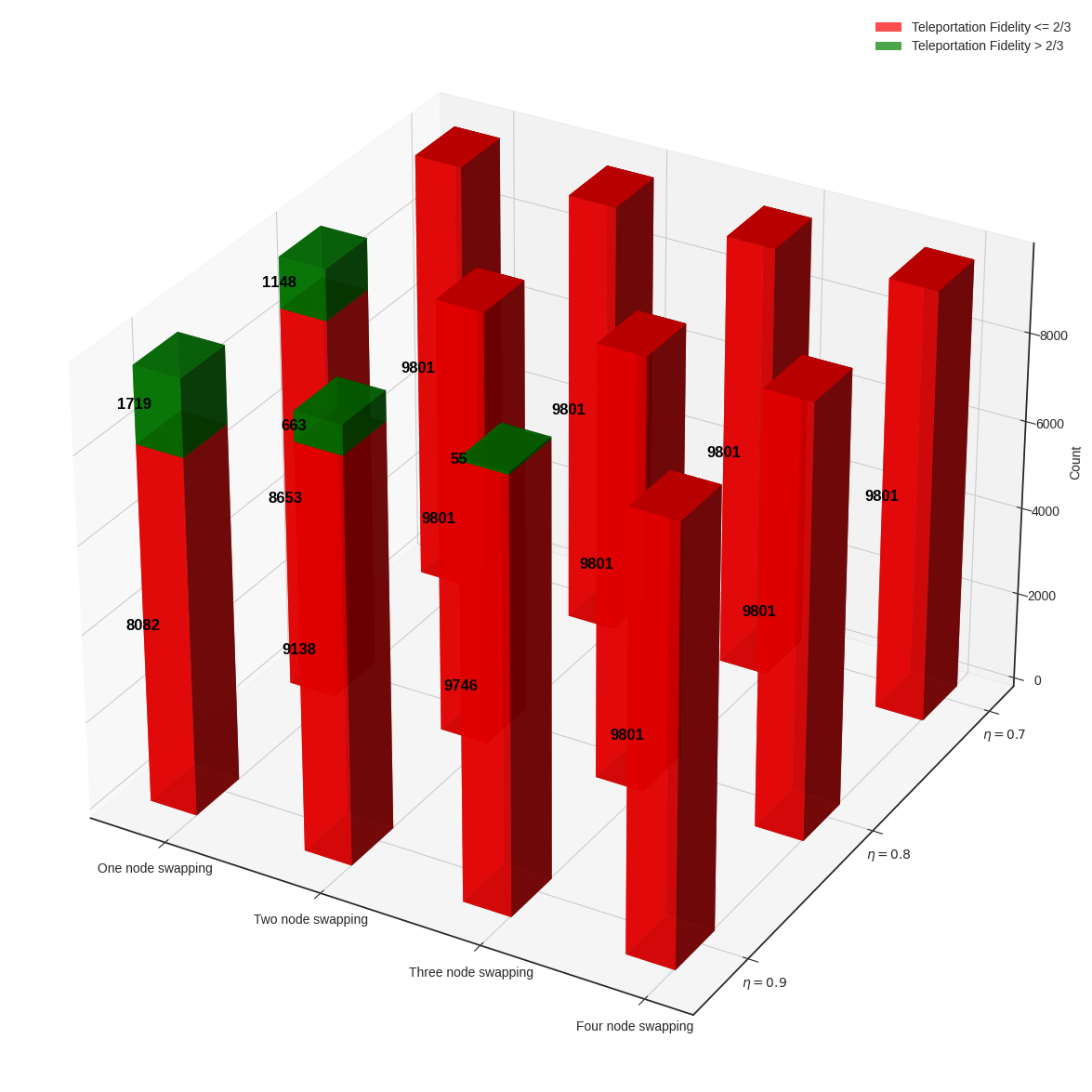}
    \caption{Werner State Multi Node Imperfect Swappings with Same Visible Parameters}
    \label{fig:label_24}
\end{figure}

\noindent In Fig.(\ref{fig:label_24}), X-axis represents the number of nodes being swapped, Y-axis represents $\eta$ from 0.0 to 1.0, and Z-axis represents the number of distinct Werner states considered for swapping. There are approximately $12 \times 10^{3}$ considered for entanglement swappings. All the red regions represent the number of output  Werner states whose fidelity is less than and equal to $\frac{2}{3}$ after entanglement swapping, and green regions represent the number of Werner states whose fidelity is greater than $\frac{2}{3}$ after entanglement swapping. \\

\noindent\textbf{Case II:} Consider $n+2$ nodes, a source node, $n$ repeater nodes, and a target node. After performing entanglement swapping at all the intermediate nodes, the final fidelity between the source and target node when the measurement is imperfect, with imperfection being different in all nodes, is given as 
\begin{eqnarray}
    &&F(\ket{\psi_{1, n+2}}) = 2^{n}{}\nonumber\\&&\left[\frac{1 +  \frac{(\prod\limits_{i = 1}^{n}\eta_{i})\left(p\right)^{n+1}}{N}}{\left({(1+p_{i})^{n+1}}\right)}\right]{}\nonumber\\&&\prod\limits_{i = 1}^{n+1}F(\ket{\psi_{i, i+1}}),
\end{eqnarray}
\noindent where the value of the normalization constant is given by,
\begin{equation}
N=(\prod\limits_{i=1}^{n}\eta_{i} + \sum\limits_{i=1}^{n}(\sum\limits_{cyc}(4^{i})(\prod\limits_{j=1}^{i}(1 - \eta_{i}))(\prod\limits_{k = i+1}^{n}\eta_{k}))).    
\end{equation}

\begin{figure}[h]
    \centering
    \includegraphics[scale = 0.3]{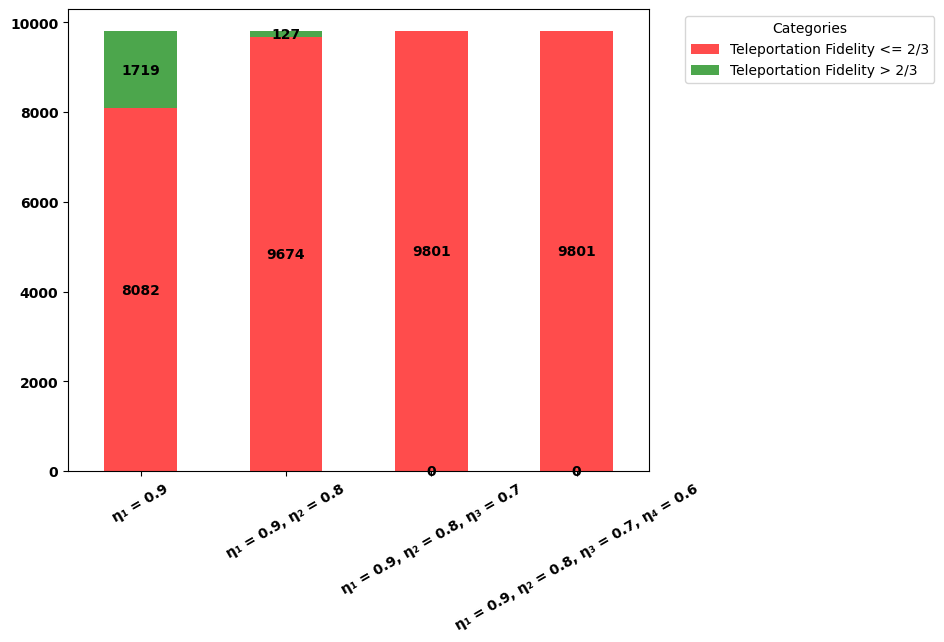}
    \caption{Werner State Multi Node Imperfect Swappings with Different Visible Parameters}
    \label{fig:label_25}
\end{figure}

\noindent  In Fig.(\ref{fig:label_25}), X-axis is clubbed with $\eta$ values, and the number of nodes that are swapped, and the Y-axis represents the number of distinct Werner states considered for swapping. There are approximately $10 \times 10^{3}$ considered for entanglement swappings. All the red regions represent the number of Werner states whose fidelity is less than or equal to $\frac{2}{3}$ after entanglement swapping, and green regions represent the number of Werner states whose fidelity is greater than $\frac{2}{3}$ after entanglement swapping. It can be observed from the graph that after two node swappings, with different $\eta$ values being 0.9, 0.8, and 0.7, all teleportation fidelity values, irrespective of visible parameter values, are less than or equal to $\frac{2}{3}$.

\subsection{Relations on Teleportation Fidelity for Bell Diagonal States}

\subsubsection{Perfect Measurements}
Consider $n+2$ nodes with a source node, $n$ repeater nodes, and a target node. After performing entanglement swapping at all the intermediate nodes, the final fidelity between the source and target node when the measurement is perfect is given as \\

\begin{eqnarray}
    &&F_{max}(\ket{\psi_{1, n+2}}) = 2^{n}{}\nonumber\\&&\left[\frac{\left(1 + {p_{1}^{n+1} + p_{2}^{n+1} +p_{3}^{n+1}}\right)}{\left((1 + {p_{1} + p_{2} + p_{3}})^{n+1}\right)}\right]{}\nonumber\\&&\prod\limits_{i}^{n+1}F(\ket{\psi_{i, i+1}}).
\end{eqnarray}

 \begin{figure}[h]
    \centering
    \includegraphics[scale = 0.38]{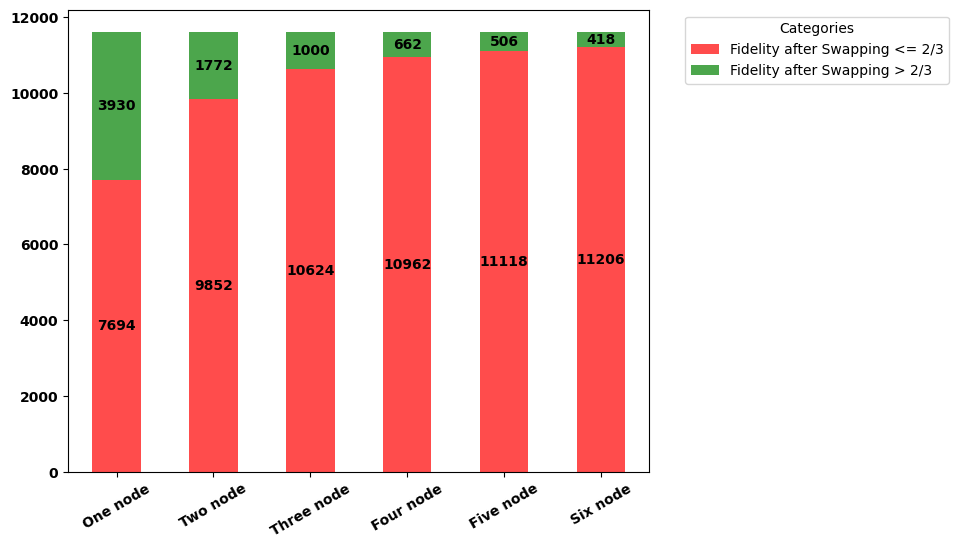}
    \caption{BDS Multi Node Perfect Swappings with Same Visible Parameters}
    \label{fig:label_26}
\end{figure}

\noindent In Fig.(\ref{fig:label_26}), X-axis represents the number of nodes being swapped, and Y-axis represents the number of distinct Bell diagonal states considered for swapping. There are approximately $12 \times 10^{3}$ considered for entanglement swappings. All the red regions represent the number of Bell diagonal states whose fidelity is less than or equal to $\frac{2}{3}$ after entanglement swapping, and green regions represent the number of output Bell diagonal states whose fidelity is greater than $\frac{2}{3}$ after entanglement swapping. \\

\subsubsection{Imperfect Measurements}

\noindent \textbf{Case I :} Consider $n+2$ nodes, with a source node, $n$ repeater nodes, and a target node. After performing entanglement swapping at all the intermediate nodes, the final fidelity between the source and target node when the measurement is imperfect ($\eta$ - the success probability of perfection), with the imperfection being the same in all nodes, is given as 
\begin{eqnarray}
    &&F_{max}(\ket{\psi_{1, n+2}}) = 2^{n} {}\nonumber\\&&\left[\frac{\left(1 + ({\eta^{n}})\frac{p_{1}^{n+1} + p_{2}^{n+1} + p_{3}^{n+1}}{N}\right)}{\left((1 + {p_{1} + p_{2} + p_{3}})^{n+1}\right)}\right]{}\nonumber\\&&\prod\limits_{i}^{n+1}F(\ket{\psi_{i, i+1}}).
\end{eqnarray}
\noindent where the normalization constant $N$ is having the value $(\eta^{n} + 4(1 - \eta)(\sum\limits_{i=0}^{n-1}(\eta^{n-1-i})(\eta + 4(1 - \eta))^{i}))$.

 \begin{figure}[h]
    \centering
    \includegraphics[scale = 0.26]{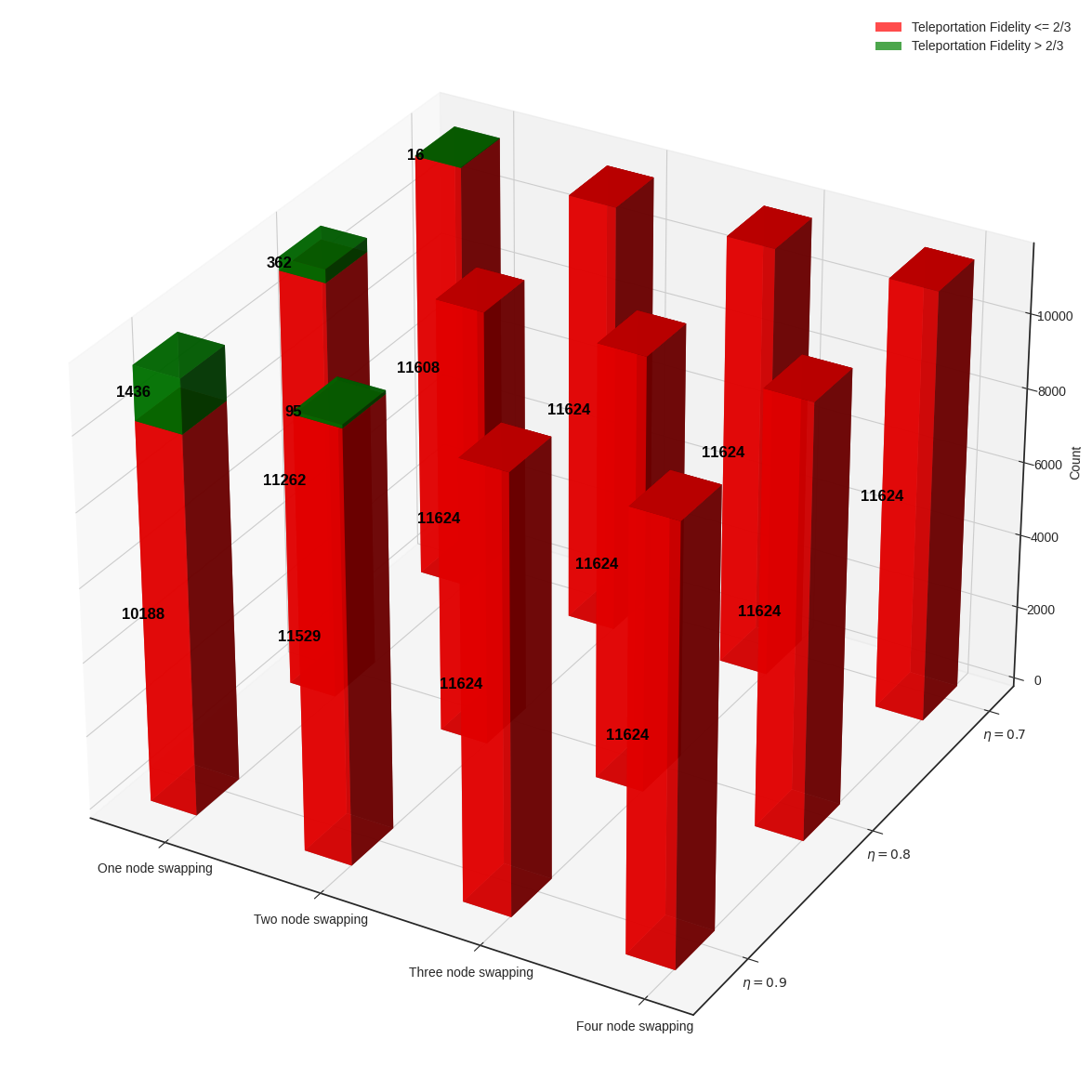}
    \caption{BDS Multi Node Imperfect Swappings with Same Visible Parameters}
    \label{fig:label_27}
\end{figure}

\noindent In Fig.(\ref{fig:label_27}), X-axis represents the number of nodes being swapped, Y-axis represents $\eta$ with values from 0.0 to 1.0, and Z-axis represents the number of distinct Bell diagonal states considered for swapping. There are approximately $12 \times 10^{3}$ considered for entanglement swappings. All the red regions represent the number of Bell diagonal states whose fidelity is less than $\frac{2}{3}$ after entanglement swapping, and green regions represent the number of Bell diagonal states whose fidelity is greater than $\frac{2}{3}$ after entanglement swapping. It can be observed from the graph that if the value of $\eta$ is less than 0.7, even for single node swapping, the value of teleportation fidelity is less than or equal to $\frac{2}{3}$. \\

\noindent \textbf{Case II :} In this case, after performing entanglement swapping at all the intermediate nodes, the final fidelity between the source and target node when the measurement is imperfect, and imperfection is different in all nodes is given as,
\begin{eqnarray}
    &&F_{max}(\ket{\psi_{1, n+2}}) = 2^{n}{}\nonumber\\&&\left[\frac{\left(1 + (\prod\limits_{i=1}^{n}{\eta_{i}})\frac{p_{1}^{n+1} + p_{2}^{n+1} +p_{3}^{n+1}}{N}\right)}{\left((1 + {p_{1} + p_{2} + p_{3}})\right)^{n+1}}\right]{}\nonumber\\&&\prod\limits_{i}^{n+1}F(\ket{\psi_{i, i+1}}).
\end{eqnarray}
\noindent where $N$  is the normalization constant and is given by,
\begin{equation}
 4(\prod\limits_{i=1}^{n}\eta_{i} + \sum\limits_{i=1}^{n}(\sum\limits_{cyc}(4^{i})(\prod\limits_{j=1}^{i}(1 - \eta_{i}))(\prod\limits_{k = i+1}^{n}\eta_{k}))).   
\end{equation}

\begin{figure}[h]
    \centering
    \includegraphics[scale = 0.32]{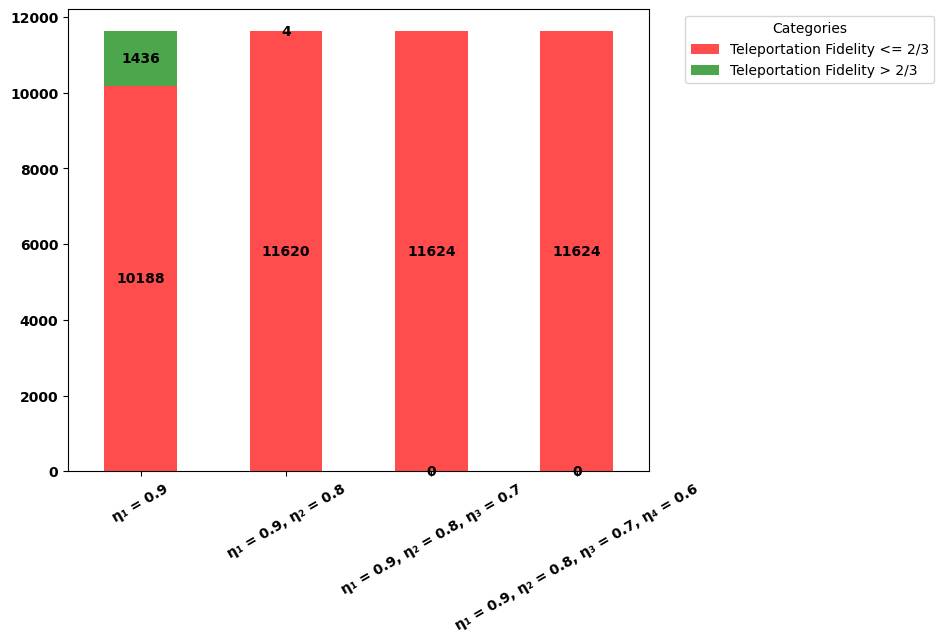}
    \caption{BDS Multi Node Imperfect Swappings with Different Imperfections}
    \label{fig:label_28}
\end{figure}

\noindent  In Fig.(\ref{fig:label_28}), X-axis represents the number of nodes used for swapping with different $\eta$ values from 0.0 to 1.0, and Y-axis represents the number of distinct Bell diagonal states considered for swapping. There are approximately $12 \times 10^{3}$ considered for entanglement swappings. All the red regions represent the number of Bell diagonal states whose fidelity is less than $\frac{2}{3}$ after entanglement swapping, and green regions represent the number of Bell diagonal states whose fidelity is greater than $\frac{2}{3}$ after entanglement swapping. It can be observed that after two node swappings, irrespective of visible parameter values, teleportation fidelity is less than or equal to $\frac{2}{3}$.

\subsection{Relations of Teleportation Fidelities for General Mixed States}

In this section, we consider the  Bloch vector representation of two qubit mixed states. We have only numerical evidence of how teleportation fidelity behaves after swapping, considering both perfect and imperfect measurement scenarios.

\subsubsection{Perfect Measurements}

\noindent Consider $n+2$ nodes, a source node, $n$ repeater nodes, and a target node. After performing entanglement swapping at all the intermediate nodes, the final fidelity between the source and target node, when the measurement is perfect, is given in the following graph.

\begin{figure}[h]
    \centering
    \includegraphics[scale = 0.32]{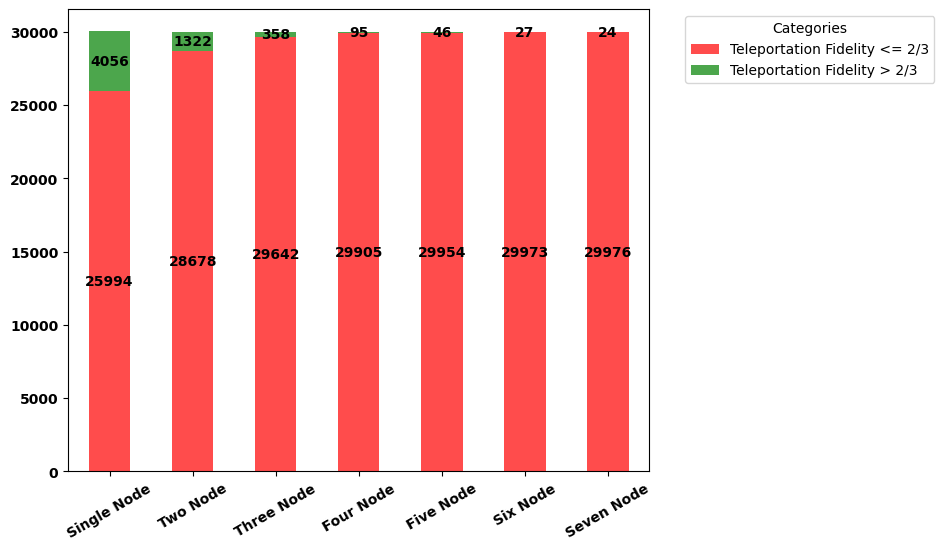}
    \caption{General Mixed State Multi Node Perfect Swappings}
    \label{fig:mixed_state_multi_node_same_parm_per_tf}
\end{figure}

\noindent In Fig.(\ref{fig:mixed_state_multi_node_same_parm_per_tf}) X-axis represents the number of nodes used for swappings, and the Y-axis represents output states. All the red regions represent output states with teleportation fidelity less than or equal to $\frac{2}{3}$, and the green regions represent output states with teleportation fidelity greater than $\frac{2}{3}$. It can be observed from the figure that after three node swappings, the teleportation fidelity of the output states is less than $\frac{2}{3}$ even though swapping is perfect for the given input states.


\subsubsection{Imperfect Measurements}
\noindent Consider the same scenario with $n+2$ nodes having a source node, $n$ repeater nodes, and a target node. After performing entanglement swapping at all the intermediate nodes, here we study the final fidelity between the source and target node when the measurement is imperfect.\\

\noindent \textbf{Case I :}  After performing entanglement swapping at all the intermediate nodes, the final fidelity between the source and target node when the measurement is imperfect,  with the same imperfections ($1-\eta$), in all nodes.\\

\begin{figure}[h]
    \centering
    \includegraphics[scale = 0.26]{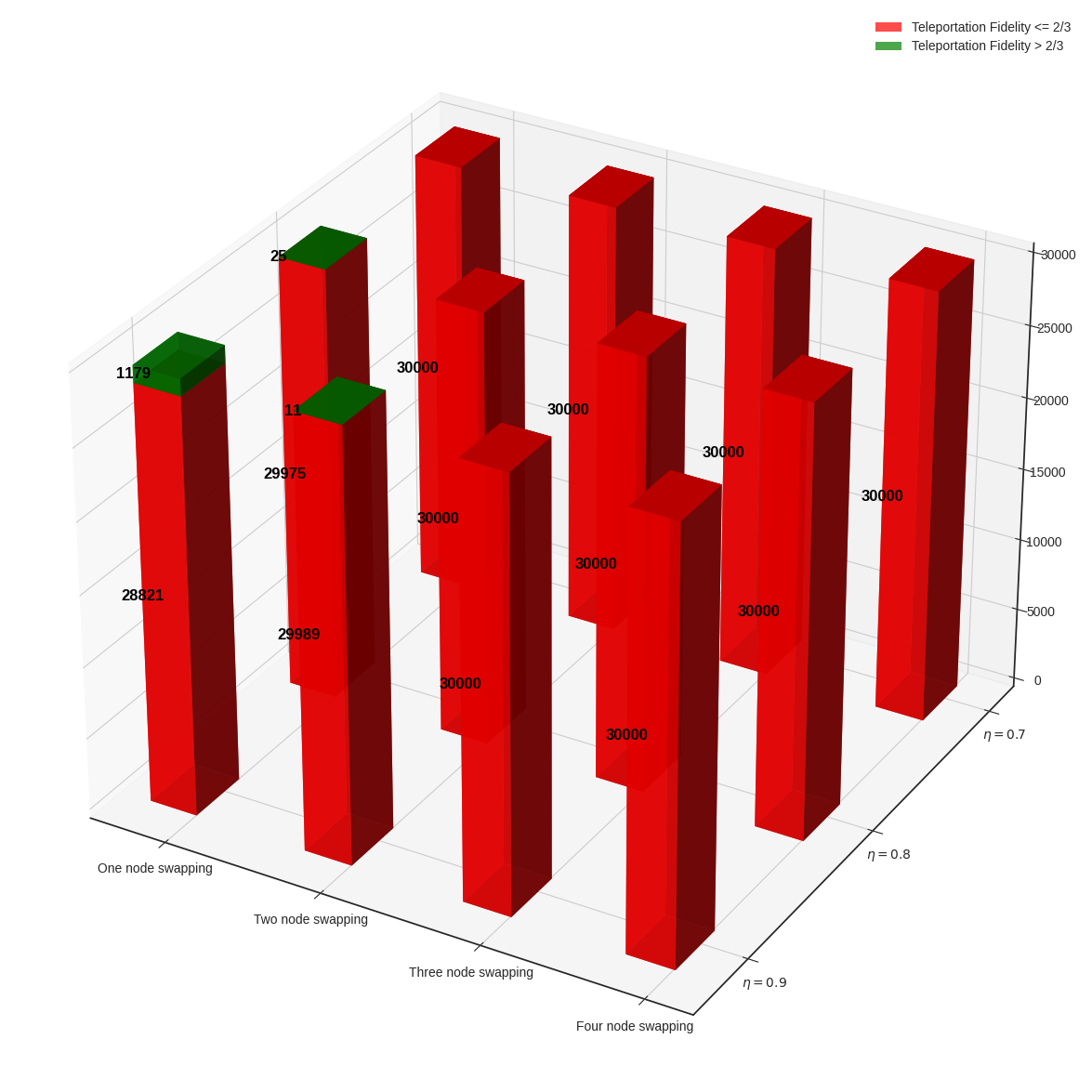}
    \caption{General Mixed State Multi Node Imperfect Swappings with same imperfections}
    \label{fig:mixed_state_multi_node_imp_tf}
\end{figure}

\noindent In Fig.(\ref{fig:mixed_state_multi_node_imp_tf})
X-axis represents the number of nodes used in swappings, Y-axis represents $\eta$, and Z-axis represents the number of general mixed states considered. All the red regions represent states whose fidelity is less than or equal to $\frac{2}{3}$ after entanglement swapping, and the green region represents states whose fidelity is greater than $\frac{2}{3}$ after entanglement swapping.


\noindent \textbf{Case II :} Here we consider the case when the measurement is imperfect, and the imperfection $1-\eta_i$ is different in each measurement.\\

\begin{figure}[h]
    \centering
    \includegraphics[scale = 0.32]{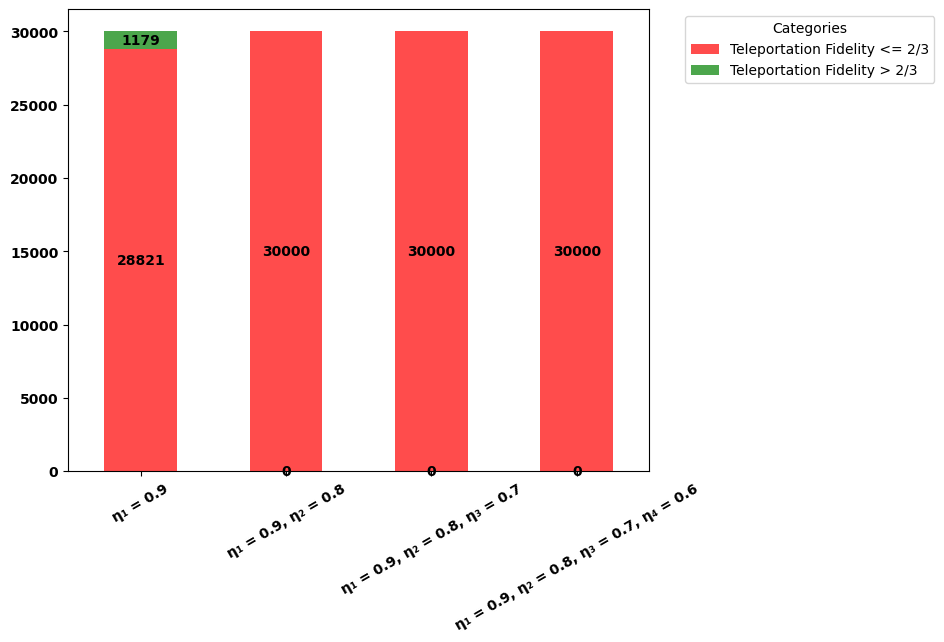}
    \caption{General Mixed State Multi Node Imperfect Swappings with different imperfections}
    \label{fig:mixed_state_multi_node_imp_diff_eta_tf}
\end{figure}

\noindent In Fig.(\ref{fig:mixed_state_multi_node_imp_diff_eta_tf}) X-axis represents multi-node swappings with different imperfections, and Y-axis represents output states after entanglement swappings. It can be observed front the graph that the teleportation fidelity is greater than $\frac{2}{3}$ only for a few output states in single node swapping with $\eta = 0.9$

\section{Relations On Teleportation Fidelity In Multipartite with different states Remote Entangled Distribution (RED)}

In this section, we once again consider a 1-D network with $(n+2)$ nodes. Just like the previous scenario, here, also we have $n$ repeater stations between the source and target nodes. Our target is to see, after $n$ nodes swapping, how the teleportation fidelity of the final state between source and target will depend upon the teleportation fidelity of the resource states. Interestingly in this section here, we consider resource states with a different set of parameters. We consider both perfect and imperfect measurement scenarios (characterized by the probability of success $\eta$). First, we consider the Werner state and then, subsequently, Bell diagonal state and general mixed state. 

\subsection{Relations of Teleportation Fidelities for Werner States}
In this section, we start with Werner states with different visible parameters. We consider two cases: 1) Perfect Measurement and 2) Imperfect Measurement, as mentioned earlier. 

\subsubsection{Perfect Measurement}
 Let us begin with a 1-D network having $n+2$ nodes. This network consists of a source node, $n$ repeater nodes, and a target node. All the resource states between the initial and target nodes are Werner states, but they are with different input/visible parameters. After performing entanglement swapping at all the intermediate nodes, the final fidelity between the source and target node, when the measurement is perfect, is given as 
 
\begin{eqnarray}
    &&F(\ket{\psi_{1, n+2}}) = 2^{n}{}\nonumber\\&&\left[\frac{1 +  \left(\prod\limits_{i = 1}^{n+1}p_{i} \right)}{\left( \prod\limits_{i = 1}^{n+1}{(1+p_{i})}\right)}\right] \prod\limits_{i = 1}^{n+1}F(\ket{\psi_{i, i+1}}).
\end{eqnarray}

\noindent We can express the final fidelity in terms of visible parameters as, 
\begin{eqnarray}
    F(\ket{\psi_{1, n+2}}) = \frac{1}{2}\left[{1 +  \left(\prod\limits_{i = 1}^{n+1}p_{i} \right)}\right].
\end{eqnarray}
\noindent From the above equation it can be observed that when the product of visible parameters is greater than $\frac{1}{3}$ then overall teleportation fidelity will be greater than $2/3$.
\begin{figure}[h]
    \centering
    \includegraphics[scale = 0.38]{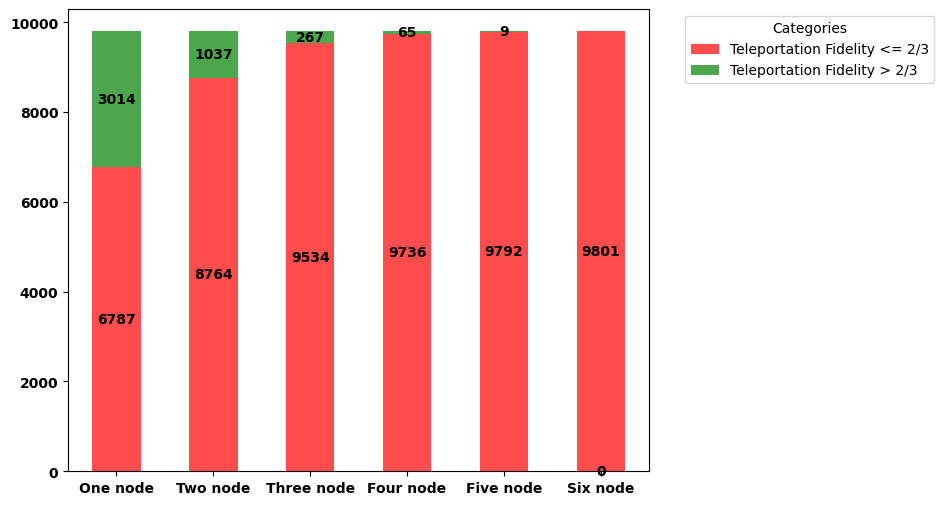}
    \caption{Werner Multi Node  Swappings with different Visible Parameters}
    \label{fig:label_29}
\end{figure}

\noindent In Fig.(\ref{fig:label_29}) X-axis represents the number of nodes used for entanglement swapping, Y-axis represents the number of distinct Werner states. There are approximately $10 \times 10^{3}$ considered for swapping, and all the red regions represent the number of final states whose final teleportation fidelity is less than equal to $\frac{2}{3}$ after swapping and green region represents the number of states whose teleportation fidelity is greater than $\frac{2}{3}$. It can be observed from the graph that for the considered Werner states, teleportation fidelity is less than $\frac{2}{3}$ for all the states from six node swappings.

\subsubsection{Imperfect Measurement with same imperfections}
Here in this subsection, we consider imperfect measurements at each of the repeater's stations instead of perfect measurements. The scenario is the same with $n+2$ nodes, having a source node, $n$ repeater nodes, and a target node. \\

\noindent \textbf{Case I:} For simplicity, first of all, we consider the measurement imperfections to be the same for each node. Here also, the resource states are again with different input parameters. After performing entanglement swapping at all the intermediate nodes, the final fidelity between the source and target node, when the measurement is imperfect, is characterized by the parameter $1-\eta$, \\

\begin{equation}
    F(\ket{\psi_{1, n+2}}) = 2^{n}\left[\frac{1 +  \frac{\eta^{n}\left(\prod\limits_{i = 1}^{n+1}p_{i}\right)}{N}}{\left({\prod\limits_{i = i}^{n+1}}(1+p_{i})\right)}\right]\prod\limits_{i = 1}^{n+1}F(\ket{\psi_{i, i+1}}).
\end{equation}
\noindent Here $\eta$ is the success probability of doing the perfect measurement and $N$ is the normalization constant with its value equal to $(\eta^{n} + 4(1 - \eta)(\sum\limits_{i=0}^{n-1}(\eta^{n-1-i})(\eta + 4(1 - \eta))^{i}))$.

 \begin{figure}[h]
    \centering
    \includegraphics[scale = 0.26]{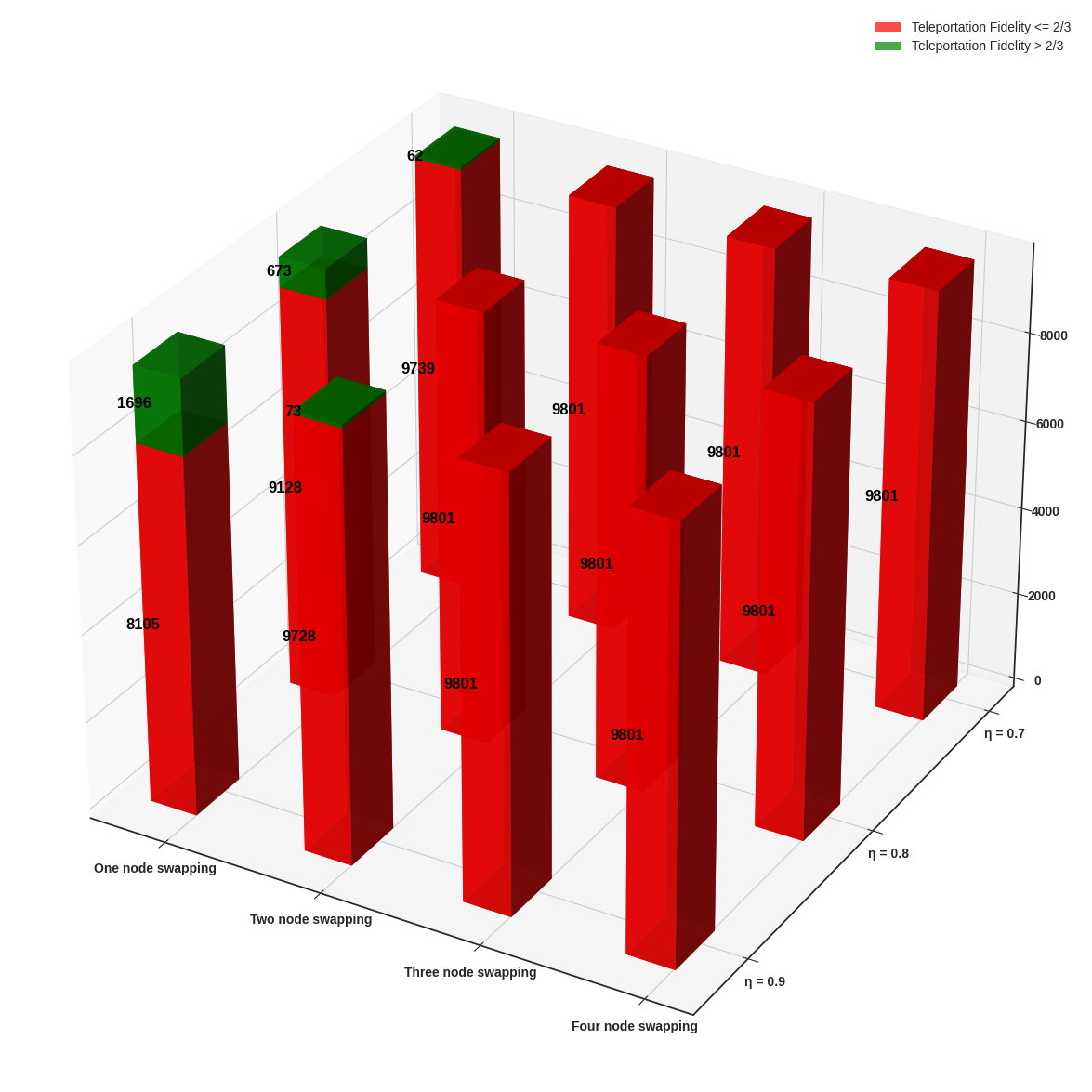}
    \caption{Werner Multi Node Imperfect Swappings with different Visible Parameters}
    \label{fig:label_30}
\end{figure}

\noindent In Fig.(\ref{fig:label_30}) X-axis represents the number of nodes used in swappings, Y-axis represents $\eta$, and Z-axis represents the number of Werner states considered. All the red regions represent states whose fidelity is less than or equal to $\frac{2}{3}$ after entanglement swapping, and the green region represents states whose fidelity is greater than $\frac{2}{3}$ after entanglement swapping.\\

\noindent \textbf{Case II :} In this case, all the measurement imperfections are different. So instead of one $1-\eta$, these will be characterized by $1-\eta_1, 1-\eta_2 \cdot 1-\eta_n$. After performing entanglement swapping at all the intermediate nodes,  the final fidelity between the source and target node in terms of the teleportation fidelities of the resource state is given as, 

\begin{eqnarray}
    &&F(\ket{\psi_{1, n+2}}) = 2^{n}{}\nonumber\\&&\left[\frac{1 +  \frac{(\prod\limits_{i = 1}^{n}\eta_{i})\left(\prod\limits_{i=1}^{n+1}p_{i}\right)}{N}}{\left({\prod\limits_{i=1}^{n+1}}(1+p_{i})\right)}\right]{}\nonumber\\&&\prod\limits_{i = 1}^{n+1}F(\ket{\psi_{i, i+1}}).
\end{eqnarray}
\noindent The normalization constant in this case is given by,
\begin{equation}
N=(\prod\limits_{i=1}^{n}\eta_{i} + \sum\limits_{i=1}^{n}(\sum\limits_{cyc}(4^{i})(\prod\limits_{j=1}^{i}(1 - \eta_{i}))(\prod\limits_{k = i+1}^{n}\eta_{k})))    
\end{equation}.

 \begin{figure}[h]
    \centering
    \includegraphics[scale = 0.36]{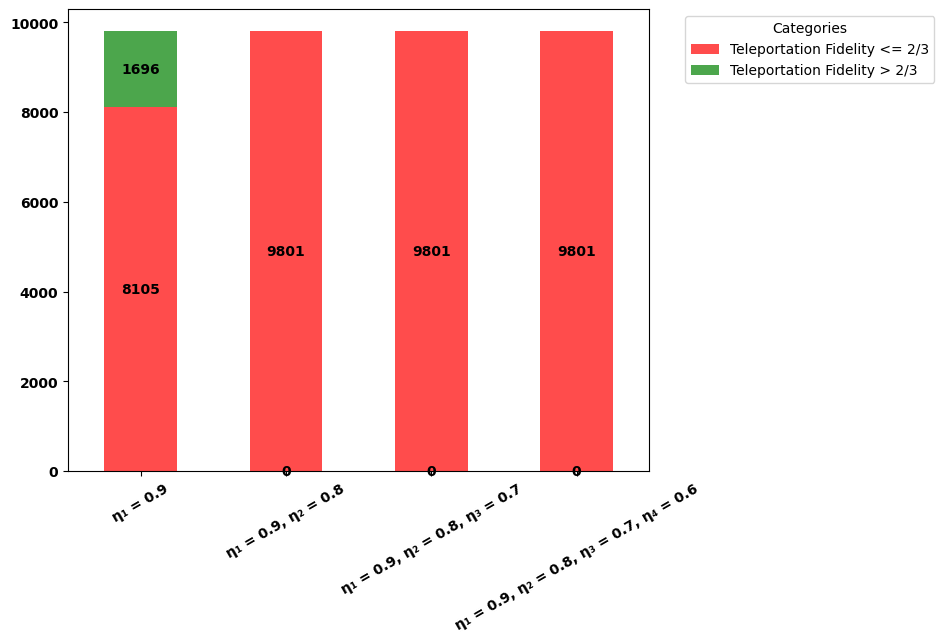}
    \caption{Werner Multi Node Imperfect Swappings with different Visible Parameters}
    \label{fig:label_31}
\end{figure}

\noindent In Fig.(\ref{fig:label_31}) X-axis represents the number of nodes used for swapping with different $\eta$ values, and Y-axis represents the number of Werner states considered for Fidelity. All the red regions represent states whose Fidelity is less than or equal to $\frac{2}{3}$, and green regions represent states whose Fidelity is greater than $\frac{2}{3}$. There are approximately $10 \times 10^{3}$ Werner states considered, and Fidelity can be observed only when $\eta = 0.9$ for some states.

\subsection{Relations of Teleportation Fidelities for Bell Diagonal States}

\subsubsection{Perfect Measurements}
Consider $n+2$ nodes, a source node, $n$ repeater nodes, and a target node. After performing entanglement swapping at all the intermediate nodes with different visible parameters, the final fidelity between the source and target node, when the measurement is perfect, is given as, 
\begin{eqnarray}
    &&F_{max}(\ket{\psi_{1, n+2}}) = 2^{n}{}\nonumber\\&&\left[\frac{\left(1 + ({\prod\limits_{i=1}^{n+1}p_{1}^{i} + \prod\limits_{i=1}^{n+1}p_{2}^{i} +\prod\limits_{i=1}^{n+1}p_{3}^{i}}\right)}{\left(\prod\limits_{i=1}^{n+1}(1 + {p_{1}^{i} + p_{2}^{i} + p_{3}^{i}})\right)}\right]{}\nonumber\\&&\prod\limits_{i}^{n+1}F(\ket{\psi_{i, i+1}}).
\end{eqnarray}

 \begin{figure}[h]
    \centering
    \includegraphics[scale = 0.38]{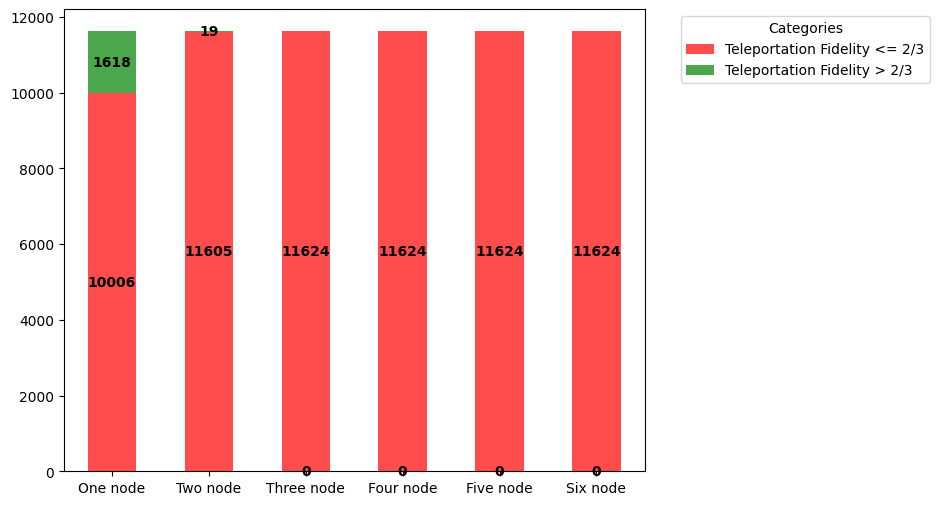}
    \caption{BDS Multi Node Perfect Swappings with different Visible Parameters}
    \label{fig:label_32}
\end{figure}

\noindent In Fig.(\ref{fig:label_32}) X-axis represents the number of nodes used for swappings, Y-axis represents the number of states considered for swappings. There are approximately $12 \times 10^{3}$ stated considered, red regions represent the number of output states whose fidelity is less than or equal to $\frac{2}{3}$ after entanglement swapping, and green regions represent the number of output states whose fidelity is greater than $\frac{2}{3}$. It can be observed from the graph that for the considered Bell Diagonal States, fidelity is greater than $\frac{2}{3}$ only upto two nodes swappings.\\

\subsubsection{Imperfect Measurements}
\noindent \textbf{Case I : } Let us again consider the scenario with  $n+2$ nodes, a source node, $n$ repeater nodes, and a target node. After performing entanglement swapping at all the intermediate nodes with different visible parameters, the final fidelity between source and target node when the measurement is imperfect with $1-\eta$ being the imperfection in measurement (same in all nodes), is given as 
\begin{eqnarray}
    &&F_{max}(\ket{\psi_{1, n+2}}) = 2^{n} {}\nonumber\\&&\left[\frac{\left(1 + ({\eta^{n}})\frac{\prod\limits_{i=1}^{n+1}p_{1}^{i} + \prod\limits_{i=1}^{n+1}p_{2}^{i} +\prod\limits_{i=1}^{n+1}p_{3}^{i}}{N}\right)}{\left(\prod\limits_{i=1}^{n+1}(1 + {p_{1}^{i} + p_{2}^{i} + p_{3}^{i}})\right)}\right]{}\nonumber\\&&\prod\limits_{i}^{n+1}F(\ket{\psi_{i, i+1}}).
\end{eqnarray}
\noindent Here the normalization constant $N$ is  
\begin{equation}
 N=  (\eta^{n} + 4(1 - \eta)(\sum\limits_{i=0}^{n-1}(\eta^{n-1-i})(\eta + 4(1 - \eta))^{i})).
\end{equation}

 \begin{figure}[h]
    \centering
    \includegraphics[scale = 0.26]{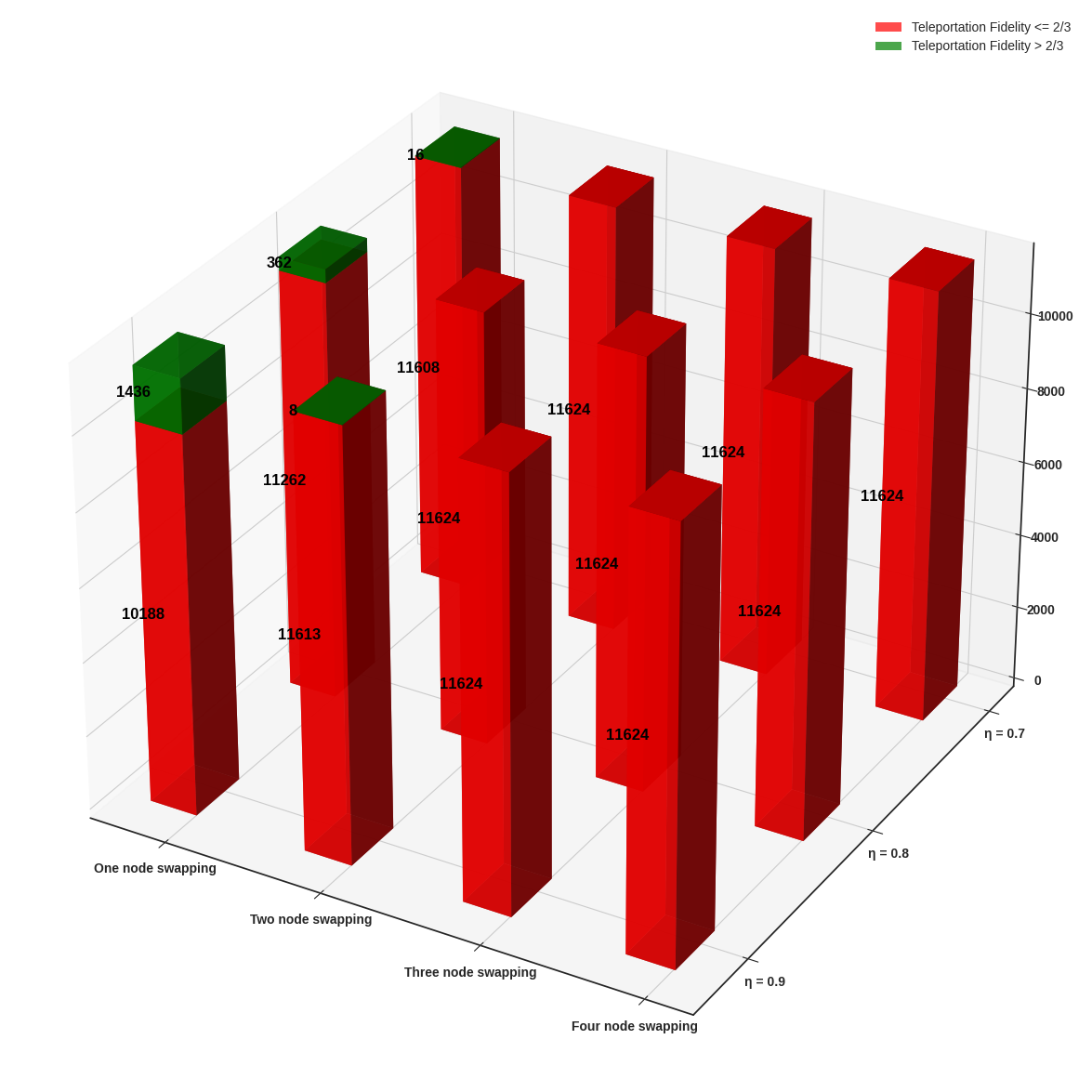}
    \caption{BDS Multi Node Imperfect Swappings with different Visible Parameters and Same Imperfection}
    \label{fig:BDS_Fidelity_diff_vis_eta}
\end{figure}

\noindent In Fig.(\ref{fig:BDS_Fidelity_diff_vis_eta}) X-axis represents the number of nodes used for swapping, Y-axis represents $\eta$ from 0.0 to 0.9, and Z-axis represents the number of states considered for entanglement swappings. All the red regions represent states whose fidelity is less than or equal to $\frac{2}{3}$ after swapping, and green regions represent output states whose fidelity is greater than $\frac{2}{3}$\\

\noindent \textbf{Case II: } Consider $n+2$ nodes, a source node, $n$ repeater nodes, and a target node. After performing entanglement swapping at all the intermediate nodes with different visible parameters, the final fidelity between source and target node when the measurement is imperfect $1-\eta_{i}$ with imperfection being different in all nodes, is given as 
\begin{eqnarray}
    &&F_{max}(\ket{\psi_{1, n+2}}) = 2^{n}{}\nonumber\\&&\left[\frac{\left(1 + (\prod\limits_{i=1}^{n}{\eta_{i}})\frac{\prod\limits_{i=1}^{n+1}p_{1}^{i} + \prod\limits_{i=1}^{n+1}p_{2}^{i} +\prod\limits_{i=1}^{n+1}p_{3}^{i}}{N}\right)}{\left(\prod\limits_{i=1}^{n+1}(1 + {p_{1}^{i} + p_{2}^{i} + p_{3}^{i}})\right)}\right]{}\nonumber\\&&\prod\limits_{i}^{n+1}F(\ket{\psi_{i, i+1}}).
\end{eqnarray}

\noindent where the normalization constant $N$ is given by
\begin{equation}
N=4(\prod\limits_{i=1}^{n}\eta_{i} + \sum\limits_{i=1}^{n}(\sum\limits_{cyc}(4^{i})(\prod\limits_{j=1}^{i}(1 - \eta_{i}))(\prod\limits_{k = i+1}^{n}\eta_{k}))).    
\end{equation}

 \begin{figure}[h]
    \centering
    \includegraphics[scale = 0.36]{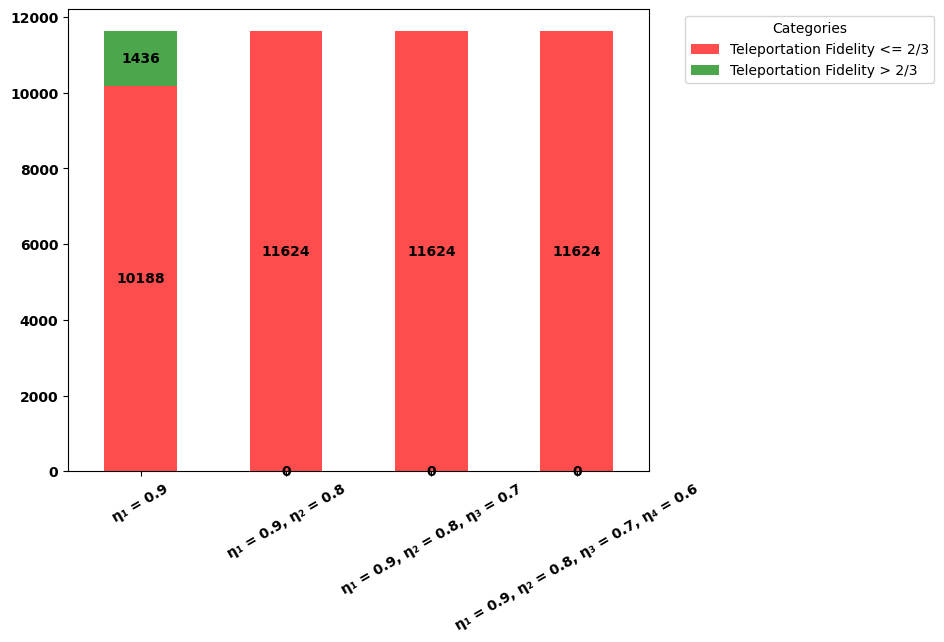}
    \caption{BDS Multi Node Imperfect Swappings with different Visible Parameters and different Imperfection}
    \label{BDS_Fidelity_diff_vis_diff_eta}
\end{figure}

\noindent In Fig.(\ref{BDS_Fidelity_diff_vis_diff_eta}) X-axis represents multi-node swappings with different $\eta$, and Y-axis represents the number of states considered. All the red regions represent states whose fidelity is less than or equal to $\frac{2}{3}$
and green regions represent states whose fidelity is greater than $\frac{2}{3}$. It can be observed from the figure that only when $\eta = 0.9$ very few states have fidelity greater than $\frac{2}{3}$ after entanglement swapping.
 
\subsection{Relations on Teleportation Fidelity for General Mixed States}
Let us begin with a 1-D network having $n+2$ nodes. This network consists of a source node, $n$ repeater nodes, and a target node. All the resource states between the nodes are general mixed states, but they are with different input parameters. In this subsection, we are only able to give numerical evidence for what happens to teleportation fidelity after entanglement swapping. 

\subsubsection{Perfect Measurements}
After performing entanglement swapping at all the intermediate nodes, the final fidelity between the source and target node, when the measurement is perfect, can be observed in the below graph.
\begin{figure}[h]
    \centering
    \includegraphics[scale = 0.34]{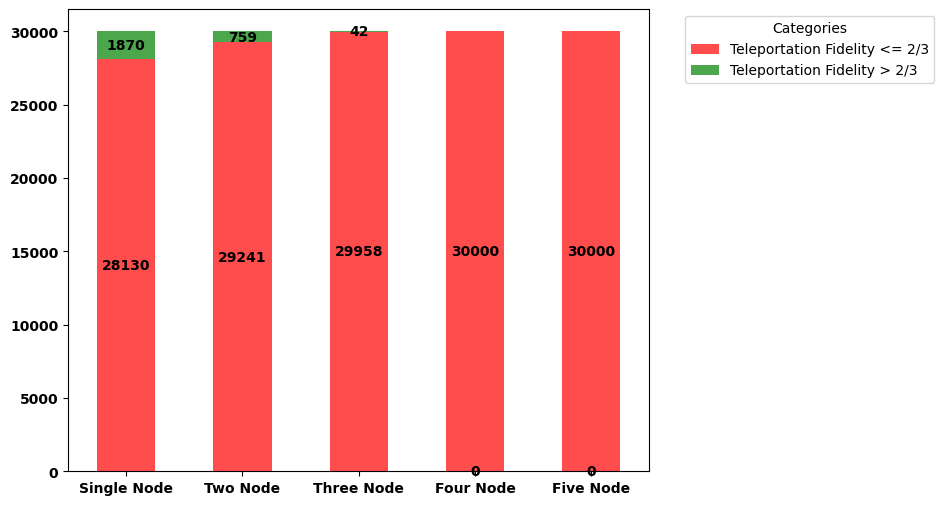}
    \caption{General Mixed State Multi Node perfect Swappings with different Parameters}
    \label{fig:mixed_state_multi_node_per_tf}
\end{figure}

\noindent In Fig.(\ref{fig:mixed_state_multi_node_per_tf}) X-axis represents the number of nodes that used swappings, and Y-axis represents the number of states considered for swapping. There are $30 \times 10^{3}$ states that are considered, and it can be observed that the teleportation fidelity with a value greater than $2/3$ (represented by the green region) can be observed till the seventh node swapping.

\subsubsection{Imperfect Measurements}

\noindent\textbf{Case I:} Consider $n+2$ nodes, a source node, $n$ repeater nodes, and a target node. After performing entanglement swapping at all the intermediate nodes with different visible parameters, the final fidelity between the source and target node, when the measurement is imperfect, can be observed in the below figure.

\begin{figure}[h]
    \centering
    \includegraphics[scale = 0.2]{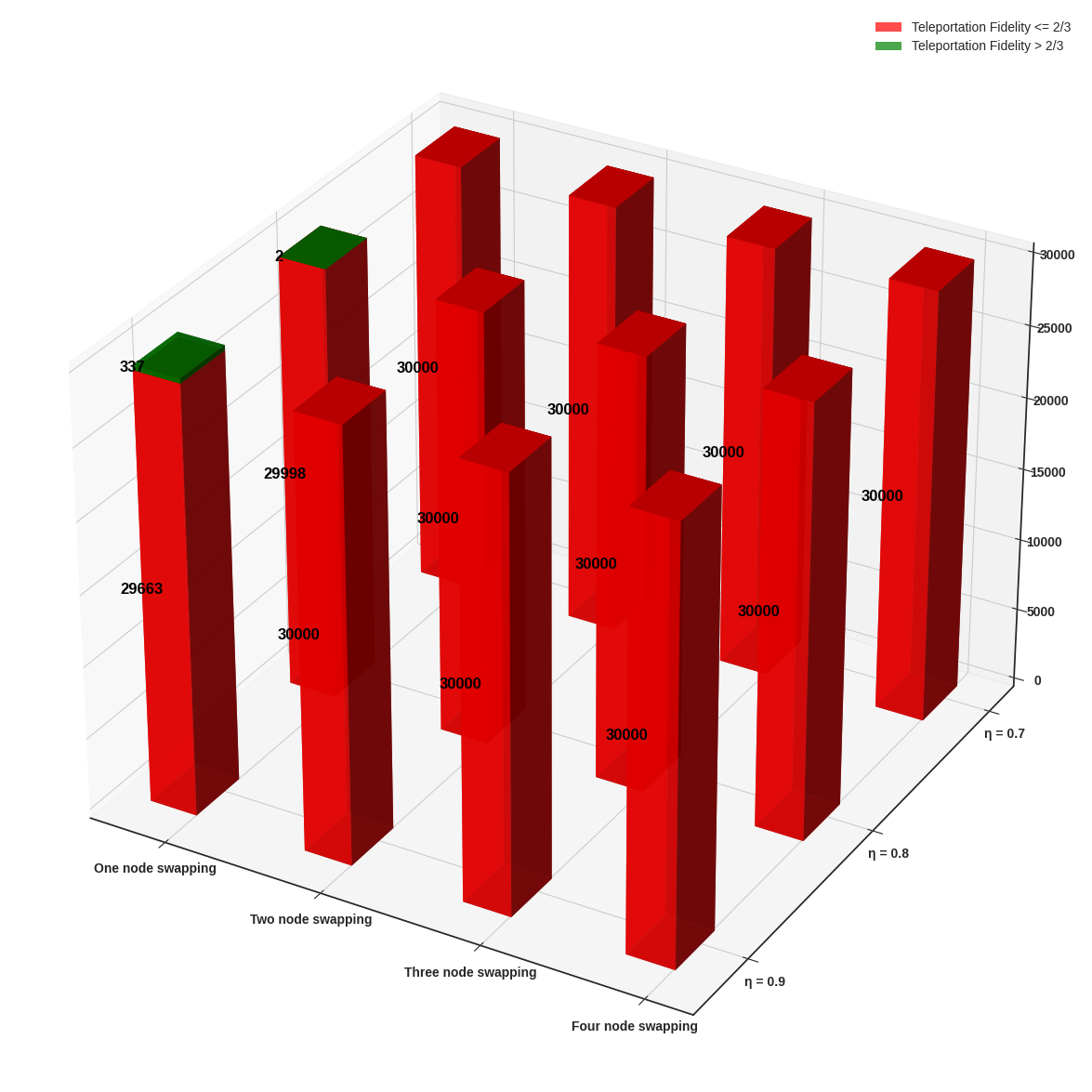}
    \caption{General Mixed State Multi Node Imperfect Swappings with same $\eta$}
    \label{fig:mixed_state_multi_same_imp_tf}
\end{figure}

\noindent In Fig.(\ref{fig:mixed_state_multi_same_imp_tf}) X-axis represents the number of nodes used for swappings, Y-axis represents $\eta$ from 0.0 to 0.9, and Z-axis represents the number of states considered for swappings. Teleportation fidelity is observed only for single node swappings for $\eta = 0.9$ and $\eta = 0.8$. For the rest of the swappings and $\eta$ fidelity is less than or equal to $\frac{2}{3}$ (represented by the red color).\\

\noindent\textbf{Case II:} Consider $n+2$ nodes, a source node, $n$ repeater nodes, and a target node. After performing entanglement swapping at all the intermediate nodes with different visible parameters, the final fidelity between the source and target node (when the measurement is imperfect and is different in different nodes) can be observed in the below figure.

\begin{figure}[h]
    \centering
    \includegraphics[scale = 0.34]{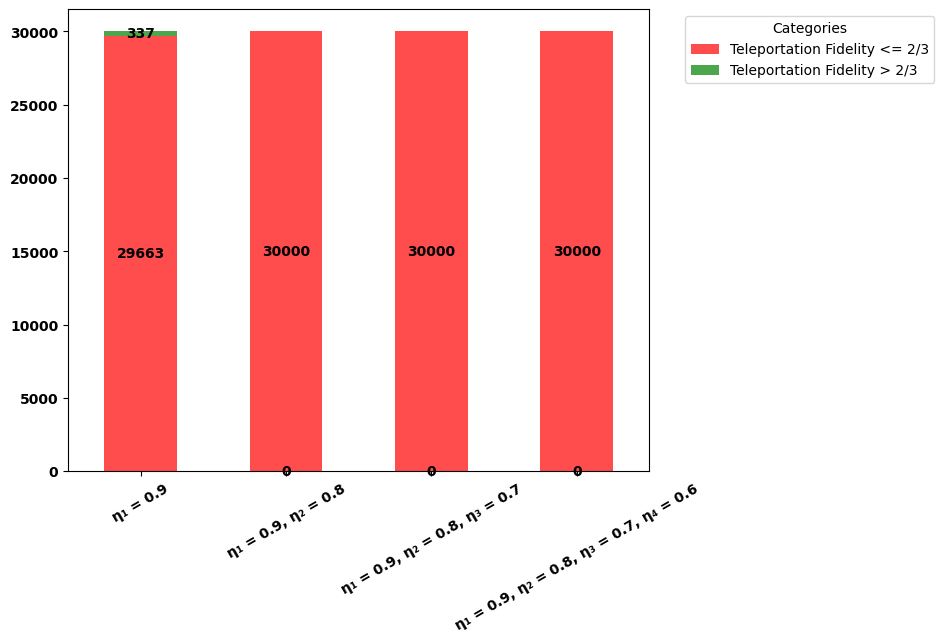}
    \caption{General Mixed State Multi Node Imperfect Swappings with different $\eta$}
    \label{mixed_state_multi_diff_eta_tf}
\end{figure}

\noindent In Fig.(\ref{mixed_state_multi_diff_eta_tf}) X-axis represents different imperfections clubbed with the number of nodes, and Y-axis represents the number of general mixed states considered for swappings. All the red regions represent the number of output states whose fidelity is less than or equal to $\frac{2}{3}$, and the green region represents the number of states whose fidelity is greater than $\frac{2}{3}$. It can be observed from the graph that only when $\eta = 0.9$ fidelity is greater than $\frac{2}{3}$ for some states.


\newpage
\section{Conclusion}
In a nutshell, here in this work, we have established how in a 1-D network with repeaters, the amount of entanglement (concurrences) present in the resource states contributes to the final entanglement between the initial and final node obtained after swapping. Not only that, we establish a vital relationship connecting the fidelities of teleportation of the resource states with the fidelity of the final state obtained as a result of entanglement swapping. We did not restrict ourselves to an idealistic scenario, where there is only perfect swapping but to the cases when we have imperfect swapping. These relations put bounds on the number of swappings and the success probability of measurements so that the final state is entangled and, at the same time, can be used as a teleportation channel. We have obtained these relationships using mixed entangled states like Werner states and Bell diagonal states as resources. Not only that, we have extended to the general two qubit mixed states to obtain a numerical prediction on how the concurrence and the teleportation fidelity change.



\end{document}